\documentclass[a4paper]{JHEP3}
\usepackage[dvips]{graphicx}
\usepackage{cite}
\usepackage{amssymb}
\usepackage{amsmath}
\usepackage{latexsym}
\usepackage{multirow}
\usepackage{xspace}
\usepackage{array}
\usepackage{url}
\usepackage{bbding}
\usepackage{afterpage}

\usepackage{tabularx}
\newcolumntype{L}[1]{>{\raggedright\arraybackslash}p{#1}} 
\newcolumntype{C}[1]{>{\centering\arraybackslash}p{#1}}

\newcommand{\SqrtS}{\sqrt{S}}
\newcommand{\alphas}{\alpha_s}
\newcommand{\mZ}{\ensuremath{M_{\mathrm{Z}}}}
\newcommand{\mW}{\ensuremath{M_{\mathrm{W}}}}
\newcommand{\TB}{\ensuremath{\tan\beta}\xspace}
\newcommand{\bbar}{\ensuremath{b\bar{b}}\xspace}
\newcommand{\mbeff}{\ensuremath{m_b^{\text{eff}}}\xspace}

\newcommand{\stau}{\ensuremath{{\tilde{\tau}}_{1}}\xspace}
\newcommand{\staui}{\ensuremath{{\tilde{\tau}}_{i}}\xspace}
\newcommand{\stauj}{\ensuremath{{\tilde{\tau}}_{j}}\xspace}

\newcommand{\tausneutrino}{\ensuremath{\tilde{\nu}_{\tau}}\xspace}
\newcommand{\supertau}{\ensuremath{\tilde\tau}\xspace}
\newcommand{\stautwo}{\ensuremath{{\tilde{\tau}}_{2}}\xspace}
\newcommand{\mstau}{\ensuremath{m_{\stau}}\xspace}
\newcommand{\mstautwo}{\ensuremath{m_{\stautwo}}\xspace}
\newcommand{\HH}{\ensuremath{H^{0}}\xspace}
\newcommand{\mHH}{\ensuremath{m_{H^{0}}}\xspace}
\newcommand{\hh}{\ensuremath{h^{0}}\xspace}
\newcommand{\mh}{\ensuremath{m_{h^{0}}}\xspace}
\newcommand{\staubar}{\ensuremath{{{\tilde{\tau}}}_1^{*}}\xspace}
\newcommand{\staustaubar}{\ensuremath{{{\tilde\tau}}_{1}^{\phantom{*}}\staubar}\xspace}
\newcommand{\hhiggs}{\ensuremath{{h}^0}\xspace}
\newcommand{\Hhiggs}{\ensuremath{{H}^0}\xspace}
\newcommand{\Ahiggs}{\ensuremath{{A}^0}\xspace}

\newcommand{\mtau}{\ensuremath{m_{\tau}}}
\newcommand{\mLL}{\ensuremath{m_{\mathrm{LL}}}\xspace}
\newcommand{\mRR}{\ensuremath{m_{\mathrm{RR}}}\xspace}

\newcommand{\thetastau}{\ensuremath{\theta_{{\tilde{\tau}}}}\xspace}
\newcommand{\Xtau}{\ensuremath{X_{\tau}}}

\newcommand{\Atau}{\ensuremath{A_{\tau}}\xspace}
\newcommand{\Atop}{\ensuremath{A_{{t}}}}
\newcommand{\Abottom}{\ensuremath{A_{{b}}}}
\newcommand{\stauR}{\ensuremath{{\tilde{\tau}_{\mathrm{R}}}}\xspace}
\newcommand{\stauL}{\ensuremath{{\tilde{\tau}_{\mathrm{L}}}}\xspace}
\newcommand{\stauROT}{\ensuremath{R_{\supertau}}\xspace}
\newcommand{\stauMAT}{\ensuremath{\mathcal{M}_{\supertau}^2}}

\newcommand{\couptriLR}[3]{\ensuremath{\tilde{C}[#1,#2,#3]}}
\newcommand{\couptri}[3]{\ensuremath{C[#1,#2,#3]}}

\newcommand{\sapb}{\ensuremath{s_{\alpha+\beta}}}

\newcommand{\sa}{\ensuremath{s_{\alpha}}}

\newcommand{\ca}{\ensuremath{c_{\alpha}}}

\newcommand{\cb}{\ensuremath{c_{\beta}}}

\newcommand{\stwoth}{\ensuremath{s_{2\thetastau}}\xspace}

\newcommand{\csqth}{\ensuremath{c^{2}_{\thetastau}}\xspace}
\newcommand{\ctwoth}{\ensuremath{c_{2\thetastau}}\xspace}
\newcommand{\stauY}{\ensuremath{Y_{\stau}}\xspace}

\newcommand{\squark}{\tilde{q}}

\newcommand{\St}{\tilde{t}}
\newcommand{\Sb}{\tilde{b}}
\newcommand{\cha}{\tilde{\chi}^{\pm}}
\newcommand{\neu}{\ensuremath{\tilde{\chi}^{0}}}
\newcommand{\gluino}{\tilde{g}}
\newcommand{\MeV}{{\rm Me\kern -1pt V}}
\newcommand{\GeV}{{\rm Ge\kern -1pt V}}
\newcommand{\TeV}{{\rm Te\kern -1pt V}}

\newcommand{\Lagrangian}{\ensuremath{\mathcal{L}}}
\renewcommand{\L}{\mathrm{L}}
\newcommand{\R}{\ensuremath{\mathrm{R}}}
\newcommand{\DY}{Drell--Yan}

\newcommand{\mhalf}{\ensuremath{m_{1/2}}\xspace}
\newcommand{\mzero}{\ensuremath{m_{0}}\xspace}
\def \msbar{\overline{\text{MS}}}
\def \drbar{\overline{\text{DR}}}

\def \BR{\text{BR}}

\newcommand{\be}{\begin{eqnarray*}}
\newcommand{\ee}{\end{eqnarray*}}
\newcommand{\bee}{\begin{eqnarray}}
\newcommand{\eee}{\end{eqnarray}}
\renewcommand{\eqref}[1]{eq.~(\ref{#1})}

\newcommand{\figref}[1]{figure~\ref{#1}}
\newcommand{\tabref}[1]{table~\ref{#1}}
\newcommand{\figsref}[2]{figures~\ref{#1} and \ref{#2}}

\newcommand{\ord}{{\cal O}}

\usepackage{cancel}
\usepackage{dcolumn}
\newcolumntype{d}[0]{D{.}{.}{-1}}

\newcommand{\eg}[0]{\textit{e.g.}\xspace}
\newcommand{\cf}[0]{\textit{cf.}\xspace}
\newcommand{\ie}[0]{\textit{i.e.}\xspace}

\newcommand{\TR}{\ensuremath{T_{\mathrm{R}}}}
\newcommand{\Tfreezeout}{\ensuremath{T_{\mathrm{f}}}}
\newcommand{\taustau}{\ensuremath{\tau_{\stau}}}
\newcommand{\Ystau}{\ensuremath{Y_{\stau}}}
\newcommand{\nstau}{\ensuremath{n_{\stau}}}
\newcommand{\seconds}{\ensuremath{\mathrm{s}}}
\newcommand{\gravitino}{\ensuremath{\widetilde{G}}}
\newcommand{\mgravitino}{\ensuremath{m_{\gravitino}}}
\newcommand{\mgluino}{\ensuremath{m_{\gluino}}}
\newcommand{\bbbar}{b\bar{b}}

\newcommand{\mstaustau}{m_{\stau\stau^*}}
\newcommand{\GammaHH}{\Gamma_{\Hhiggs}}
\newcommand{\MPlanck}{M_\mathrm{Pl}}
\newcommand{\fPQ}{f_\mathrm{PQ}}
\newcommand{\fbarn}{\mathrm{fb}}

\newcommand{\CL}{\mathrm{CL}}

\newcommand{\SPheno}[0]{\texttt{SPheno~3.0}\xspace}
\newcommand{\Prospino}[0]{\texttt{Prospino~2}\xspace}
\newcommand{\FeynArts}[0]{\texttt{FeynArts~3.6}\xspace}
\newcommand{\FormCalc}[0]{\texttt{FormCalc~7.0}\xspace}
\newcommand{\LoopTools}[0]{\texttt{LoopTools~2.6}\xspace}
\newcommand{\FeynHiggs}[0]{\texttt{FeynHiggs~2.7.4}\xspace}
\newcommand{\SuperISO}[0]{\texttt{SuperISO~3.0}\xspace}
\newcommand{\micromegas}[0]{\texttt{micrOMEGAs~2.4}\xspace}
\newcommand{\pT}{\ensuremath{p^\mathrm{T}}\xspace}

\newcommand{\OmegaDM}{\Omega_{\mathrm{DM}}}
\newcommand{\OmegaNeu}{\Omega_{\neu_1}}
\newcommand{\axino}{\widetilde{a}}
\newcommand{\maxino}{m_{\axino}}

\newcommand{\MGUT}{M_\mathrm{GUT}}

\newcommand{\monetwo}{m_{1/2}}
\newcommand{\tanbeta}{\tan\beta}
\newcommand{\meters}{\mathrm{m}}

\title{Direct stau production at hadron colliders\\
in cosmologically motivated scenarios}
\author{Jonas M.~Lindert, Frank D.~Steffen\\
 Max-Planck-Institut f\"ur Physik, 
 F\"ohringer Ring 6, 
 D-80805 M\"unchen, Germany\\
 Email: \email{lindert@mpp.mpg.de}, \email{steffen@mpp.mpg.de}}
\author{Maike K.~Trenkel\\
 Phenomenology Institute, Department of Physics, 
 University of Wisconsin-Madison, 
 1150 University Avenue, 
 Madison, Wisconsin 53706, USA\\
 Email: \email{trenkel@hep.wisc.edu}}
\abstract{
  We calculate dominant cross section contributions for stau pair
  production at hadron colliders within the MSSM, taking into account
  left-right mixing of the stau eigenstates.
  We find that $b$-quark annihilation and gluon fusion can enhance the
  cross sections by more than one order of magnitude with respect to
  the Drell--Yan predictions.
  These additional production channels are not yet included in the
  common Monte Carlo analysis programs and have been neglected in
  experimental analyses so far.
  For long-lived staus, we investigate differential distributions and
  prospects for their stopping in the collider detectors.
  New possible strategies are outlined to determine the mass and width
  of the heavy CP-even Higgs boson $H^0$.
  Scans of the relevant regions in the CMSSM are performed and
  predictions are given for the current experiments at the LHC and the
  Tevatron.
  The obtained insights allow us to propose collider tests of
  cosmologically motivated scenarios with long-lived staus that have
  an exceptionally small thermal relic abundance.
}
\preprint{arXiv:1106.4005 \\ MADPH-11-1571 \\ MPP-2011-67}

\begin{document}

\section{Introduction}
\label{sec:intro}

The ongoing experiments at the Large Hadron Collider (LHC) allow us to
enter new terrain at the $\TeV$ scale and to search for new physics in
an unprecedented way. In fact, due to the remarkable properties of
supersymmetric (SUSY) extensions of the Standard Model
(SM)~\cite{Wess:1992cp,Nilles:1983ge,Haber:1984rc,Martin:1997ns,Drees:2004jm,Baer:2006rs},
there are high hopes to discover superpartners of the SM fields in the
mass range probed by the LHC experiments. Such a discovery would be a
major breakthrough with far reaching consequences also for our
understanding of cosmology and the early Universe.  With the lightest
SUSY particle (LSP) being a promising dark matter (DM)
candidate~\cite{Jungman:1995df,Baltz:2006fm,Steffen:2008qp}, collider
studies may help us to clarify the origin and identity of DM and to
probe the early thermal history even prior to primordial
nucleosynthesis.

In this work we study the direct pair production of the lighter stau
$\stau$ at hadron colliders.  The \stau is the lighter one of the two
scalar partners of the tau lepton~$\tau$ and often the lightest
slepton within the minimal SUSY extension of the SM (MSSM).  Among the
potential SUSY discovery channels, the production of color-charged
SUSY particles, squarks and gluinos, is typically assumed to play a
key role since they can be produced via the strong interaction;
\cf~\cite{Martin:1997ns,Drees:2004jm,Baer:2006rs} and references
therein.  However, recent searches for signals with jets and missing
energy at the LHC~\cite{Khachatryan:2011tk,daCosta:2011qk} disfavor
very light squarks and gluinos.  In case these searches keep on in
setting new limits in the near future, the viable mass range for
squarks and gluinos will soon be pushed towards and above
$1~\TeV$~\cite{Bechtle:2011dm}, where the associated production cross
sections drop sharply, especially during the early run of the LHC with
a center-of-mass energy of $\SqrtS = 7~\TeV$.  However, color-singlet
SUSY particles such as sleptons, neutralinos, and charginos could
still be light enough to be produced in large numbers at colliders.
Direct stau production could thereby allow for SUSY discoveries in the
near future. On the other hand, non-observation of the direct
production channels will allow us to infer exclusion limits on subsets
of the SUSY parameter space, independent of the colored sector.

The lighter stau can play a crucial role not only for collider
phenomenology but also in cosmology. While the $\stau$ with its
electric charge cannot be DM, it can be the next-to-lightest SUSY
particle (NLSP) in an R-parity conserving realization of SUSY.  If the
lightest neutralino $\neu_1$ is the LSP, a stau NLSP that is almost as
light as the $\neu_1$ can participate via coannihilation in the
primordial freeze-out of the $\neu_1$. In fact, such a coannihilation
scenario might be the key for an agreement of the relic $\neu_1$
density $\OmegaNeu$ with the DM density $\OmegaDM$;
\cf~\cite{Jungman:1995df,Ellis:1999mm,Drees:2004jm,Baer:2006rs,Baltz:2006fm,Steffen:2008qp}
and references therein. The direct pair production of such light
$\stau$'s can then have a sizeable cross section at hadron colliders
even if the colored sparticles are substantially more massive. Since
each of those $\stau$'s will decay rapidly into a $\neu_1$, an excess
in missing transverse energy is the expected signature of such a
scenario.

There are other well-motivated scenarios in which the lighter stau
$\stau$ is central for both cosmology and phenomenology. In fact, in
large parts of the parameter space of many constrained SUSY models,
such as the constrained MSSM (CMSSM), the lighter stau $\stau$ is the
lightest SUSY particle in the MSSM spectrum, to which we refer as the
lightest ordinary SUSY particle (LOSP). While restrictive upper limits
exist on the abundance of a stable charged massive particle
(CHAMP)~\cite{Nakamura:2010zzi}, the $\stau$ LOSP becomes a viable
possibility in scenarios with broken
R-parity~\cite{Akeroyd:1997iq,Allanach:2003eb,Buchmuller:2007ui,Dreiner:2008rv,Desch:2010gi}
or in R-parity-conserving scenarios in which the LSP is an extremely
weakly interacting particle (EWIP), such as the
gravitino~\cite{Ambrosanio:1997rv,Feng:1997zr,Martin:1998vb,Ambrosanio:2000ik,Buchmuller:2004rq,Steffen:2006hw,Ellis:2006vu}
or the axino~\cite{Covi:2004rb,Brandenburg:2005he,Freitas:2011fx}. The
$\stau$ LOSP can then be long-lived and as such appear in the collider
detectors as a quasi-stable muon-like particle. Such scenarios will
come with distinctive signatures that are very different from those in
the $\neu_1$ LSP
case~\cite{Hamaguchi:2004df,Ellis:2006vu,Ishiwata:2008tp,Biswas:2009rba,Feng:2009bd,Ito:2009xy,Heckman:2010xz,Kitano:2010tt,Ito:2010xj,Ito:2010un,Asai:2011wy}.
In fact, direct $\stau$-pair production events will be easier to
identify experimentally if the $\stau$ is quasi-stable. It may even be
possible to stop initially slow staus within the main
detectors~\cite{Martyn:2006as,Asai:2009ka,Pinfold:2010aq,Freitas:2011fx}
or in some additional dedicated stopping
detectors~\cite{Goity:1993ih,Hamaguchi:2004df,Feng:2004yi,Hamaguchi:2006vu} for
analyses of their late decays.

The cosmological implications of a long-lived $\stau$ depend on its
lifetime~$\taustau$, its mass~$\mstau$, and its primoridal relic
abundance prior to decay~$\Ystau\equiv\nstau/s$, where $\nstau$ is its
comoving number density prior to decay and $s$ the entropy density.
For example, in the R-parity conserving case with an EWIP LSP, each
$\stau$ NLSP decays into one LSP which contributes to the DM density
$\OmegaDM$.  Moreover, $\taustau$ can exceed $1$--$100~\seconds$ in
scenarios with the gravitino LSP~\cite{Buchmuller:2004rq} or the axino
LSP~\cite{Covi:2004rb,Brandenburg:2005he,Freitas:2011fx}.  Long-lived
$\stau$'s can then decay during or after big bang nucleosynthesis
(BBN) and the emitted SM particles can reprocess the primordial
abundances of deuterium, helium, and
lithium~\cite{Cyburt:2002uv,Kawasaki:2004qu,Jedamzik:2006xz,Kawasaki:2008qe}.
Negatively charged $\stau$'s may even form bound states with the
primordial nuclei leading to catalyzed BBN (CBBN) of lithum-6 and
beryllium-9~\cite{Pospelov:2006sc,Cyburt:2006uv,Hamaguchi:2007mp,Pradler:2007is,Pospelov:2007js,Pospelov:2008ta}.
The DM abundance and the observationally inferred primordial
abundances of the light elements thereby impose $\taustau$-dependent
upper limits on the yield $\Ystau$ which translate into cosmological
constraints on the parameter space of the respective SUSY
models; \cf~\cite{Cyburt:2006uv,Steffen:2006wx,Pradler:2006hh,Pradler:2007is,Kersten:2007ab,Pradler:2007ar,Kawasaki:2008qe,Pospelov:2008ta,Bailly:2008yy,Freitas:2009fb,Freitas:2011fx} and references therein.

The cosmological constraints on scenarios with long-loved staus
$\supertau_\heartsuit$ are often quite restrictive with potentially
severe implications for collider phenomenolgy and cosmology.  For
example, the CBBN constraints on gravitino LSP scenarios can point to
heavy colored sparticles, such as a gluino with mass
$\mgluino\gtrsim 2.5~\TeV$~\cite{Cyburt:2006uv,Pradler:2006hh,Pradler:2007is,Pradler:2007ar},
so that direct stau production would be particularly important for
SUSY discoveries at the LHC.  Moreover, together with $\OmegaDM$
limiting the thermally produced EWIP
density~\cite{Bolz:2000fu,Brandenburg:2004du,Pradler:2006qh,Pradler:2006hh,Rychkov:2007uq,Strumia:2010aa,Bae:2011jb}
from above, the CBBN constraints can restrict the post-inflationary
reheating temperature to
$\TR \ll 10^9~\GeV$~\cite{Pradler:2006hh,Freitas:2009fb},
which disfavors the viability of thermal leptogenesis with
hierarchical heavy Majorana neutrino
masses~\cite{Fukugita:1986hr,Davidson:2002qv,Buchmuller:2004nz,Blanchet:2006be,Antusch:2006gy}.
In light of those findings, it is remarkable that parameter regions
exist in which one finds an exceptionally small relic $\stau$
abundance~\cite{Ratz:2008qh,Pradler:2008qc} which may respect the
(C)BBN limits on $\Ystau$ so that the above restrictions do no longer
apply. Such exceptional yields can result from efficient primordial
annihilation of staus with $\mstau\lesssim 200~\GeV$ via enhanced
stau-Higgs couplings~\cite{Ratz:2008qh,Pradler:2008qc} and/or stau
annihilation at the resonance of the CP-even heavy Higgs boson
$\HH$~\cite{Pradler:2008qc}. Here in this paper we address the
question whether such a scenario with efficient primordial stau
annihilation can be identified by considering direct $\stau$-pair
production at hadron colliders.

Our paper provides predictions for direct $\stau$-pair production
cross sections and kinematical distributions at hadron colliders in
the framework of the R-parity conserving MSSM. Our calculations
include \DY\ processes as well as $b$-quark annihilation and gluon fusion, 
where diagrams with s-channel \hh or \HH exchange can be become dominant.
All third generation mixing effects are taken into account. 
While the obtained cross sections are independent of the stau lifetime, we
interpret our findings with a special focus on scenarios with
long-lived staus. For such scenarios, we address collider prospects
for the SUSY parameter determination, the stopping of slow staus in
the detectors, and viability tests of exceptionally small relic stau
abundances.

Theoretical predictions for slepton-pair production via the \DY\ 
channel are well known since many years and include next-to-leading
order (NLO) corrections from QCD and
SUSY-QCD~\cite{Eichten:1984eu,Baer:1997nh,Beenakker:1999xh} and 
resummation improved results at the next-to-leading logarithmic (NLL) 
level~\cite{Bozzi:2006fw,Bozzi:2007qr,Bozzi:2007tea}. The NLO QCD and 
SUSY-QCD results are available via the software package \Prospino
\cite{Prospino}. Also contributions from \bbar annhilation and from
gluon fusion were studied in
refs.~\cite{delAguila:1990yw,Bisset:1996qh,Borzumati:2009zx} which
focussed on the limit of no left-right mixing. While this limit is
usually a good approximation for first and second generation sleptons,
the mixing between the (third generation) stau gauge eigenstates can
be substantial. Moreover, the $\bbar$ and $gg$ channels are not
included in the stau-pair production cross section predictions
provided by Monte Carlo simulation codes such as \texttt{Pythia}
\cite{Sjostrand:2000wi} or \texttt{Herwig} \cite{Corcella:2000bw}.
This provides additional motivation for our present study which
considers for the first time all three of the mentioned direct stau
production mechanisms with particular emphasis on potentially sizeable
left-right mixing.

The outline of this paper is as follows. In the next section we
elaborate on the motivation for exploring direct stau production and
the considered parameter regions. Section~\ref{sec:production}
presents our calculation of direct stau production at hadron
colliders. Here we show our cross section results for the Tevatron and
the LHC and compare the \DY\ contributions to those from
$\bbbar$-annihilation and gluon fusion.
Section~\ref{sec:phenolonglived} concentrates on scenarios with
long-lived staus and associated prospects for the SUSY parameter
determination and the stopping of staus. In section~\ref{sec:cmssm} we
consider the CMSSM for parameters for which exceptionally small
$\Ystau$ occur. Here we study representative CMSSM benchmark scenarios
also to explore the relative importance of direct stau production with
respect to the stau production in cascade decays.  In
section~\ref{sec:cmssm_exceptional} we explore ways to probe the
viability of an exceptionally small relic stau abundance at colliders.
As exemplary models we consider the CMSSM and a model with
non-universal Higgs masses (NUHM) to illustrate our main points. We
summarize our findings in section~\ref{sec:conclusion}.

\section{Motivation}
\label{sec:motivation}

In this section we continue to review the various implications of a
stau NLSP in collider phenomenology and cosmology. Focussing on
gravitino and axino dark matter scenarios with long-lived staus, we
address the appearance of such staus in the detectors, existing
limits, and the conceivable stopping of staus for studies of their
late decays. We describe the typical primordial stau LOSP freeze-out
and associated cosmological constraints. Considering the CMSSM with
the gravitino LSP, the constraints can point to heavy colored
sparticles and one may have to rely on direct stau production for a
SUSY discovery at the LHC.

We discuss the possibility of particularly efficient primordial stau
annihilation leading to an exceptionally small thermal relic stau
abundance such that restrictive cosmological constraints can be
evaded. We summarize the conditions required for such a behavior and
emphasize that the potential occurrence of a (color and) charge
breaking (CCB) vacuum, B-physics observables, and Higgs searches can
give restrictions in the relevant parameter regions.

Before proceeding let us comment briefly on direct stau production
events in the neutralino dark matter case. As already mentioned in the
introduction, the $\stau$ NLSP is an attractive possibility in the
$\neu_1$ LSP case and primordial $\neu_1$--$\stau$-coannihilation
processes may even turn out to be crucial for
$\OmegaNeu\simeq\OmegaDM$. In an R-parity conserving realization of
SUSY, we expect each directly produced $\stau\stau^*$ pair to
decay into a pair of unlike-sign $\tau$ leptons and a $\neu_1$ pair.
The emitted $\tau^+\tau^-$ pair would then help to identify direct
$\stau\stau^*$ production events experimentally. This identification
however requires sophisticated studies since the $\neu_1$'s and
the neutrinos from the rapidly decaying $\tau$'s will lead to missing
transverse energy. Leaving such studies for further work, we focus in
the following more on scenarios with a long-lived $\stau$ LOSP.
Nevertheless, our calculations of direct $\stau\stau^*$ production
cross sections (see section~\ref{sec:production}) apply also to
$\neu_1$ dark matter scenarios with the $\stau$ NLSP.

\subsection{Gravitino/axino dark matter scenarios with a long-lived stau NLSP}
\label{sec:longlivedstau}

Long-lived staus occur naturally in SUSY extensions of the SM in which
the LSP is an EWIP such as the gravitino or the axino. Both of those
EWIPs are viable DM candidates which can be produced thermally in the
hot primordial plasma with the right relic density,
$\Omega_{\mathrm{EWIP}}\simeq\OmegaDM$,
depending on the reheating temperature $\TR$ after
inflation~\cite{Bolz:2000fu,Brandenburg:2004du,Pradler:2006qh,Pradler:2006hh,Rychkov:2007uq,Strumia:2010aa,Bae:2011jb}.
Both are not part of the MSSM (and thus cannot be the LOSP)
but are still very well-motivated:%
\footnote{For simplicity, we discuss scenarios in which only the
  gravitino or only the axino is lighter than the stau.  There is also
  the possibility that the gravitino and axino are both simultaneously
  lighter than the stau, and we refer to
  refs.~\cite{Tajuddin:2010dr,Baer:2010gr,Cheung:2011mg,Freitas:2011fx} for studies of the
  cosmological and phenomenological implications.}
\begin{itemize}
\item The gravitino $\gravitino$ is the gauge field of local SUSY
  transformations and an unavoidable implication of SUSY theories
  including gravity~\cite{Wess:1992cp,Nilles:1983ge}. In the course of
  SUSY breaking, it acquires a mass $\mgravitino$ that can naturally
  be smaller than the one of the $\stau$ LOSP, for example, in
  gauge-mediated and gravity-mediated SUSY breaking
  scenarios~\cite{Martin:1997ns,Drees:2004jm,Baer:2006rs}.  The stau
  lifetime $\taustau$ is then governed by the two-body decay
  $\stau\to\tau\gravitino$ which involves a supergravity vertex and
  thereby the Planck scale
  $\MPlanck=2.4\times10^{18}~\GeV$~\cite{Buchmuller:2004rq}. Moreover,
  $\taustau$ depends sensitively on $\mgravitino$ and $\mstau$ and can
  easily exceed $100~\seconds$, \eg, for $\mgravitino\sim 1~\GeV$ and
  $\mstau\lesssim 300~\GeV$; \cf~eq.~(4.1) and figure~6 in
  ref.~\cite{Pospelov:2008ta}.
\item The axino $\axino$ is the fermionic superpartner of the axion
  and appears once the MSSM is extended by the Peccei--Quinn (PQ)
  mechanism in order to solve the strong CP problem.  Its mass
  $\maxino$ is model dependent and can be smaller than $\mstau$.  It
  is then the two-body decay $\stau\to\tau\axino$ which
  governs~$\taustau$ such that it depends on the axion model, the
  Peccei--Quinn scale
  $\fPQ\gtrsim 6\times 10^{8}~\GeV$~\cite{Raffelt:2006cw,Nakamura:2010zzi},
  $\maxino$, and
  $\mstau$~\cite{Covi:2004rb,Brandenburg:2005he,Freitas:2009fb,Freitas:2011fx}.
  Again $\taustau\gtrsim 100~\seconds$ is possible, \eg, for $\fPQ\sim
  10^{12}~\GeV$ and $\mstau\lesssim 300~\GeV$; \cf~eq.~(22) and
  figure~3 in ref.~\cite{Freitas:2009fb}.
\end{itemize}
In such settings, a directly produced stau will usually appear as a
stable particle in the detector. In fact, already a $\stau$ LOSP
lifetime as short as
$\taustau\approx 10^{-6}~\seconds$
is associated with a decay length of
$c\taustau\approx 300~\meters$
for which typically only a small fraction of the produced staus will
decay within the collider detectors. Such quasi-stable staus will look
like heavy muons with distinctive signatures in the collider
detectors~\cite{Drees:1990yw,Nisati:1997gb,Ambrosanio:1997rv,Feng:1997zr,Martin:1998vb,Fairbairn:2006gg,Feng:2010ij}
(see section~\ref{sec:phenolonglived}).

Direct searches for long-lived staus at the Large Electron
Positron (LEP) collider at CERN
set currently the following model-independent limit~\cite{LEP2staulimits,Nakamura:2010zzi}:%
\footnote{Note that~(\ref{eq:mstauLEPlimit}) is a conservative limit.
  Also the LEP limit $\mstau\gtrsim 97.5~\GeV$ can be
  found~\cite{LEP2staulimits,Nakamura:2010zzi}.}
\begin{align}
        \mstau\gtrsim 82~\GeV \ .
\label{eq:mstauLEPlimit}
\end{align}
Moreover, searches for long-lived staus in proton-antiproton
collisions with $\SqrtS=1.96~\TeV$ at the
Tevatron~\cite{Abazov:2008qu,Aaltonen:2009kea} have led to the
following upper limit on the stau production cross
section~\cite{Aaltonen:2009kea}:
\begin{align}
        \sigma(\SqrtS=1.96~\TeV) \lesssim 10~\fbarn \ ,
\label{eq_tevatronlimit}
\end{align}
with which we will compare our cross section results in
sections~\ref{sec:kinematicalcuts} and~\ref{sec:cmssm}.

\subsection*{Stopping of long-lived staus and studies of their late decays}

Quasi-stable staus may allow for other intriguing non-standard
collider phenomenology. Because of ionisation energy loss, staus will
be slowed down when traversing the detector material. In this way,
staus that are produced with a relatively small initial velocity of
\begin{align}
p_{\stau}/\mstau=\beta\gamma\lesssim 0.45
\label{eq:betagammacut}
\end{align}
are expected to get trapped, \eg, in the calorimeters of the ATLAS
detector~\cite{Asai:2009ka} or in some additional dedicated stopping
detector outside of the CMS detector~\cite{Hamaguchi:2006vu}. This can
allow for experimental studies of the stau decays. Measurements of
$\taustau$ could then probe the coupling strength that governs the
stau decays and thereby $\mgravitino$~\cite{Buchmuller:2004rq} or the
Peccei--Quinn scale $\fPQ$~\cite{Brandenburg:2005he,Freitas:2011fx}.
Moreover, one may be able to determine the mass of the EWIP LSP by
analyzing the kinematics of the mentioned two-body
decays~\cite{Buchmuller:2004rq,Brandenburg:2005he,Martyn:2006as,Hamaguchi:2006vu,Freitas:2011fx}.
In the gravitino LSP case, a kinematically determined $\mgravitino$
would allow us to test the Planck scale $\MPlanck$
microscopically~\cite{Buchmuller:2004rq} and to probe the reheating
temperature $\TR$ at colliders and thereby the viability of thermal
leptogenesis~\cite{Pradler:2006qh}. Also studies of three-body stau
decays are conceivable, which could give further insights into the
nature of the EWIP
LSP~\cite{Buchmuller:2004rq,Brandenburg:2005he,Hamaguchi:2006vu,Freitas:2011fx}.
The success of such studies will depend sensitively on the number of
staus that can be stopped for the analysis of their decays and thereby
on the initial velocity distribution. This motivates us to consider
such distributions in section~\ref{sec:stopping} below.

\subsection{Thermal relic abundances of long-lived staus and cosmological constraints}

Let us now turn to long-lived staus in the early Universe, the
potential cosmological implications and associated constraints. The
relic stau abundance $\Ystau$ prior to decay depends on $\mstau$, the
left-right mixing of the $\stau$ and other details of the SUSY model,
and on the early thermal history of the Universe. For a standard
thermal history with a reheating temperature $\TR$ that exceeds
$\mstau/20$, there was a period in which the $\stau$ LOSP was in
thermal equilibrium with the primordial plasma. At the freeze-out
temperature $\Tfreezeout\lesssim\mstau/20$, at which the $\stau$
annihilation rate equals the Hubble rate, the by then non-relativistic
staus $\stau$'s decouple from the thermal plasma. Taking into account
all possible (co-)annihilation channels, the associated Boltzmann
equation can be solved
numerically~\cite{Belanger:2001fz,Belanger:2010gh}. For a dominantly
right-handed stau, $\stau\simeq\stauR$, the resulting yield is found
to be governed mainly by $\mstau$,
\begin{align}
        \Ystau 
        \simeq 
        (0.4-2.0)\times 10^{-13}
        \left(\frac{\mstau}{100~\GeV}\right)
\label{eq:YstauR} , 
\end{align}
where larger values of the prefactor account for possible mass
degeneracies and associated effects such as stau-slepton or
stau-neutralino
coannihilation~\cite{Asaka:2000zh,Fujii:2003nr,Pradler:2006hh,Berger:2008ti}.

Confronting the yield~(\ref{eq:YstauR}) with the CBBN constraints
(shown \eg\ in figure~5 of ref.~\cite{Pospelov:2008ta}), the following
upper limit on the stau lifetime
emerges~\cite{Pospelov:2006sc,Pospelov:2008ta}
\begin{align}
\taustau \lesssim 5\times 10^{3}~\seconds \ .
\label{eq:taustauCBBN}
\end{align}
For the cases with the gravitino LSP and the axino LSP, this implies
$\mstau$-dependent upper limits on the gravitino mass $\mgravitino$
(\cf\ figure~6 in ref.~\cite{Pospelov:2008ta}) and the PQ scale $f_a$
(\cf\ figure~5 in ref.~\cite{Freitas:2009jb}), respectively. In
particular, a PQ scale $\fPQ$ around the scale of grand
unification $\MGUT\simeq 2\times 10^{16}\,\GeV$ is in conflict with
the CBBN constraint for the $\mstau$ range accessible at the
LHC~\cite{Freitas:2009jb}. Moreover, the $\mgravitino$ limit disfavors
the kinematical $\mgravitino$ determination~\cite{Steffen:2006wx} and
thereby both the mentioned $\MPlanck$
determination~\cite{Buchmuller:2004rq} and the probing of $\TR$ at
colliders~\cite{Pradler:2006qh}. These CBBN limits on $\mgravitino$
and $\fPQ$ tighten also the upper limits on the reheating temperature
$\TR$ imposed by
$\Omega_{\mathrm{EWIP}}\leq\OmegaDM$~\cite{Pradler:2006hh,Freitas:2009fb}.
The resulting $\TR$ limits can then be in considerable tension with
$\TR\gtrsim 10^9~\GeV$ required for viable thermal leptogenesis with
hierarchical heavy Majorana neutrino
masses~\cite{Fukugita:1986hr,Davidson:2002qv,Buchmuller:2004nz,Blanchet:2006be,Antusch:2006gy}.

For CMSSM scenarios with the gravitino LSP, the CBBN constraints have
been found to be particularly
restrictive~\cite{Cyburt:2006uv,Pradler:2006hh,Pradler:2007is,Kersten:2007ab,Pradler:2007ar}.
In the framework of the CMSSM, the gaugino masses, the scalar masses,
and the trilinear scalar interactions are assumed to take on the
respective universal values $\monetwo$, $\mzero$, and $A_0$ at
$\MGUT$. Specifying those values in addition to the mixing angle in
the Higgs sector $\tanbeta$ and the sign of the Higgs-higgsino-mass
parameter $\mu$, the low energy mass spectrum is given by the
renormalization group running from $\MGUT$ downwards. Here the $\stau$
LOSP case occurs in a large part of the parameter space and in
particular for $\mzero^2\ll\monetwo^2$. The CBBN
constraint~(\ref{eq:taustauCBBN})---which emerges for $\Ystau$ given
by~(\ref{eq:YstauR})---can then be translated into the following
$\mgravitino$-dependent limits~\cite{Pradler:2007is,Pradler:2007ar}:
\begin{eqnarray}
        & \monetwo 
        & \geq \,\, 
        0.9~\TeV \left(\frac{\mgravitino}{10~\GeV}\right)^{2/5} \ ,
\label{eq:monetwolimit}
\\
        & \TR 
        & \leq \,\, 
        4.9\times 10^7\,\GeV \left(\frac{\mgravitino}{10~\GeV}\right)^{1/5} \ ,
\label{eq:TRlimit}
\end{eqnarray}
where the latter accounts also for $\Omega_{\gravitino}\leq\OmegaDM$.
For $\mgravitino$ at the $\GeV$ scale, the lower
limit~(\ref{eq:monetwolimit}) then implies heavy colored sparticles
that will be difficult to probe at the LHC. As already mentioned
above, this provides additional motivation for this work since a SUSY
discovery could still be possible via direct stau pair production.

\subsection{Exceptionally small thermal relic stau  abundances}
\label{sec:exceptionalYstau}

The $\TR$ limit~(\ref{eq:TRlimit}) illustrates the mentioned tension
with thermal leptogenesis being a viable explanation of the baryon
asymmetry in the Universe.%
\footnote{In scenarios that are less constrained than the CMSSM,
the $\TR$ limit will be more relaxed if the ratio of the masses  
$\mgluino$ and $\mstau$ is smaller than $\mgluino/\mstau>6$ 
encountered in the $\stau$ LOSP region of the 
CMSSM~\cite{Steffen:2006wx}.}
In fact, this tension has motivated studies of scenarios with
non-standard thermal history in which $\Ystau$ is diluted by
significant entropy production after decoupling and before
BBN~\cite{Pradler:2006hh,Hamaguchi:2007mp,Hasenkamp:2010if} and
scenarios with R-parity violation~\cite{Buchmuller:2007ui} such
that~(\ref{eq:taustauCBBN}) is respected. Nevertheless, with a
standard thermal history and R-parity conservation, it has also been
found that SUSY models with enhanced stau-Higgs
couplings~\cite{Ratz:2008qh,Pradler:2008qc} and/or the pattern
$2\mstau\simeq\mHH$ between the mass of the stau \stau and the mass of
the heavy neutral CP-even Higgs $\HH$~\cite{Pradler:2008qc} can lead
to particularly efficient primordial stau annihilation and thereby to
exceptionally small yields
that may respect the CBBN constraint~\cite{Pospelov:2008ta}:%
\footnote{There are other cosmological constraints on $\Ystau$ in
  addition to the considered CBBN constraints, as briefly mentioned in
  section~\ref{sec:intro}. The $\Ystau$ limits imposed by these other
  constraints are typically at most equally restrictive and are
  usually evaded also for $\Ystau$
  satisfying~(\ref{eq:exceptionalYstau}); \cf\ section~1 in
  ref.~\cite{Pradler:2008qc}.}
\begin{align}
\Ystau\lesssim 2\times 10^{-15} \ .
\label{eq:exceptionalYstau}
\end{align}
For such a yield, the $\taustau$ limit~(\ref{eq:taustauCBBN}) is no
longer applicable and the larger values of $\mgravitino$ or $\fPQ$ are
not disfavored such that $\TR\gtrsim 10^9~\GeV$ and standard thermal
leptogenesis may be viable. For the $\gravitino$ LSP case, also the region 
$0.1\mstau\lesssim\mgravitino<\mstau$,
in which the kinematical $\mgravitino$
determination~\cite{Buchmuller:2004rq} is expected to be viable, is no
longer disfavored.

In light of these appealing features, let us recall some aspects of
the conditions that lead to enhanced stau-Higgs couplings; see also
refs.~\cite{Ratz:2008qh,Pradler:2008qc}. (For details and notations of
couplings we refer to appendix~\ref{app:staus}.) The stau-Higgs
couplings are governd by $\tanbeta$, $\mu$, and the trilinear coupling
$\Atau$ in the stau sector. These parameters determine also the
admixture of the left-handed and right-handed gauge eigenstates,
$\stauL$ and $\stauR$, in the lighter stau mass eigenstate
\begin{align}
        \stau=\cos\thetastau\stauL+\sin\thetastau\stauR \ .
\label{eq:StauMixing}
\end{align}
Thereby, there is a relation between the size of the stau-Higgs
couplings and the stau mixing angle $\thetastau$. This becomes most
explicit in the decoupling limit~\cite{Gunion:2002zf} in which the
light CP-even Higgs boson $\hh$ is much lighter than $\Hhiggs$ and the
CP-odd Higgs boson $\Ahiggs$. Here one finds that the $\stau\stau\hh$
coupling is proportional to $\sin 2\thetastau$ and the off-diagonal
term in the stau-mass squared matrix, $\Xtau=\Atau-\mu\tanbeta$, while
the $\stau\stau\Hhiggs$ coupling is found to be proportional to
$(\Atau\tanbeta-\mu)\sin 2\thetastau$. Thus, the absolute value of
these couplings becomes maximal for $\thetastau\to\pi/4$, which
corresponds to maximal left-right mixing in~(\ref{eq:StauMixing}), and
sizeable for large $\tanbeta$ and large absolute values of $\mu$
and/or $\Atau$. In the corresponding parameter regions, on which we
focus in this work, one then finds enhanced stau-Higgs couplings and
the mentioned efficient stau annihilation that can lead
to~(\ref{eq:exceptionalYstau}).

Here one has to stress that additional theoretical constraints might
become important in regions with large stau-Higgs couplings. These
regions can be associated with unwanted CCB minima in the scalar MSSM
potential~\cite{Pradler:2008qc,Ratz:2008qh,Endo:2010ya,Hisano:2010re}.
Our electroweak vacuum is then only a local minimum and as such
metastable. This will still be a viable scenario if the quantum
transition rate to the CCB minimum is so small that the lifetime of
our electroweak vacuum exceeds the age of the Universe. By studying
the decay of the electroweak vacuum with the usual `bounce
method'~\cite{Linde:1981zj} in an effective potential
approach~\cite{Espinosa:1996qw}, it has been found from a fit in the
relavant paramter space that a viable scenario has to respect the
following metastability condition~\cite{Hisano:2010re}:
\begin{align}
        \mu \TB 
        < 
        76.9\,\sqrt{m_{\tilde L_3}m_{\tilde E_3}} 
        + 38.7\,(m_{\tilde L_3} + m_{\tilde E_3}) 
        - 1.04\times 10^{-4}~\GeV \, , 
\label{eq:CCB}
\end{align}
where $m_{\tilde L_3}$ and $m_{\tilde E_3}$ are respectively the
left-handed and right-handed stau soft-breaking masses. We have
checked this condition by explicitly constructing the bounce action
for several parameter points and agree within the uncertainty given
in~\cite{Hisano:2010re}. However, this condition is not as rigid as,
for example, bounds from direct SUSY particle searches or flavor
changing decays since only the exponential contribution to the decay
of the electroweak vacuum can be evaluated easily, while a calculation
of the full width of the decay into the CCB minimum is highly
non-trivial. Nevertheless, scenarios in which an exceptional
yield~(\ref{eq:exceptionalYstau}) results from enhanced stau-Higgs
couplings only~\cite{Ratz:2008qh,Pradler:2008qc}, can be disfavored by
the CCB constraint~(\ref{eq:CCB}) if taken at face value (\cf\ 
sections~\ref{sec:cmssm} and~\ref{sec:cmssm_exceptional} and ref.~\cite{Endo:2010ya}).%
\footnote{In a recent updated study \cite{Endo:2011uw} the authors of
  \cite{Endo:2010ya} also included collider implication of the
  cosmological motivated $2\mstau\approx\mHH$ resonance region.
  However, they did not consider the additional $\bbar$ and $\gg$ direct production
  channels discussed in this work.}

In scenarios with $2\mstau\simeq\mHH$, primordial stau annihilation
can proceed efficiently via the $\Hhiggs$ resonance and thereby lead
to an exceptionally small $\Ystau$. Here the annihilation channel
$\stau\stau^*\to\bbbar$ turned out to be the most relevant one, which
also benefits from enhanced stau-Higgs couplings. However, because of
the $\Hhiggs$ resonance, an exceptional
yield~(\ref{eq:exceptionalYstau}) is already possible with more
moderate values of $\tanbeta$, $|\mu|$, and
$|\Atau|$~\cite{Pradler:2008qc}. Thereby, such scenarios can lead
to~(\ref{eq:exceptionalYstau}) and still respect the discussed CCB
constraints.

There are additional constraints from B-physics observables and Higgs
searches, which can become relevant in parameter regions with sizeable
$\tanbeta$. In particular, the non-observation of the decay
$B_s\to\mu^+\mu^-$ provides an upper limit on the corresponding
branching ratio \cite{Asner:2010qj}
\begin{align}
\BR(B_s \to \mu^+\mu^-) < 4.3\times 10^{-8}~\mbox{@}~95\%~\CL \ ,
\label{eq:Bmumu}
\end{align}
which sets stringent limits on the relevant parameter space. Also the
measurement of \cite{Asner:2010qj}
\begin{align}
\BR(b \to s \gamma) = (3.55 \pm 0.33) \times 10^{-4}
\label{eq:b2sgamma}
\end{align}
%
can give relevant constraints. Furthermore, there are constraints on
the Higgs sector of the MSSM in scenarios with large \TB and small
$m_{A^0}$ from Higgs searches in the $\tau\bar{\tau}$ and $\bbar$
channels. Most stringent limits are set recently by the LHC
experiments~\cite{Schumacher:2011jq,Chatrchyan:2011nx}. The study in
ref.~\cite{Chatrchyan:2011nx}, \eg, excludes $m_{A^0}\lesssim
280~\GeV$ for $\TB\gtrsim 50$ in the $m_h^{\mathrm{max}}$ benchmark
scenario (defined \eg\ in~\cite{Chatrchyan:2011nx}) in the $\tau\bar{\tau}$ channel. 
However, as shown in section~\ref{sec:cmssm_exceptional}, scenarios 
with resonant primordial stau annihilation leading 
to~(\ref{eq:exceptionalYstau}) can respect these B-physics and collider 
constraints as well.

In this paper, we investigate whether it can be possible to find
manifestations of an exceptionally small
yield~(\ref{eq:exceptionalYstau}) when studying the direct production
of quasi-stable staus in current collider experiments. As we will see
in the next sections, for this purpose it is crucial to consider not
only the \DY\ process but to also include the additional channels from
$\bbar$ annihilation and $gg$ fusion in the cross section calculation.

\section{Direct production of stau pairs at hadron colliders}
\label{sec:production}
\label{sec:stauprod}

In this section we calulate the cross section for direct stau pair
production at hadron colliders. We describe the relevant production
channels and the methods used in our calculations. Numerical results
are shown to illustrate the dependence on the SUSY parameters and to
provide predictions for the Tevatron and the LHC. The obtained cross
sections are independent of the stau lifetime.

Within the MSSM, stau pairs can be produced directly at hadron
colliders,
\begin{align}
pp(p\bar{p}) \to \staui^{}\stauj^*,
\end{align}
where $\tilde{\tau}_{i,j}$ denotes any of the two stau mass
eigenstates. After electroweak symmetry breaking the soft-breaking
terms in the MSSM Lagrangian induce a mixing among the particles of
identical color and electric charge. In the sfermion sector,
left-handed and right-handed gauge eigenstates mix to form mass
eigenstates, see appendix~\ref{app:staus}. The mixing is proportional
to the mass of the SM partner fermion and can thus be sizeable for
sleptons of the third generation. In the following, we concentrate on
the production of the lighter $\staustaubar$ pairs.  Results for
${\tilde\tau}_2^{\phantom{*}} {\tilde\tau_2}^*$ and
${\tilde\tau_1}^{\phantom{*}} {\tilde\tau_2}^*$ production can be
obtained in close analogy. Their production cross sections, however,
are suppressed by the heavier ${\tilde\tau_2}$ mass.
%

\subsection[Direct $\tilde{\tau}_1 \tilde{\tau}_1^*$ production channels]
{\boldmath Direct $\stau\stau^*$ production channels}
\label{subsec:overview}

At hadron colliders typically the leading contribution to direct
$\staustaubar$ production arises from the $q\bar{q}$ induced \DY\ type
process at $\ord(\alpha^2)$, see \figref{fig:feynman_dybb}\,(a).
%
\FIGURE[t]{
\includegraphics{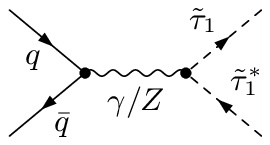}
\hspace*{2.5cm}
\includegraphics{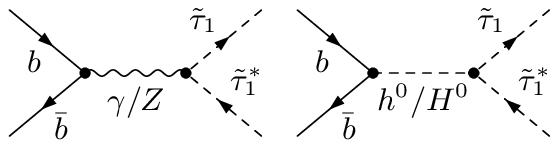}\\[1ex]
\small (a) \hspace*{6cm} (b) \hspace*{2cm}
\caption{Feynman diagrams for stau pair production 
  (a)~via the \DY\ process and (b)~via \bbar annihilation. Here, $q=
  u,d,c,s$.}
\label{fig:feynman_dybb}
}
%
The \DY\ production cross section depends only on the stau
mass $\mstau$ and the stau mixing angle $\thetastau$.

Stau pairs can also be produced from \bbar annihilation, mediated by
the neutral gauge bosons ($\gamma$, $Z$) and by the neutral CP-even
Higgs bosons ($h^0$, $H^0$), at the same order of perturbation theory.
The corresponding Feynman diagrams are displayed in
\figref{fig:feynman_dybb}\,(b). This channel is suppressed by the low
bottom-quark density inside protons, but can be enhanced by on-shell
Higgs propagators and by the bottom-Higgs and the stau-Higgs couplings
in certain regions of the SUSY parameter space. Note that the CP-odd
Higgs and Goldstone bosons, $A^0$ and $G^0$, do not couple to a
diagonal $\staui^{}\staui^*$ pair at tree-level and, in absence of CP
violating effects in the MSSM, there is also no induced mixing between
the CP-even and the CP-odd Higgs boson states at higher orders of
pertubation theory. The $A^0$ and $G^0$ bosons thus do not enter our
calculation.

Gluon-induced $\staustaubar$ production is only possible at the
one-loop level, mediated by a quark or squark loop, as shown in
\figref{fig:feynman_gluglu}.
%
\FIGURE[t]{
\includegraphics{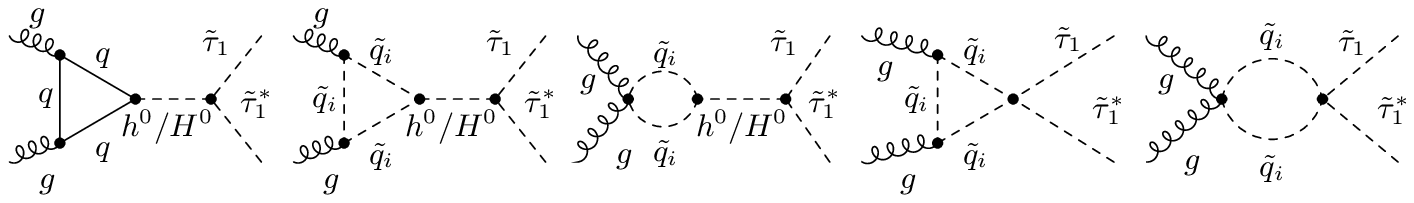}
\caption{Feynman diagrams for the gluon fusion contribution to stau pair production. 
  The quarks~$q$ and squarks $\squark_i$, $i=1,2$, running in the
  loops can be of any flavor.}
\label{fig:feynman_gluglu}
}
%
Even though these contributions are formally of higher orders,
$\ord(\alpha_s^2 \alpha^2)$, they can give sizeable contributions at
the proton-proton machine LHC at high center-of-mass energies where
the $gg$ luminosity is significantly higher than the
$q\bar{q}$~luminosity.

Let us note again that the additional $\bbar$ and $gg$ channels are
not included in the general purpose Monte Carlo event generators like
\texttt{Pythia} \cite{Sjostrand:2000wi} or \texttt{Herwig}
\cite{Corcella:2000bw}. We use the programs
\FeynArts~\cite{Hahn:2000kx} and \FormCalc with
\LoopTools~\cite{Hahn:1998yk} to generate and calculate the amplitudes
corresponding to the Feynman diagrams of
\figsref{fig:feynman_dybb}{fig:feynman_gluglu}. The Higgs boson masses
and the $H^0$ width are computed with \FeynHiggs~\cite{Frank:2006yh}.
We include QCD and SUSY-QCD corrections at NLO for the \DY\ channel
predictions calculated with \Prospino, by scaling our cross sections
with the respective $K$-factors,
$K\equiv\sigma_{\mathrm{NLO}}/\sigma_{\mathrm{LO}}$.
Furthermore, we use a resummed effective $\bbar h^0/\bbar H^0$ vertex
for the gluon fusion and \bbar contributions, as explained below and
in appendix~\ref{app:resum_b}.

We do not include higher-order QCD and SUSY-QCD corrections to the
Higgs-mediated channels. These are expected to be positive and similar
to the results for on-shell Higgs production,
see~\cite{Dittmaier:2011ti} and references therein. In this way our
analysis gives a conservative estimate of the enhancement effects from
the \bbar and gluon fusion production channels. Note also that
additional contributions to the direct $\staustaubar$ production can
arise from $W^{+}W^{-}$ fusion. Those however would be smaller by at
least one order of magnitude compared to the other
channels~\cite{delAguila:1990yw} and are not included in our analysis.

As motivated in section~\ref{sec:motivation}, we are particularly
interested in parameter regions with enhanced stau--Higgs couplings
and thus typically in scenarios with large \TB. It has been known for
a long
time~\cite{Hall:1993gn,Hempfling:1993kv,Carena:1994bv,Pierce:1996zz,Carena:1999py}
that radiative corrections to the $\bbar h^0/\bbar H^0$ vertex can be
important especially for large \TB and drive down the cross section
compared to the tree-level result. As shown in
\cite{Carena:1999py,Heinemeyer:2004xw} the leading \TB-enhanced
corrections can be resummed to all orders in pertubation theory by
using an appropriate effective bottom-quark mass, \mbeff, and
effective $\bbar h^0/\bbar H^0$ couplings. We adopt this approach, as
explained in detail in appendix \ref{app:resum_b}.

At hadron colliders, the gluon-fusion and \bbar-annhilation processes
with an $s$-channel Higgs boson can become resonant in regions of the
SUSY parameter space in which the Higgs boson is heavier than the two
produced staus. For intermediate $\stau$ masses respecting the robust
LEP limit $\mstau\geq 82~\GeV$~\cite{Nakamura:2010zzi}, the lighter
CP-even Higgs boson, $h^0$, is expected to be too light to go on-shell
($m_{h^0} < 140~\GeV$, \eg, \cite{Djouadi:2005gj}).  This is different
for the heavier $H^0$ boson. In parameter regions with $m_{H^0} \ge 2
\mstau$ we therefore include the total decay width of the $\HH$ boson,
$\Gamma_\HH$ in the propagator,
\be
\frac{1}{p^2-\mHH^2} \longrightarrow
\frac{1}{p^2-\mHH^2+i\mHH\Gamma_{H^0}} \, .
\label{eq:higgswidth}
\ee
%

\subsection{Numerical results}
\label{subsec:numerical}

Let us now investigate direct $\staustaubar$ production at hadron
colliders numerically.  Our focus is on the impact of the
$\bbar$-annihilation and the gluon-fusion channels in comparison to
the \DY\ process.
 
The cross section for direct stau production depends mainly on
$\mstau$, $m_{H^0}$, $\TB$, and on $\thetastau$ (or equivalently on
$\mu$ and $A_{\tau}$). It also depends on the \HH~boson width,
$\Gamma_{\HH}$, and thus indirectly on the SUSY mass spectrum. In
addition, squark masses enter indirectly via the loops in the
gluon-fusion channel and, as does the trilinear coupling $A_t$ in the
stop sector, via the effective bottom couplings.

We use the following input parameters in our numerical study. As a
starting point, we choose a $\stau$-LOSP scenario with moderate squark
masses and a large stau--Higgs coupling, fixed by the following
soft-breaking parameters at the low scale:
\begin{align}
\begin{split}
M_1 &= M_2 = M_3 = 1.2~\TeV, \qquad
\Atop=\Abottom=\Atau=600~\GeV,
\\
m_{\tilde Q_i} &= m_{\tilde U_i} = m_{\tilde D_i} = 1~\TeV, \qquad
m_{\tilde L_{1/2}} = m_{\tilde E_{1/2}} =  500~\GeV,
\label{eq_SUSYinputs}
\end{split}
\end{align}
where $M_i$ denote the gaugino mass parameters associated with the SM
gauge groups U$(1)_{\mathrm{Y}}$, SU$(2)_{\mathrm{L}}$, and SU$(3)_c$,
$m_{\tilde Q_i}$ ($m_{\tilde U_i}$ and $m_{\tilde D_i}$) the
left-handed (right-handed) squark soft-breaking masses, $m_{\tilde
  L_{1/2}}$ ($m_{\tilde E_{1/2}}$) the left-handed (right-handed)
slepton soft-breaking masses for the first two generations, and
$\Abottom$ the trilinear coupling in the sbottom sector. If not stated
otherwise, we choose
\begin{align}
\begin{split}
\thetastau &= 45^{\circ}, \qquad 
\mstau = 200~\GeV, \qquad
\\ 
\TB &= 30, \qquad~
\mu = 500~\GeV, \qquad
m_A = 400~\GeV,
\end{split}
\label{eq_STAUinputs}
\end{align}
as inputs for the third-generation sleptons, as discussed in
appendix~\ref{app:staus}, and for the Higgs-sector, where $m_A$
denotes the mass of the CP-odd Higgs boson $A^0$. With these
parameters, the considered scenario falls into the decoupling limit of
the MSSM~\cite{Gunion:2002zf} where $\mHH \approx m_A$. Note that the
chosen value of \TB (and also of $\mu$) is rather moderate compared to
the cosmologically motivated scenarios considered in
ref.~\cite{Pradler:2008qc} and discussed in
section~\ref{sec:exceptionalYstau}.

From the input parameters~(\ref{eq_SUSYinputs})
and~(\ref{eq_STAUinputs}), we calculate the physical MSSM parameters
using tree-level relations for sfermions, neutralinos, and charginos.
Physical masses are then passed to \Prospino to calculate the \DY\ 
$K$-factors at NLO in QCD and SUSY-QCD. The NLO corrections to the \DY\ 
channel typically amount to $20$--$40\%$ in the considered parameter
space.

SM input parameters are chosen according to~\cite{Nakamura:2010zzi}
\begin{align}
M_Z & = 91.1876~{\GeV},& M_W & = 80.4248~\GeV,& G_F & =1.1664\times10^{-5}~\GeV,
\nonumber\\
m_b^{\msbar}(M_Z) & = 2.936~\GeV,&
m_t & = 173.1~{\GeV},&  m_{\tau} & = 1.776~{\GeV}.
\label{eq_SMinputs}
\end{align}
We include the MSTW08LO \cite{Martin:2009iq} set of parton
distribution functions (PDFs) and use the running strong coupling
constant $\alphas(\mu_R)$ they provide.  Factorization and
renormalization scales are set to the mass of the produced stau
$\mu_R=\mu_F=\mstau$.

At this point we want to mention again that our calculation of the
$\bbar$ and $gg$ channels is formally at LO.  By using an effective
bottom-quark mass, \mbeff, and effective $\bbar h^0/\bbar H^0$
couplings, however, the dominant \TB enhanced corrections are included
in our results. Nevertheless, we do not consider non-\TB enhanced
higher-order corrections, and the remaining renormalization and
factorization scale dependence yields a possibly large theoretical
uncertainty to our cross section predictions.  A more detailed study
at NLO would be desirable, taking also uncertainties due to the
dependence on the PDF set, and the bottom-quark PDF in particular,
into account.
 
In \figref{fig:crosssections1} we show the direct production cross
section for $\stau\stau^*$ pairs at the LHC with $\SqrtS=14~\TeV$ as a
function of (a)~\mstau, (b)~\thetastau, (c)~\mHH, and (d)~\TB.
%
\FIGURE[t]{
\includegraphics[width=.49\textwidth,]{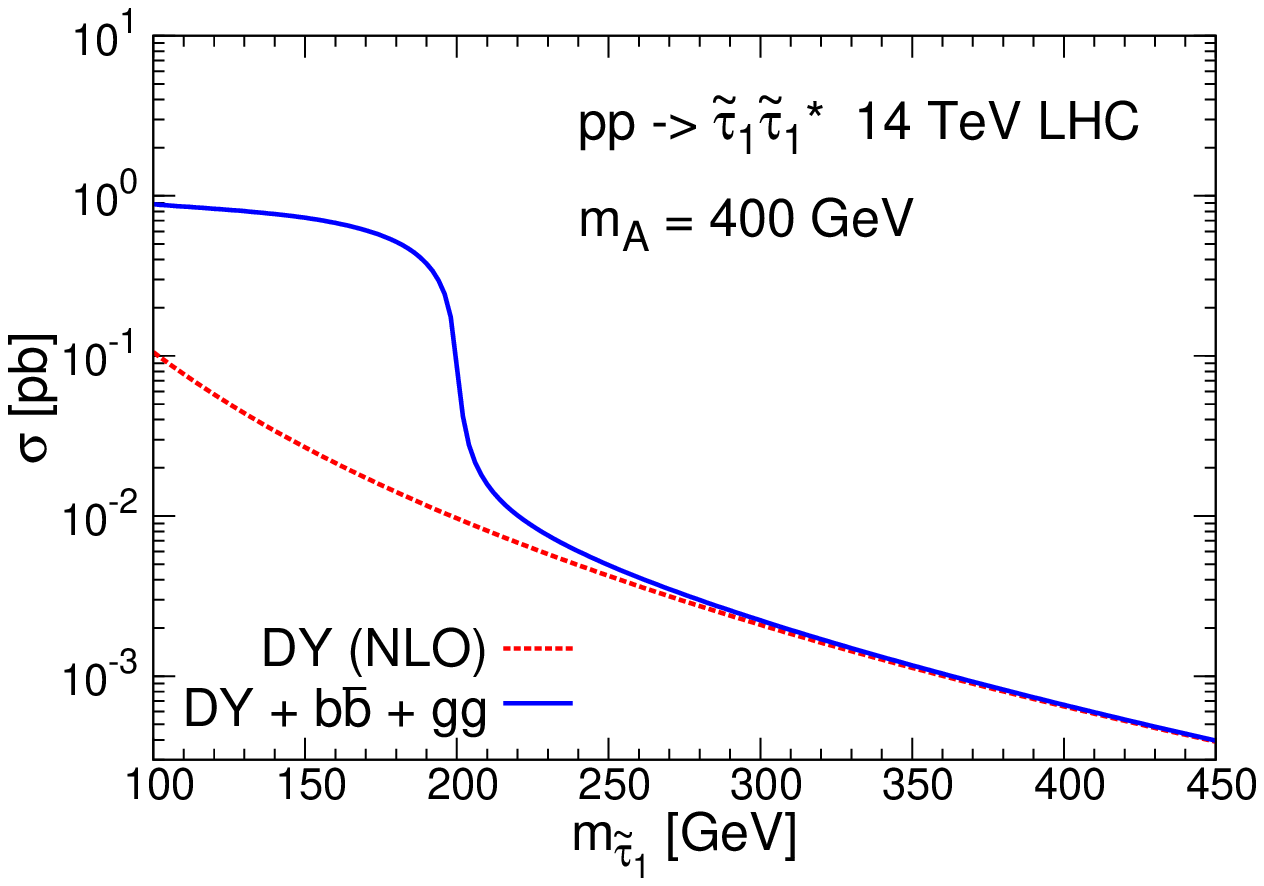}
\includegraphics[width=.49\textwidth,]{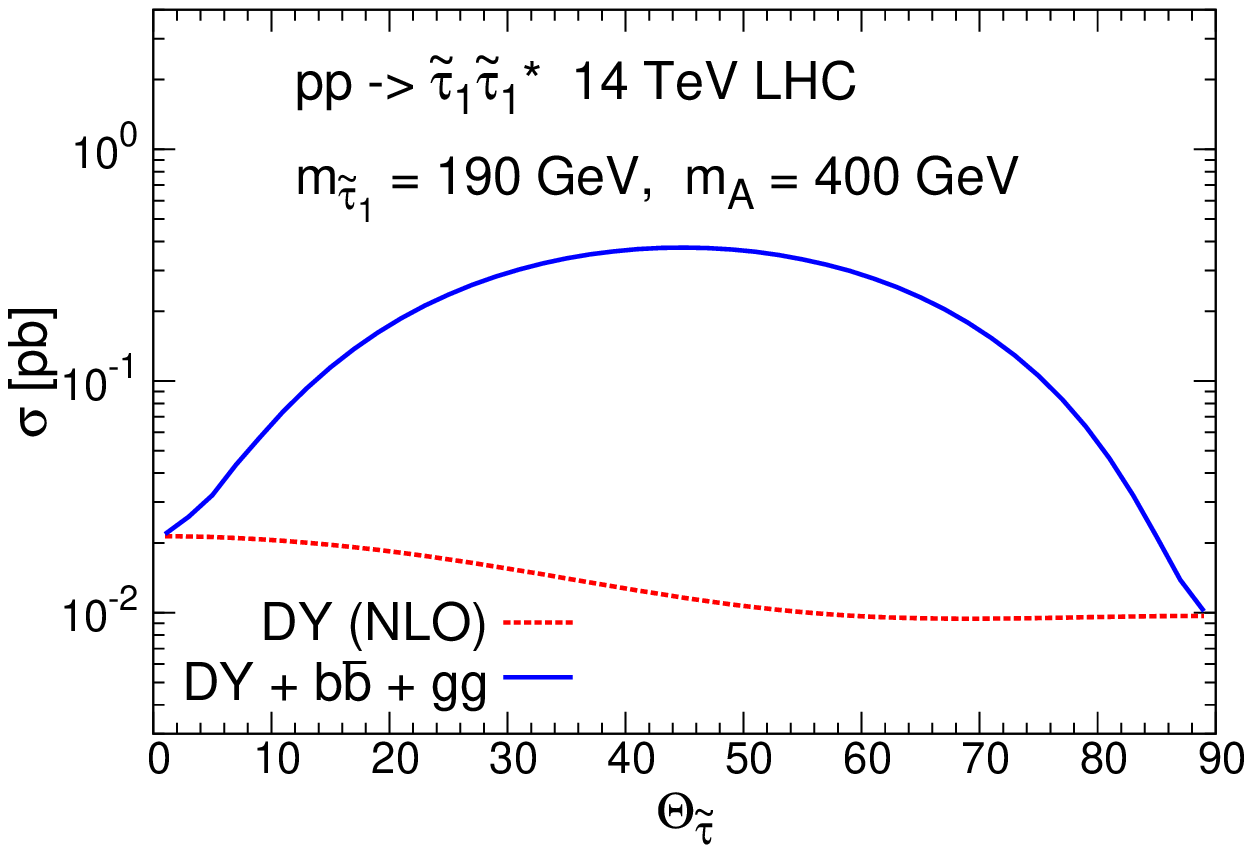}\\[.5ex]
{\small \hspace*{2cm} (a) \hspace*{.47\linewidth} (b)} \\[2.5ex]
\includegraphics[width=.49\textwidth,]{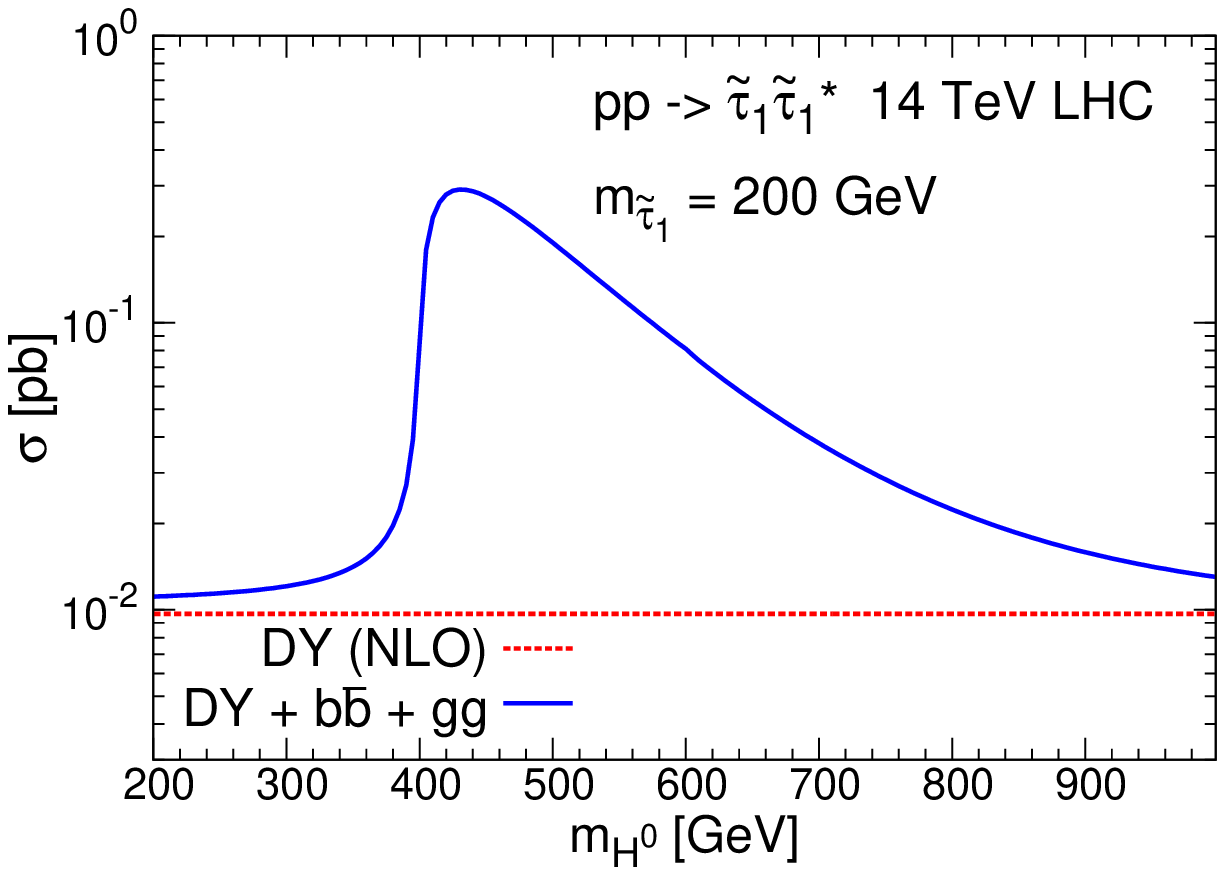}
\includegraphics[width=.49\textwidth,]{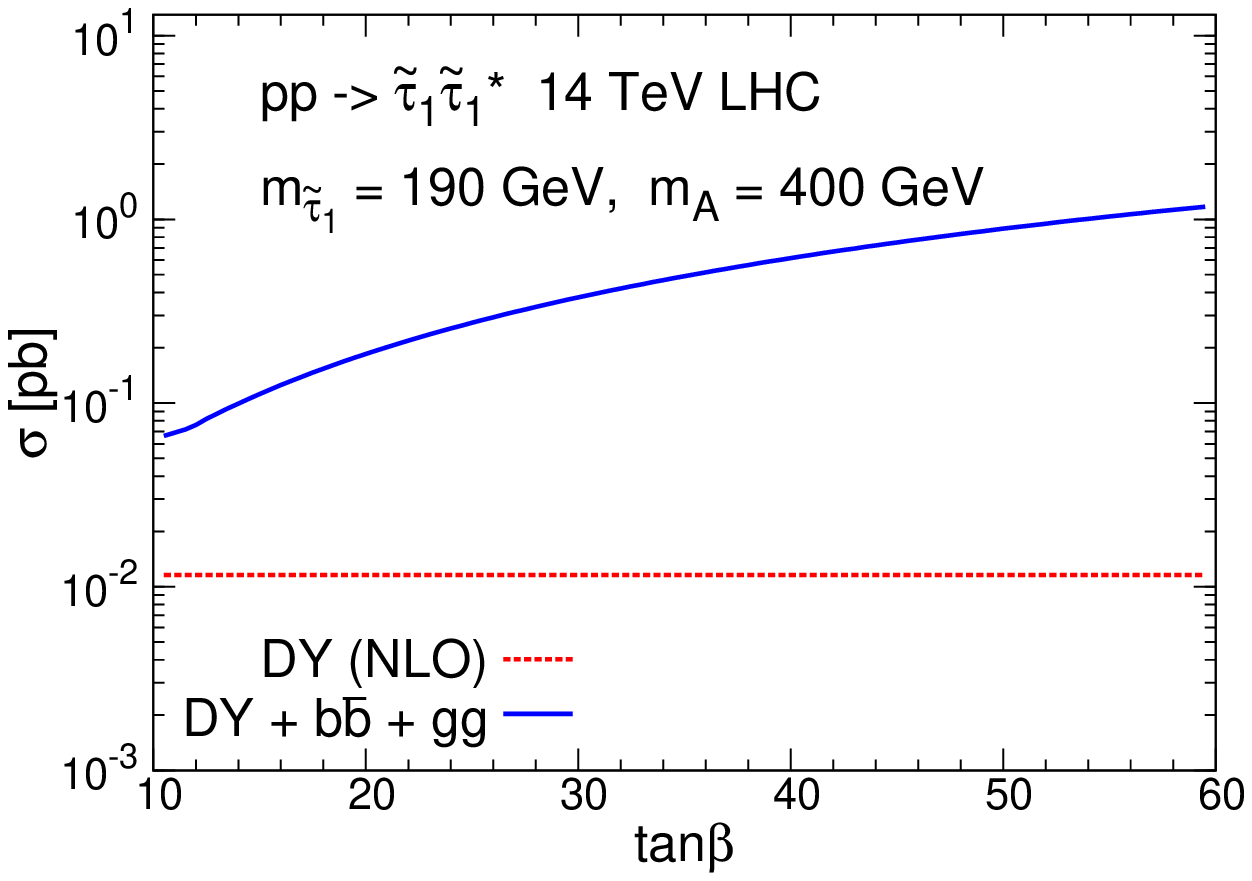}\\[.5ex]
{\small \hspace*{2cm} (c) \hspace*{.47\linewidth} (d)} \\[-1ex] 
\caption{Cross section of direct $\stau\stau^*$-pair production (solid lines, blue) 
  and the \DY\ prediction (dashed lines, red) at the LHC with
  $\SqrtS=14~\TeV$ as a function of (a)~\mstau, (b)~\thetastau,
  (c)~\mHH, and (d)~\TB. No kinematical cuts are applied. SUSY input
  parameters are as given in~(\ref{eq_SUSYinputs}) and
  (\ref{eq_STAUinputs}) if not varied or unless stated otherwise in
  the legend of the respective panel. Note that $\mHH \approx m_A$
  (decoupling limit) holds in most of the shown regions.}
\label{fig:crosssections1}
}
%
The remaining SUSY parameters are basically fixed according to
(\ref{eq_SUSYinputs}) and (\ref{eq_STAUinputs}). In figures
\ref{fig:crosssections1}\,(b) and~(d), we move to $\mstau=190~\GeV$,
where stau production is possible via on-shell \HH exchange. The
dashed (red) lines show the \DY\ (DY) cross section at NLO, whereas
the solid (blue) lines include the additional $\bbar$ and
$gg$~contributions.  The \DY\ cross section depends only on \mstau and
\thetastau, as already mentioned above. It decreases strongly for
increasing $\stau$ masses and varies with $\thetastau$ by a factor
that can be at most slightly larger than $2$. As shown in
\figref{fig:crosssections1}\,(b), this factor takes on its largest
value for $\thetastau \approx 0$, \ie, for an almost left-handed
\stau. The factor of $\approx2$ difference between the \DY\ cross
sections at $\thetastau \approx 0$ and $\thetastau \approx \pi/2$ is
determined mainly by different gauge couplings of the left-handed and
right-handed states.  It is almost independent of kinematics
and hardly effected when going from $\SqrtS=14~\TeV$ to $7~\TeV$.

The impact of the $\bbar$ and $gg$ channels depends strongly on the
mass hierarchy between \stau and \Hhiggs, as can clearly be seen in
figures~\ref{fig:crosssections1}\,(a) and~(c). If $\mHH > 2 \mstau$,
these additional channels can change the direct production cross
section by more than one order of magnitude with respect to the \DY\ 
result. At the threshold $\mHH=2\mstau$, the $\bbar$ and $gg$
contributions drop steeply and are only marginally important for
$\mHH\ll2\mstau$.

Figures~\ref{fig:crosssections1}\,(b) and (d) illustrate the
dependence of the total direct production cross section on the
parameters $\thetastau$ and $\TB$ that govern the stau-Higgs-coupling
strength. Here, the total direct production cross section is dominated
by on-shell Higgs production. Thus, the dependence on $\thetastau$ and
$\TB$ reflects the $\stau\stau^*\HH$ coupling, discussed in
section~\ref{sec:exceptionalYstau} and given in appendix
\ref{app:stauhiggs_couplings}. As shown there and illustrated here,
this coupling is basically proportional to $\sin2\thetastau$ and also
to $\TB$ (or more precisely to $\Atau\TB$). The additional
contributions from the $\bbar$ and $gg$ channels are tiny in cases of
very small mixing, $\thetastau\to 0, \pi$.
The exact position of the minimum depends on the relative importance of the 
$\stau\staubar\HH$ and $\stau\stau\hh$ couplings,
and can be slightly above/below $\thetastau = 0, \pi$, see also (\ref{eq:higgses-stau-stau-couplings-PHYS}).

In turn, they
become most important for maximal mixing, \ie, at
$\thetastau\approx\pi/4$. There, the additional contributions push up
the total direct production cross section by up to two orders of
magnitude for very large \TB and are already sizeable for small \TB.

Let us now turn to a scenario where the \HH is very heavy and thus
almost decoupled, $\mHH=1~\TeV$. We again investigate the dependence
of the total cross section on \thetastau and \TB, shown in
\figref{fig:crosssections2}.
%
\FIGURE[t]{
\includegraphics[width=.49\textwidth,]{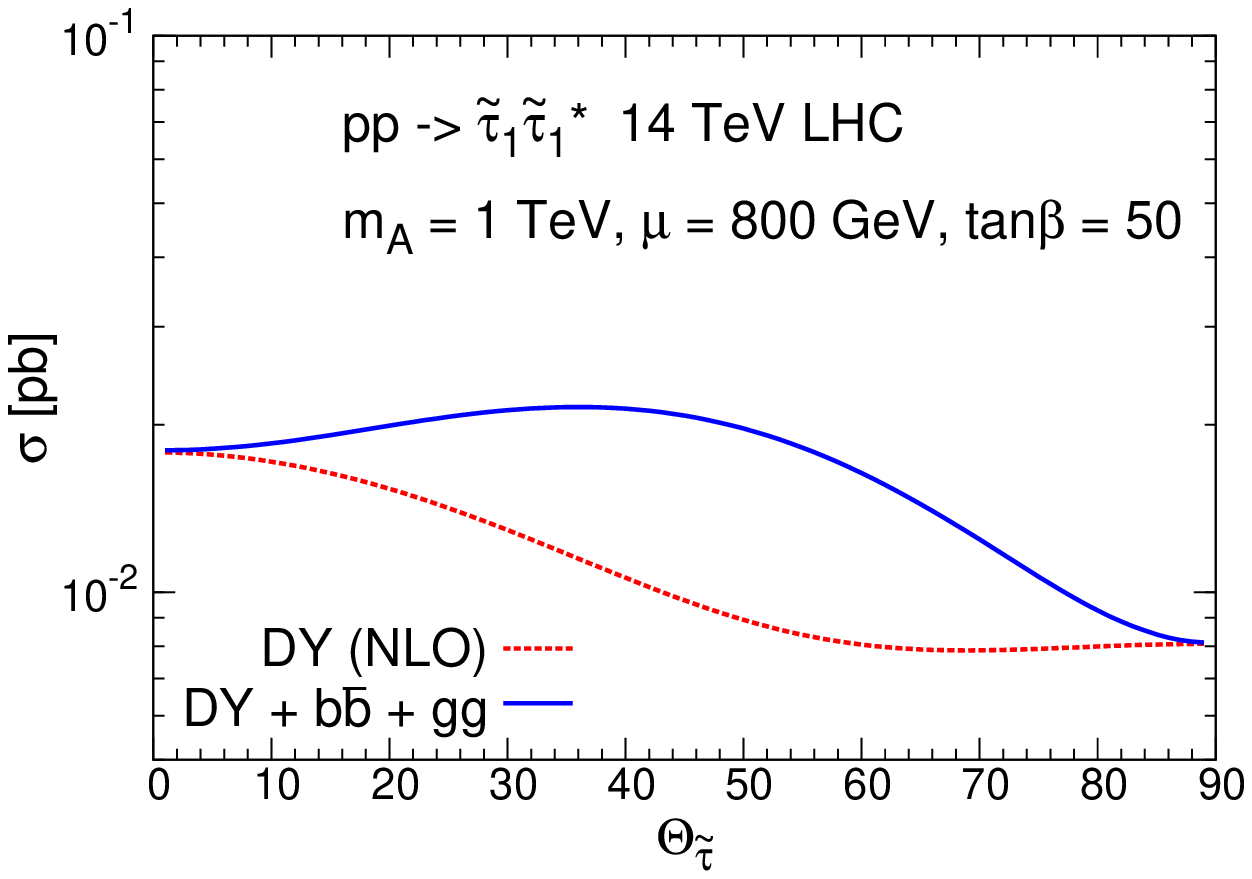}
\includegraphics[width=.49\textwidth,]{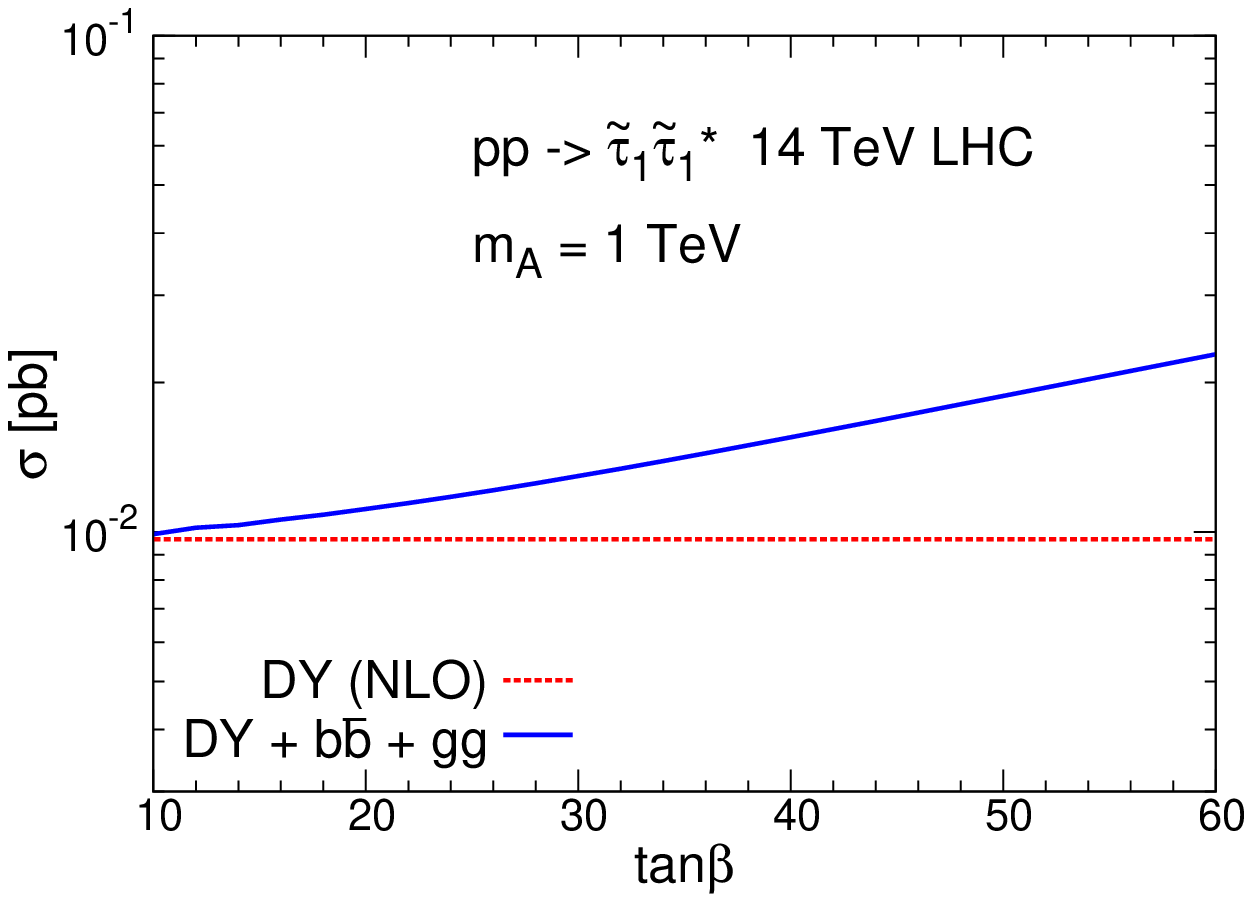}\\[.5ex]
{\small \hspace*{2cm} (a) \hspace*{.47\linewidth} (b)} \\[-1ex] 
\caption{Cross section of direct $\stau\stau^*$-pair production (solid lines, blue) 
  and the \DY\ prediction (dashed lines, red) at the LHC with
  $\SqrtS=14~\TeV$ as a function of (a)~\thetastau and (b)~\TB. No
  kinematical cuts are applied.  SUSY input parameters as given in
  (\ref{eq_SUSYinputs}) and (\ref{eq_STAUinputs}) if not varied or
  stated otherwise. Note that here $\mHH \approx m_A = 1~\TeV$
  (decoupling limit).}
\label{fig:crosssections2}
}
%
We focus on parameters that allow for enhanced $\stau\stau^*\hh$
couplings, \ie\ large values for $\mu$ and $\TB$.
In \figref{fig:crosssections2}\,(a) we consider $\mu=800~\GeV$ and
$\TB=50$ while the other parameters are fixed according to
(\ref{eq_SUSYinputs}) and (\ref{eq_STAUinputs}). Again, the
contribution from the additional $\bbar$ and $gg$ channels can be
sizeable. The enhancement amounts to a factor between two and three
when considering very large values of \TB and maximal mixing
$\thetastau \approx \pi/4$.  Here dominant contributions come
mainly from off-shell \hh exchange together with large stau-Higgs
couplings.  Thus, the relative importance between the $gg$ channel and
the $\bbar$ channel can be different compared to situations with
dominant contributions from on-shell \HH exchange since the two
Higgses couple differently to the quark and squark loops. Clearly, for a larger
contribution of the loop-induced $gg$ channel, a stronger dependence
on the squark masses is introduced in our calculation. This does not
only concern the overall mass scale but also the mass splitting
between the squarks, as is well known for on-shell Higgs production
via gluon fusion, see \cite{Djouadi:2005gj} and references therein.
For example, contributions from squark loops get small when squarks
within one generation are almost degenerate.  Additionally, slight
enhancements in the $gg$ channel can appear at thresholds where the
resonance requirement $2m_{\tilde q} \approx \mHH$ is fullfilled
between squarks in the loops and the heavy Higgs \HH, as shown in ref.
\cite{Borzumati:2009zx}. We want to note that, despite the large
couplings, all parameter points considered above are in agreement with
the CCB constraint~(\ref{eq:CCB}).

We summarize the potential impact of the $\bbar$ and $gg$ channels for
the SUSY scenario defined in (\ref{eq_SUSYinputs}) and (\ref{eq_STAUinputs})
again in \figref{fig:inclusive_compare},
%
\FIGURE[t]{
\includegraphics[width=.75\textwidth,]{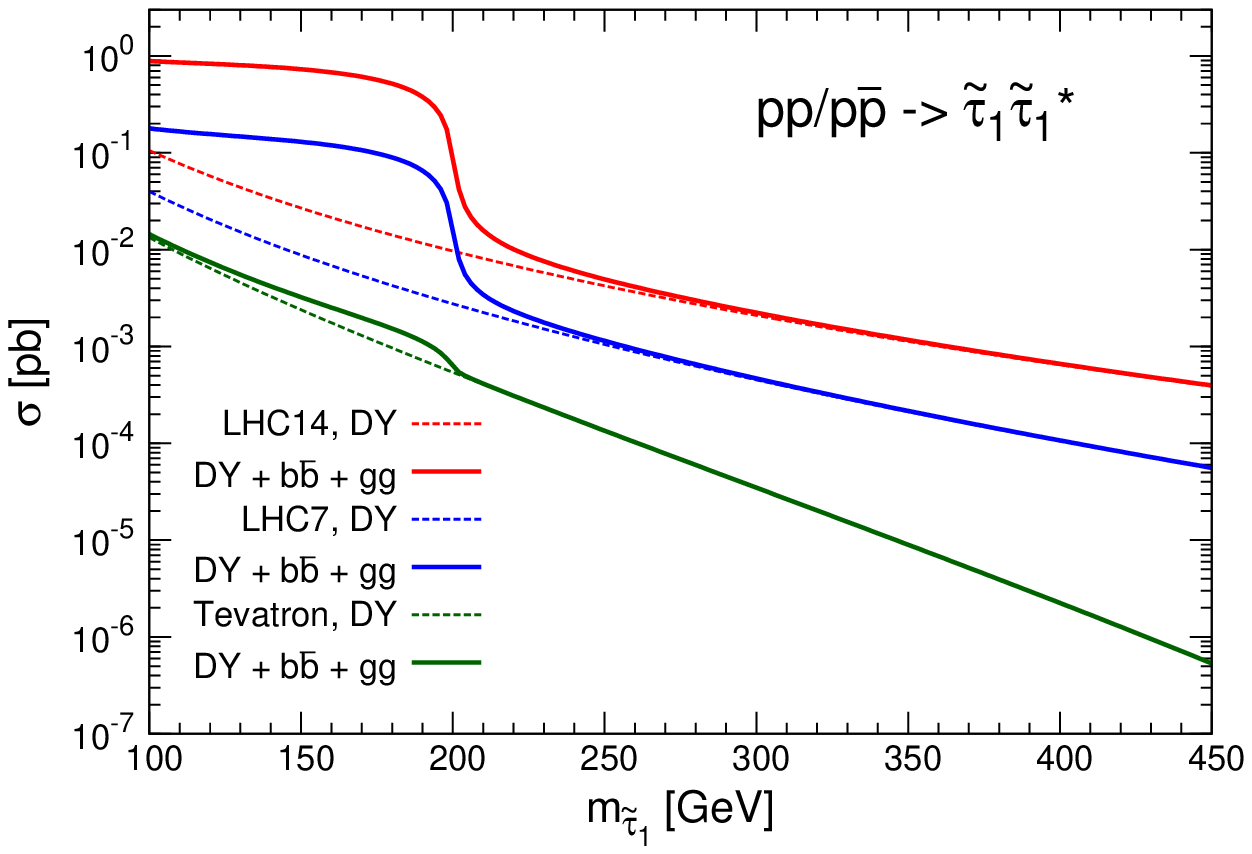}
\caption{Direct $\stau\stau^*$-pair production cross sections as a function 
  of $\mstau$ for the SUSY scenario with (\ref{eq_SUSYinputs}) and
  (\ref{eq_STAUinputs}) at the LHC with $\SqrtS=14~\TeV$ (top, red)
  and $7~\TeV$ (middle, blue) and the Tevatron with $\SqrtS=1.96~\TeV$
  (bottom, green). The \DY\ predictions are shown by the dashed lines
  and the full cross section, including $\bbar$ and $gg$ channels, by
  the solid lines.}
\label{fig:inclusive_compare}
}
%
where the \DY\ predictions (dashed lines) and the full cross sections
(solid lines) are shown for $\staustaubar$ production at the LHC for
$\SqrtS=14~\TeV$ (top, red) and for $7~\TeV$ (middle, blue). When
going down from $14~\TeV$ to $7~\TeV$, the cross section decreases by
up to about a factor of 5. The relative importance of the $\bbar$ and
$gg$ channels however are similarly important in the region where
on-shell \HH exchange contributes. Thus, for both $\SqrtS=7~\TeV$ and
$14~\TeV$, the $\bbar$ and $gg$ channels should not be neglected in a
precise cross section prediction. We also show the direct
$\staustaubar$ production cross section expected at the Tevatron with
$\SqrtS=1.96~\TeV$ (bottom, green). Due to the higher parton momentum
fractions $x$ needed at the Tevatron, the gluon and the $\bbar$
luminosities are reduced compared to the LHC case.  Accordingly, the
respective $gg$ and $\bbar$ contributions to direct stau production
are only small, as illustrated.

\section{Collider phenomenology with directly produced long-lived staus}
\label{sec:phenolonglived}

In this section we focus on scenarios with long-lived staus, \ie,
$\tau_{\stau}\gtrsim 10^{-6}~s$. At collider experiments, pair
production of long-lived staus will give a clear CHAMP signal in the
detectors if kinematical cuts are applied to discriminate between
signal and muon background. Here, we study the impact of these
kinematical cuts on the direct production cross section prediction and
differential distributions. We show that experimental observation of
direct stau production could provide important insights into the SUSY
model realized in nature. For particularly well-motivated cosmological
scenarios, we find that relatively large numbers of staus are expected
to get stopped already in the main detectors at the LHC. Thereby,
analyses of stau decays may become a viable tool to identify
the LSP and/or to probe high scales such as the Planck scale
$\MPlanck$ or the Peccei--Quinn scale $\fPQ$ in the laboratory.

\subsection{Kinematical cuts}
\label{sec:kinematicalcuts}

For a realistic experimental analysis, we need to include kinematical
cuts on the phase space of the staus to reduce possible backgrounds to
the CHAMP signal. The signature of a CHAMP traversing a detector is a
slowly moving minimal ionising particle with high transverse-momentum
$\pT$. In the experiments, this results in a long time-of-flight (TOF)
and an anomalously large ionization-energy loss rate
($dE/dx$)~\cite{Raklev:2009mg}. Since the CHAMP loses energy primarily
through low-momentum-transfer interactions, it will be highly
penetrating and will likely be reconstructed as a
muon~\cite{Aaltonen:2009kea}. At hadron colliders, experimental CHAMP
searches have been performed by the CDF~\cite{Aaltonen:2009kea} and
the D0~\cite{Abazov:2008qu} collaborations at the Tevatron and are
planned at the LHC in the near future~\cite{CMS-PAS-EXO-08-003}. In
accordance with those analyses, we apply the following kinematical
cuts on the produced staus:
\begin{align}
\begin{split}
\label{eq:cuts}
 \pT > 40~\GeV,\qquad &0.4 < \beta < 0.9,
\\
\vert \eta \vert < 0.7 ~\text{(Tevatron)},\qquad  
&\vert \eta \vert < 2.4 ~\text{(LHC)},
\end{split}  
\end{align} 
where the cuts have to be fulfilled by at least one of the \stau's.
Here $\eta = -\ln(\tan\theta/2)$ is the pseudo-rapidity and $\beta=
|\mathbf{p}|/E$ the stau velocity. This should reduce the background
from very slow moving muons to a negligible
level~\cite{DeRoeck:2005bw}.

For our theoretical predictions, we use the same inputs as in
section~\ref{sec:production}. In particular, we include the inclusive
NLO $K$-factors provided by \Prospino for the \DY\ channel, also when
cuts are applied. Since QCD and SUSY-QCD corrections only effect the
hadronic part of the considered \staustaubar production, the cut
dependence of the $K$-factors is expected to be small. Note that we
furthermore assume here direct $\staustaubar$ production to be the
dominant $\stau$ production source and do not include $\stau$'s
resulting from cascade decays in our signal definition. Otherwise, an
additional jet and/or lepton veto can be used to separate directly
produced staus from ones produced at the end of a decay chain. We will
briefly investigate the relative importance of these concurrent
production mechanisms for representative CMSSM benchmark scenarios in
section~\ref{sec:cascade_decays}.

In \figref{fig:cuts_compare} 
%
\FIGURE[t]{
\includegraphics[width=.75\textwidth]{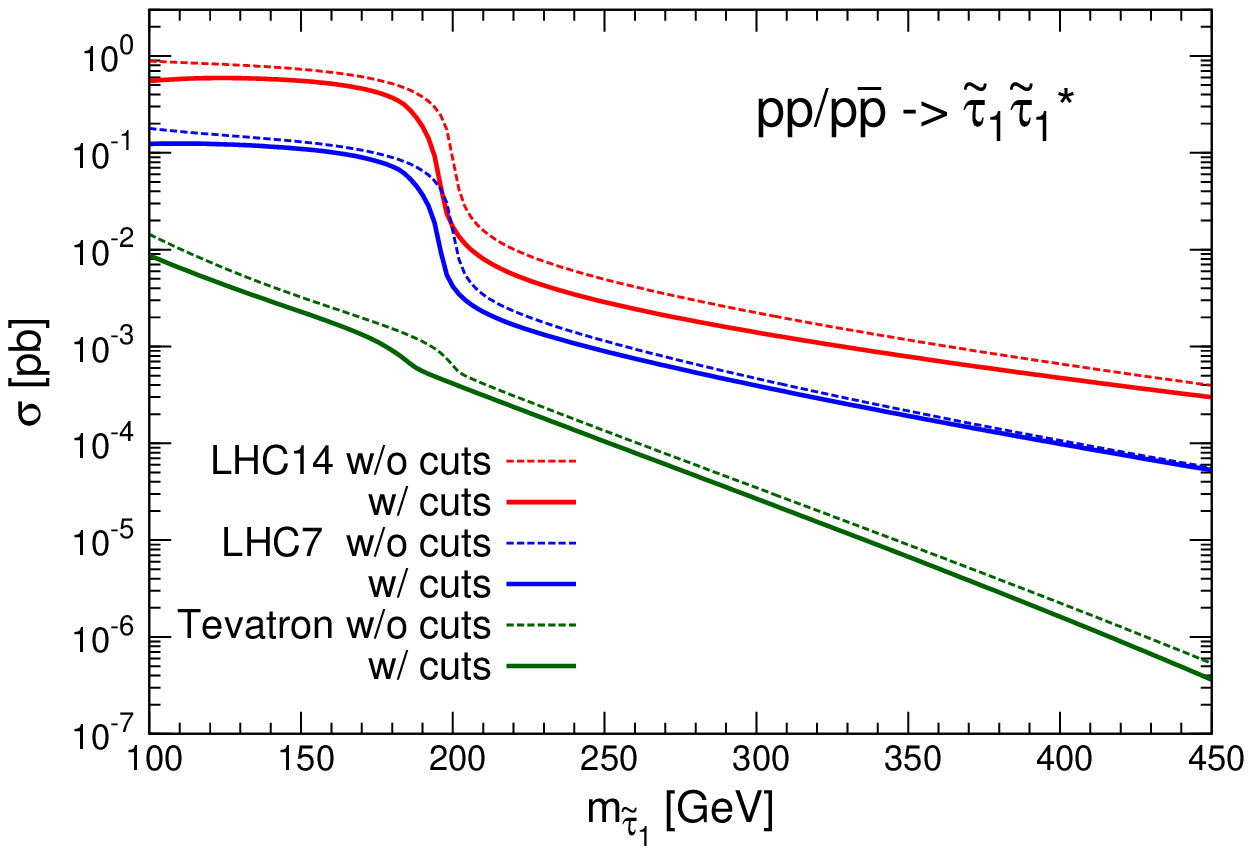}
\caption{Direct $\stau\stau^*$-pair production cross sections before 
  (dashed lines) and after (solid lines) application of the
  kinematical cuts~(\ref{eq:cuts}) as a function of $\mstau$ for the
  SUSY scenario with (\ref{eq_SUSYinputs}) and (\ref{eq_STAUinputs})
  at the LHC with $\SqrtS=14~\TeV$ (top, red) and $7~\TeV$ (middle,
  blue) and the Tevatron with $\SqrtS=1.96~\TeV$ (bottom, green).}
\label{fig:cuts_compare}
}
%
we compare the full direct production cross sections with (solid
lines) and without (dashed lines) the kinematical cuts~(\ref{eq:cuts})
applied as a function of $\mstau$ for the LHC with $\SqrtS =14~\TeV$
(top, red) and $7~\TeV$ (middle, blue) and the Tevatron with
$\SqrtS=1.96~\TeV$ (bottom, green). At the LHC, the cuts shift the
excess slightly away from the $\HH$ threshold and towards smaller
values of $\mstau$ and reduce the overall cross section by some tens
of percent. The reduction is stronger at the Tevatron, where in
particular the $\bbar$ and $gg$ channel contributions get cut
significantly, so that the \DY\ channel provides a good approximation
for the full cross section.

Assuming the produced \stau's to be stable on the scale of the
detectors, our results for the Tevatron shown in
\figref{fig:cuts_compare} can directly be compared with the
CHAMP cross-section limit from the CDF
collaboration~\cite{Aaltonen:2009kea} given
in~(\ref{eq_tevatronlimit}). This comparison does not allow to exclude
any $\mstau > 100~\GeV$ for the considered parameters.  Nevertheless,
smaller masses, allowed by the conservative LEP
limit~(\ref{eq:mstauLEPlimit}), $82~\GeV<\mstau<100~\GeV$, can be in
tension with this limit. However, any robust exclusion would require a
full simulation including detector effects, which we do not perform
here. Additional indirect stau-pair production mechanisms are not
expected to alter this statement considerably since, as discussed
above, they would contribute to a different signal region containing
jets/leptons.

Dedicated studies of the LHC experiments are announced for the near
future. Because of the increased cross section at the LHC, they will
probe in detail large parts of the small-\mstau parameter space, where
the stau can be produced via on-shell \HH exchange with only a
relatively small amount of integrated luminosity already for
$\SqrtS=7~\TeV$. In fact, the experiments at the LHC have already
performed searches for stable massive particles~\cite{Aad:2011yf,
  Khachatryan:2011ts}.  However, those studies have searched for
stable massive particles in the trackers and calorimeters. In those
parts of the detectors, the sensitivity to color-singlet particles,
such as the \stau, is reduced~\cite{Aad:2011yf}, and findings have
only been interpreted for colored massive particles~\cite{Aad:2011yf,
  Khachatryan:2011ts}.

So far we have concentrated on the integrated cross section. 
%
\FIGURE[t]{
\includegraphics[width=.49\textwidth]{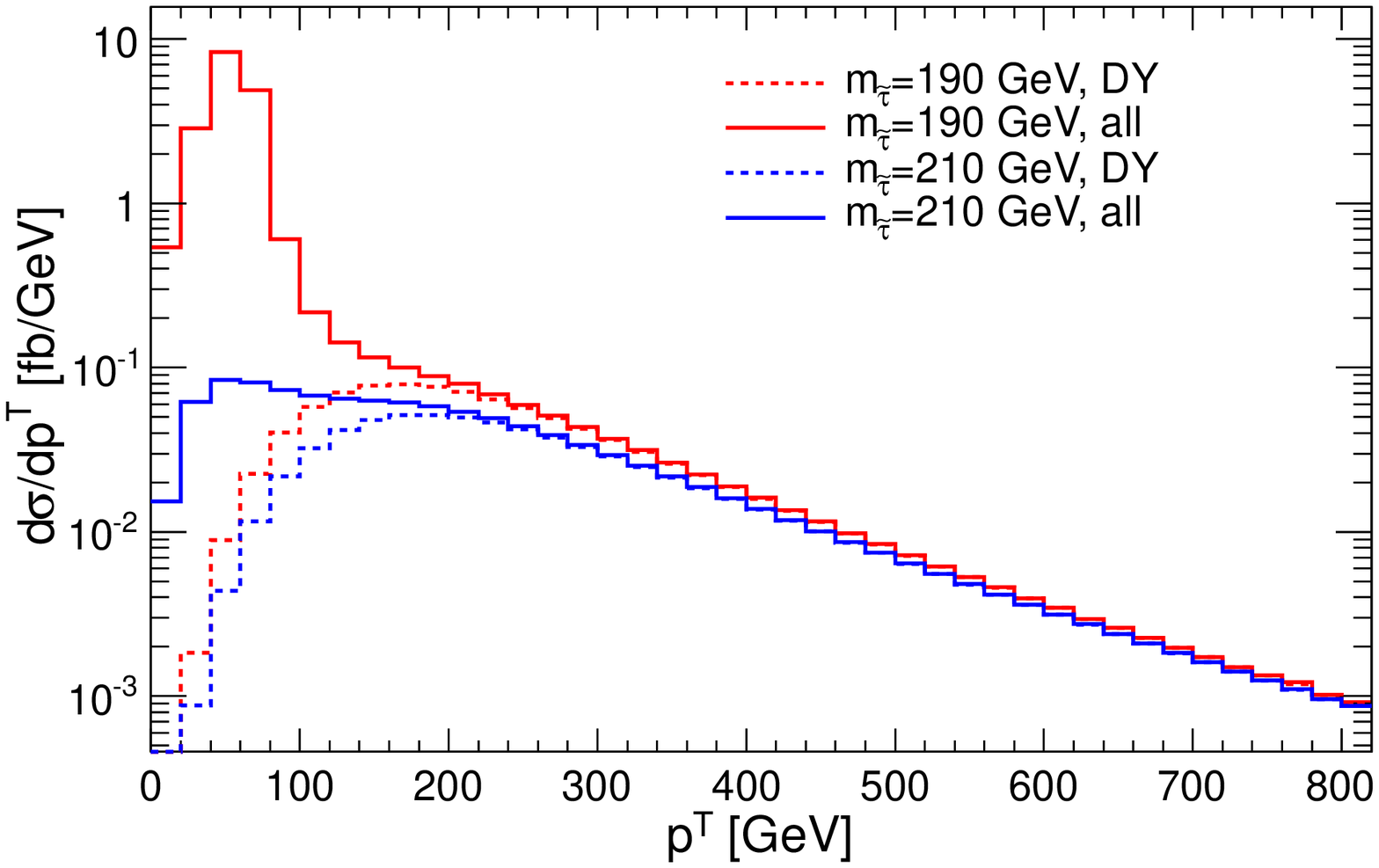}%
\includegraphics[width=.49\textwidth]{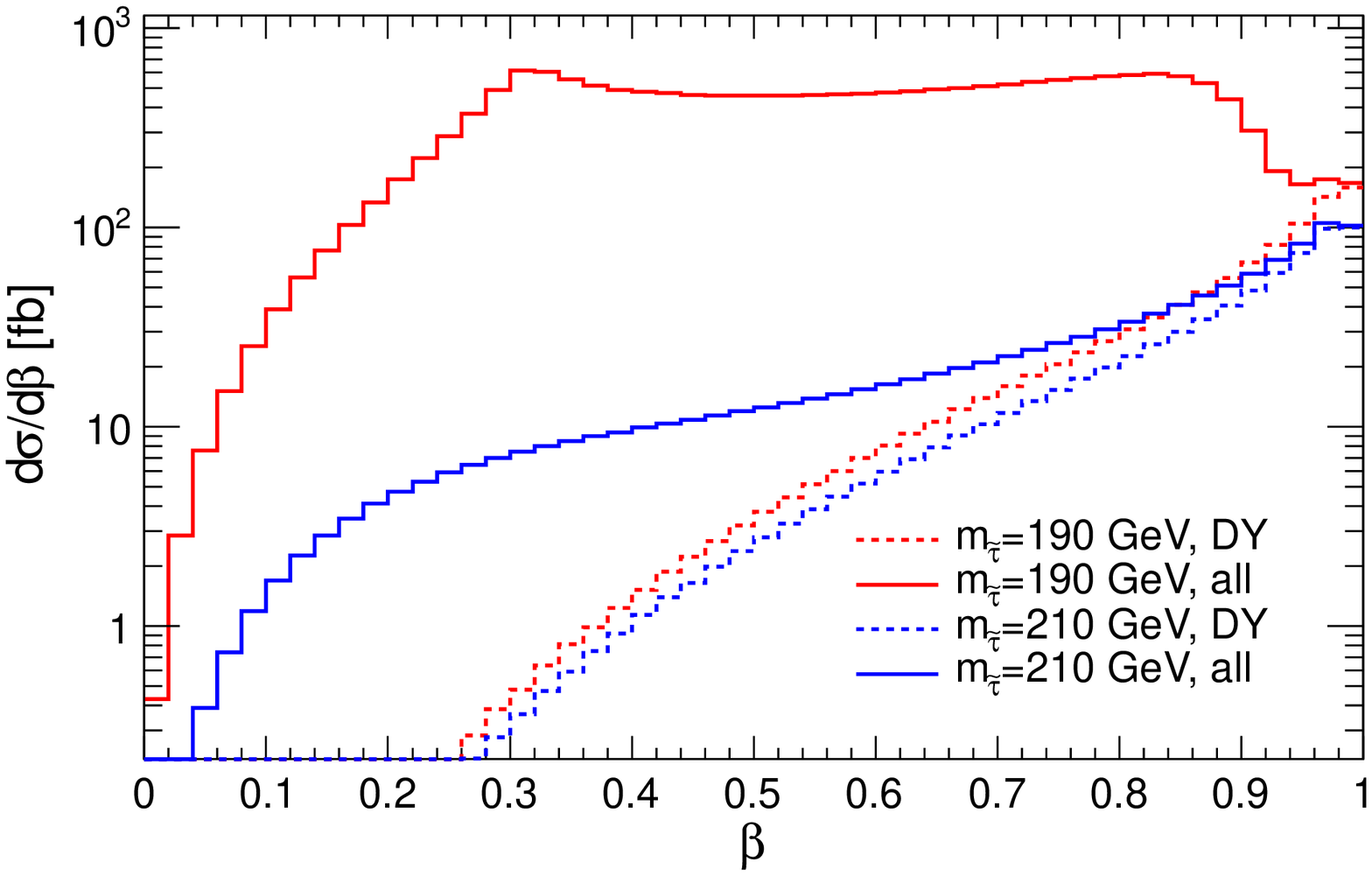}\\[.5ex]
{\small \hspace*{1cm} (a) \hspace*{.45\linewidth} (b)} 
\caption{(a)~Transverse momentum $p^T$ and (b)~velocity 
  $\beta=\vert\mathbf{p}\vert/E$ distributions of direct
  $\stau\stau^*$-pair production at the LHC with $\SqrtS = 14~\TeV$.
  The \DY\ predictions are shown by dashed lines and the full cross
  sections, including $\bbar$ and $gg$ channels, by solid lines. We
  consider $\mstau =190~\GeV$ (red line) and $\mstau =210~\GeV$ (blue
  line), whereas the other SUSY input parameters are as given
  in~(\ref{eq_SUSYinputs}) and~(\ref{eq_STAUinputs}).}
\label{fig:distributions_pt_beta}
}
%
To further illustrate the importance of the $\bbar$ and $gg$ channel
contributions and to investigate the impact of the
cuts~(\ref{eq:cuts}) on the different channels, we show the
differential distributions with respect to transverse momentum $\pT$
and the velocity $\beta$ of the directly produced staus in
\figref{fig:distributions_pt_beta}. We give results for the LHC with
$\SqrtS=14~\TeV$ only, however, results for the LHC with
$\SqrtS=7~\TeV$ are qualitatively identical. We use the basic
parameter inputs~(\ref{eq_SUSYinputs}) and~(\ref{eq_STAUinputs}) and
consider two distinct scenarios, $\mstau =190~\GeV$ (red line) and
$\mstau =210~\GeV$ (blue line). Here, it is $\mHH=400~\GeV$, \ie, in
the first scenario the intermediate \HH~boson can go on-shell while it
can only be produced off-shell in the second scenario. We apply the
cut $|\eta|<2.4$ on the pseudo-rapidity of one of the staus to ensure
that not both of the pair-produced staus leave the detector outside of
the sensitive region. Cuts on $\pT$ and $\beta$ are not applied to be
independent of a specific choice of cuts. Also, their potential
individual impact can be inferred directly from
\figref{fig:distributions_pt_beta}.

We show the \DY-cross-section prediction (dashed lines) now without
any $K$-factors and the full cross section including also the $\bbar$
and $gg$ channel contributions (solid lines). Clearly, in both
scenarios, staus produced in $\bbar$ and $gg$ channels are softer and
slower than their counterparts from the \DY-type production. The
$\pT$~distributions for the \DY\ channel peak at around $\mstau$ and
only few staus are produced with very low $\pT$. In contrast, the
$\bbar$ and $gg$ channel contributions peak for low $\pT$ and fall off
rapidly. The velocity distributions for pure \DY\ production rise
towards fast moving staus, $\beta\approx 1$, while adding the $\bbar$
and $gg$ channels results in a rather flat distribution for
intermediate values of $\beta$. This behavior is more pronounced for
scenarios in which an on-shell $\HH$ exchange is possible (red lines)
since here the relative importance of the $\bbar$ and $gg$ channels
with respect to the \DY\ channel is higher. In the case with
$2\mstau\gtrsim\mHH$ with $2\mstau$ still very close to $\mHH$ (blue
lines), the $\bbar$ and $gg$ channels still contribute significantly
and generate a large amount of events with very slow staus.

At this point we want to comment again on scenarios with a heavy,
decoupled $\HH$~boson such as those considered in
\figref{fig:crosssections2}. The contributions from \HH-boson-mediated
processes are suppressed in this case, and the same is true if the
\HH~boson is far off-shell ($\mHH \ll 2\mstau$). Still, the $\bbar$
and $gg$ channels can be sizeable if \hh-mediated processes are
important. For large $\stau\stau^*\hh$ couplings (large left-right
mixing, large $|\mu|\TB$ and/or large $|\Atau|$), we find that the
additional production mechanisms, and predominantly the $gg$-fusion
channel, can give cross section contributions of the same order of
magnitude as the \DY\ induced process. The shapes of the corresponding
$\pT$ and $\beta$ distributions are found to be similiar to those of
the \DY\ prediction but have a slightly softer $\pT$ spectrum.

To summarize, cuts on $\pT$ or $\beta$ affect the three production
channels to a very different extent and thus change the relative
importance of each of the contributions considerably.  If the cuts
(\ref{eq:cuts}) are applied, large parts of the additional $\bbar$ and
$gg$ channel contributions can be lost. We therefore recommend to
relax these cuts in future experimental searches to increase the
sensitivity especially for cosmological motivated scenarios with
$2\mstau \approx \mHH$, even when this implies that increased
backgrounds have to be taken into account. A more detailed study of
the interplay between different cut values and expected
backgrounds would be required but is beyond the scope of this paper.%
\footnote{Here we would like to refer to the recent paper on direct
  stau production at the LHC by Heisig and
  Kersten~\cite{Heisig:2011dr} which appeared during the final stage
  of our work.  In ref.~\cite{Heisig:2011dr}, the authors focus on the
  \DY-production mechanism and study the LHC discovery potential by
  performing a Monte Carlo analysis of the signal and the main dimuon
  background.}

\subsection{Prospects for SUSY parameter determination}
\label{sec:parameter_determination}
%
In this subsection we want to demonstrate how one could use the fact
that the \DY\ production cross section depends only on the stau mass
and mixing angle, whereas the $\bbar$ and $gg$ channels also depend on
other SUSY parameters. In fact, the interplay of the \DY\ cross
section and the additional channels might turn out to be helpful to
determine SUSY parameters.

If signals of a quasi-stable stau are observed at the LHC, one of the
first measurements will be the determination of the stau mass $\mstau$
by using TOF data from the muon chambers. The expected accuracy of
this $\mstau$ measurement has been estimated to be
$<1\%$~\cite{Ambrosanio:2000ik,Ellis:2006vu}.
With such a precise knowledge of $\mstau$, there are high hopes that
further SUSY parameters can be extracted from the cross section and
differential distributions by comparing experimental results and
theoretical predictions. Clearly, these measurements are possible in
the case of decay chains with several MSSM particles
involved~\cite{Ellis:2006vu,Ishiwata:2008tp,Biswas:2009rba,Feng:2009bd,Ito:2009xy,Heckman:2010xz,Kitano:2010tt,Ito:2010xj,Ito:2010un,Asai:2011wy}.
However, in the following we present ideas how such measurements
could also be possible just from direct production. To disentangle
those channels, appropriate jet and/or lepton vetos are assumed.
Moreover, the measurements require more integrated luminosity than
needed for a potential stau discovery.

First of all, once the stau mass is known, the \DY\ production cross
section can be given as a function of the stau mixing angle
$\thetastau$ alone. If also the direct stau production cross section
can be identified experimentally, this will allow us to determine
$\thetastau$ in a scenario in which the stau pair production cross
section is governed by the \DY\ channel. As shown in
figures~\ref{fig:crosssections1}\,(b) and
\ref{fig:crosssections2}\,(a) and as already discussed, the \DY\ cross
section is maximal in case of an almost purely left-handed \stau and
minimal for $\thetastau \approx \pi/2$. An excess of the
experimentally obtained cross section over the \DY\ expectation at
$\thetastau\approx\pi/2$ would thus point to $\thetastau<\pi/2$ and
non-negligible mixing between the left-handed and right-handed
eigenstates in \DY-channel-dominated scenarios. Furthermore, a
sizeable deviation from $\thetastau\approx\pi/2$ may support also the
hypothesis that the observed CHAMP is indeed a stau and not a
quasi-stable dominantly right-handed selectron or smuon.

However, in general, a larger experimentally obtained cross section
compared to the minimal \DY\ expectation for a certain mass could
imply both $\thetastau<\pi/2$ or also sizeable contributions from
$\bbar$ annihilation and $gg$ fusion; \cf\ 
figures~\ref{fig:crosssections1}\,(b) and
\ref{fig:crosssections2}\,(a). On the other hand, a significant excess
of the measured cross section over the maximal \DY\ cross section
prediction may provide a first hint for the importance of the
$\bbar$-annihilation and $gg$-fusion processes calculated in this
work; see also section~\ref{sec:cmssm_exceptional} below.

These possible ambiguities in the interpretation of experimental
findings on the integrated cross section could be resolved by studying
also the differential distributions. As we have seen above, the $\pT$
and $\beta$ distributions differ strongly from the \DY\ prediction
when the $\bbar$ and $gg$ channels are important. From the shape of
the experimentally measured distributions, one could then be able to
determine whether the \DY\ channel or the other channels give the
dominant contribution to the production cross section. Also the
distribution with respect to the invariant mass of the produced stau
pair, $\mstaustau$, can be helpful for this purpose. In fact, the
invariant mass distribution might allow even for the determination of
the mass $\mHH$ and the width $\GammaHH$ of the \HH boson: If
$2\mstau<\mHH$, \ie, if the $\HH$~boson can go on-shell in the $\bbar$
and $gg$~channels, there is the possibility to see the resonance of
the $\HH$~boson in the invariant mass distribution of the staus at
$\mstaustau\simeq\mHH$ with a width given by $\GammaHH$.

To illustrate this procedure, we consider four benchmark points,
$\alpha$, $\beta$, $\gamma$, and $\epsilon$ within the framework of
the CMSSM with parameters listed in \tabref{tab:benchmarks}.
%
\TABULAR[t]{lcC{1.6cm}C{1.6cm}C{1.6cm}C{1.6cm}}{
\hline\hline
\multicolumn{2}{c}{Benchmark point} & $\alpha$ & $\beta$ & $\gamma$ & $\epsilon$  \\
\hline\hline    
\mhalf                          &[\GeV] &       $600$   &       $1050$  & $600$  & $440$ 
\\
\mzero                          &[\GeV] &       $800$   &       $30$    & $600$ & $20$ 
\\
\TB                             &       &       $55$    &       $55$    & $55$  & $15$ 
\\
$A_0$                           &[\GeV] &       $1600$  &       $60$    & $1200$&  $-250$ 
\\ 
\hline

\mstau                          &[\GeV] & $193$         &       $136$   & $148$ & $153$ 
\\
\thetastau                      &       & $81^{\circ}$  &   $73^{\circ}$& $77^{\circ}$  & $76^{\circ}$  
\\
\mHH                            &[\GeV] & $402$         &       $763$   & $413$ & $613$ 
\\
$\Gamma_{\HH}$                  &[\GeV] & $15$          &       $26$    & $16$  & $2.2$ 
 \\
$m_{\gluino}$                   &[\GeV] & $1397$        &       $2276$  & $1385$& $1028$ 
\\
avg. $m_{\tilde{q}}$            &[\GeV] & $1370$        &       $1943$  &  $1287$  & $894$ 
\\
$\mu$                           &[\GeV] & $667$         &       $1166$  & $648$ & $562$ 
\\
$A_{\tau}$                      &[\GeV] & $515$         &       $-143$  & $351$ & $-275$ 
\\ 
 \hline
$\text{BR}(b \to s\gamma)$      & [$10^{-4}$]   & $3.08$ & $3.03$& $2.94$ & $3.00$ 
\\
$\text{BR}(B^0_s \to \mu^+\mu^-)$&[$10^{-8}$]   & $1.65$ & $1.04$& $2.44$ & $0.30$    
\\  
$a_{\mu}$                       &[$10^{-10}$]   & $13.2$ & $11.5$& $16.8$ & $18.7$   
\\
CCB \cite{Hisano:2010re}        & & $\checkmark$ & -- & $\checkmark$ & $\checkmark$    
\\ 
\hline  
\stauY &[$10^{-15}$] & $3.5$ & $2.5$ & $37.7$ & $164$
\\
\hline\hline }{Benchmark CMSSM scenarios $\alpha$, $\beta$, $\gamma$,
and $\epsilon$ defined by the given values of $\mhalf$, $\mzero$,
$\TB$, and $A_0$. For all points, $\mu>0$. Low-scale masses and
parameters are calculated using \SPheno. We also provide the
quantities that are subject to the constraints discussed in
section~\ref{sec:motivation}, as obtained with \SuperISO, and the
thermal relic stau yield $\Ystau$, as obtained with \micromegas.  The
CCB constraint~(\ref{eq:CCB}), as obtained in~\cite{Hisano:2010re}, is
respected by the scenarios $\alpha$, $\gamma$, and $\epsilon$, whereas
point $\beta$ is in tension with this constraint.
\label{tab:benchmarks}
}
%
The low-energy SUSY spectrum is obtained using
\SPheno~\cite{Porod:2003um}, while the Higgs sector is recalculated
with \FeynHiggs~\cite{Frank:2006yh}. We also refer to the constraints
discussed in section~\ref{sec:motivation}, evaluated with
\SuperISO~\cite{Arbey:2011zz}, and provide the thermal relic stau
yield $\Ystau$ as calculated with \micromegas~\cite{Belanger:2010gh}.
Benchmark points $\alpha$ and $\beta$ are similiar to points $B$ and
$C$ of ref.~\cite{Pradler:2008qc}, respectively, where we have
adjusted $\mzero$ for point $\alpha$ and $\mhalf$ for point $\beta$ so
that \SPheno provides low-energy spectra that are similar to the ones
of those points $B$ and $C$. Point $\gamma$ is very similar to point
$\alpha$ but has a much larger stau yield. Point $\epsilon$ was
already introduced in ref.~\cite{DeRoeck:2005bw}. Here, we are mainly
interested in the ratio of \mstau and \mHH. In all four benchmark
scenarios, stau production via an on-shell \HH~exchange is possible.
We have $2\mstau\approx\mHH$ for point $\alpha$, $2\mstau<\mHH$ for
point $\gamma$, and $2\mstau \ll \mHH$ for points $\beta$ and
$\epsilon$.  The stau-Higgs couplings are smallest for point
$\epsilon$, where \TB is relatively small.

In \figref{fig:distributions_minv} we display the invariant mass
distributions for the four benchmark points at the LHC with
$\SqrtS=14~\TeV$.
%
\FIGURE[t]{
\includegraphics[width=.495\textwidth,]{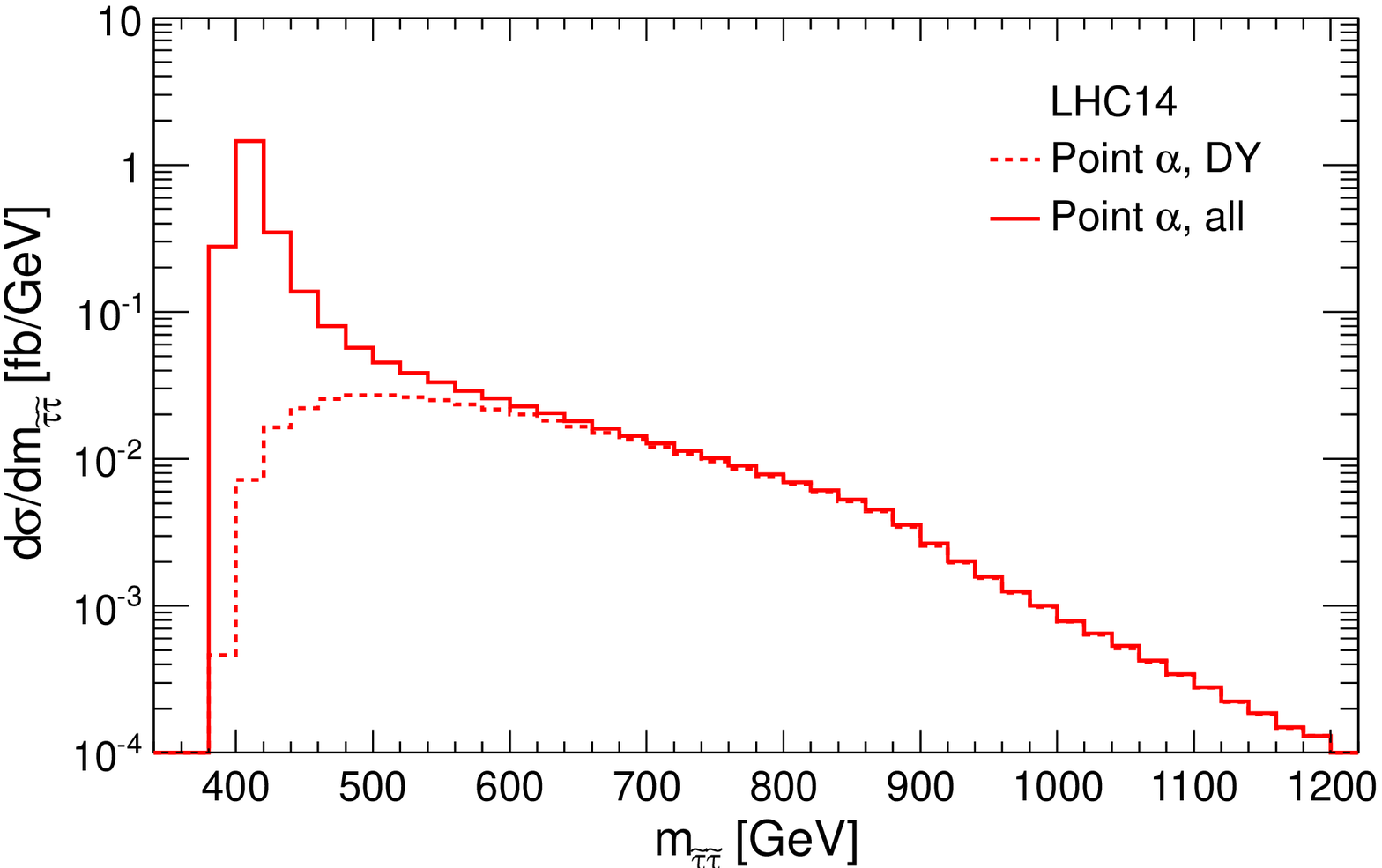}%
\includegraphics[width=.495\textwidth]{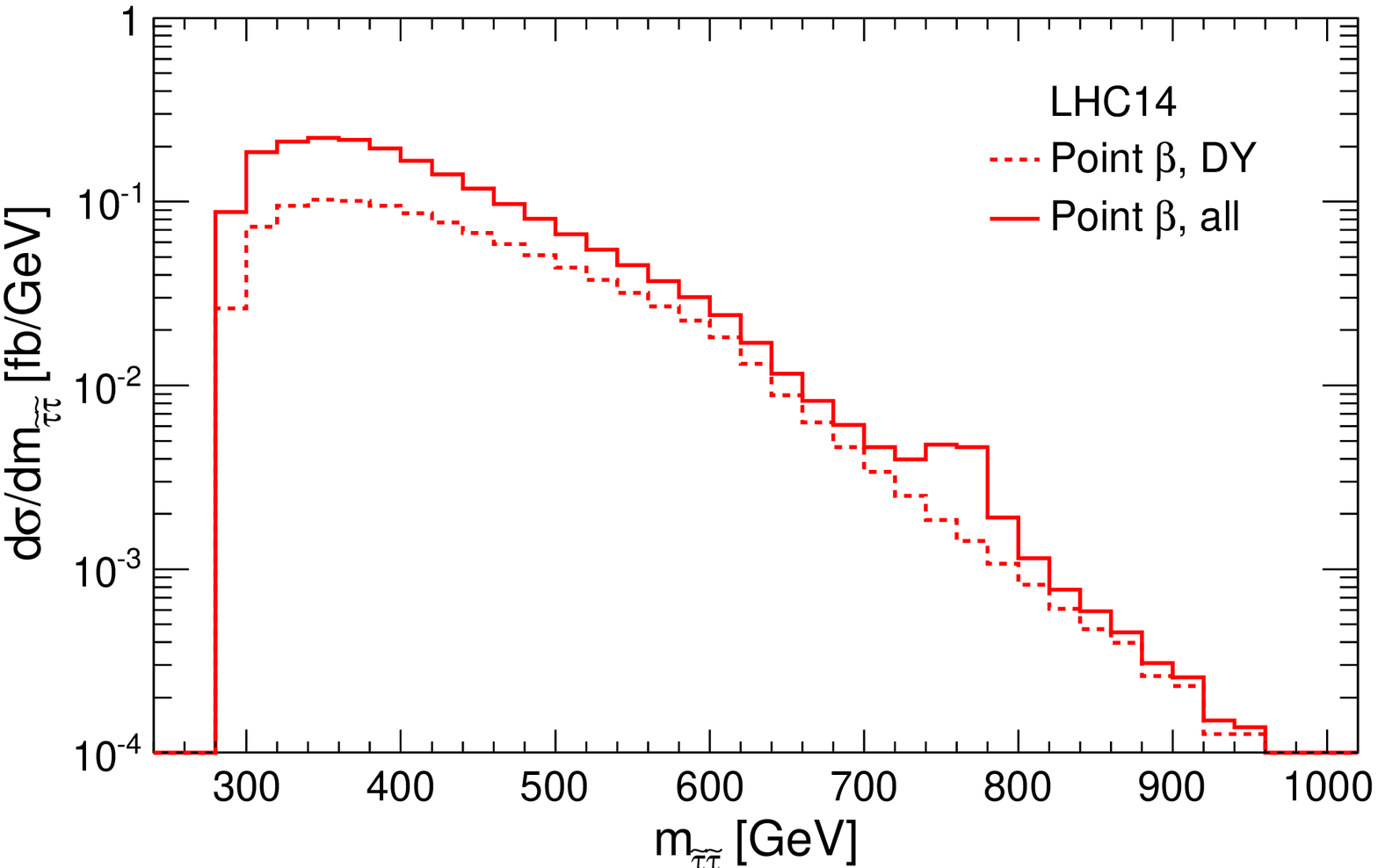}\\[.5ex]
{\small \hspace*{2cm} (a) \hspace*{.45\linewidth} (b)} \\[2.5ex]
\includegraphics[width=.495\textwidth,]{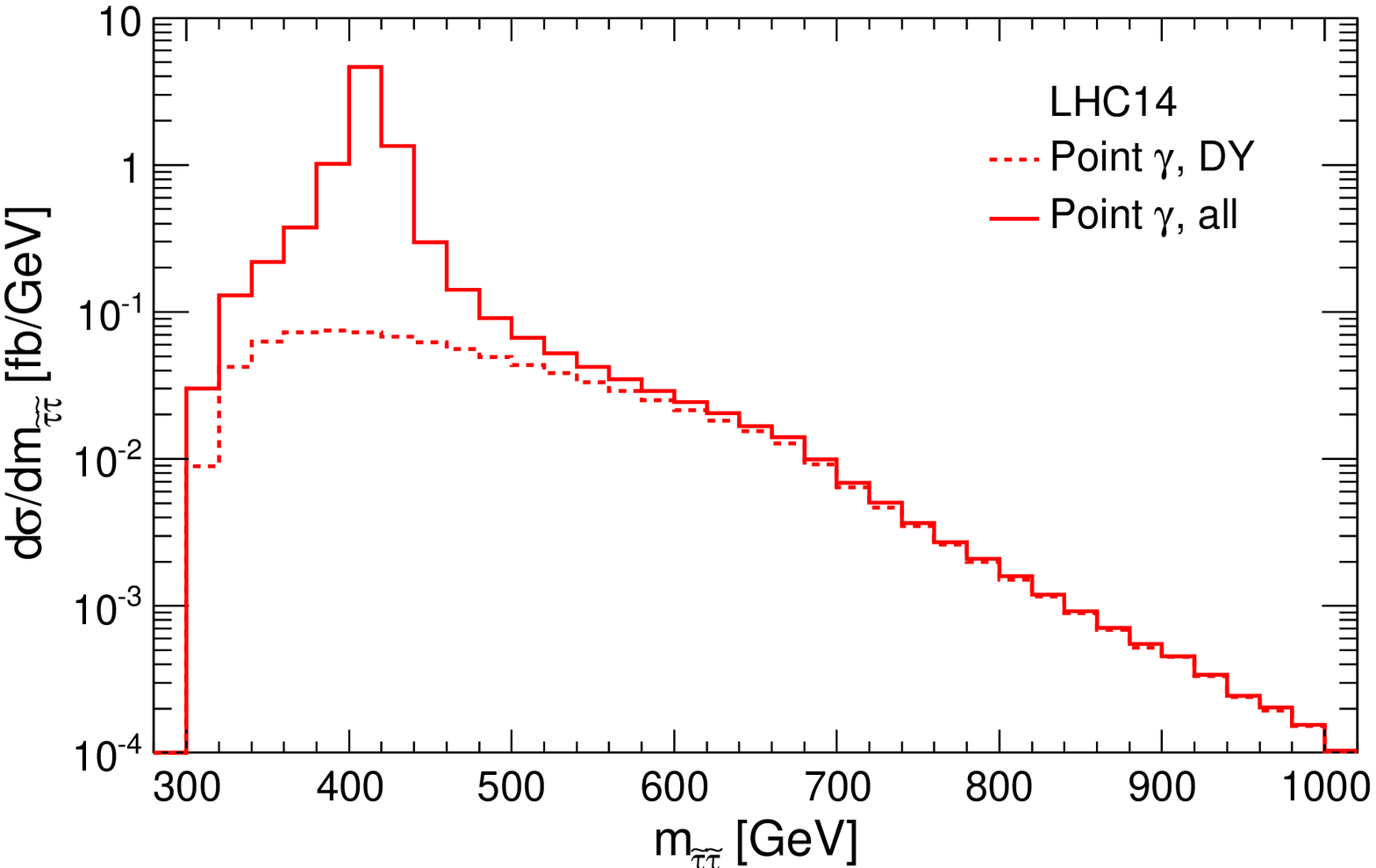}%
\includegraphics[width=.495\textwidth]{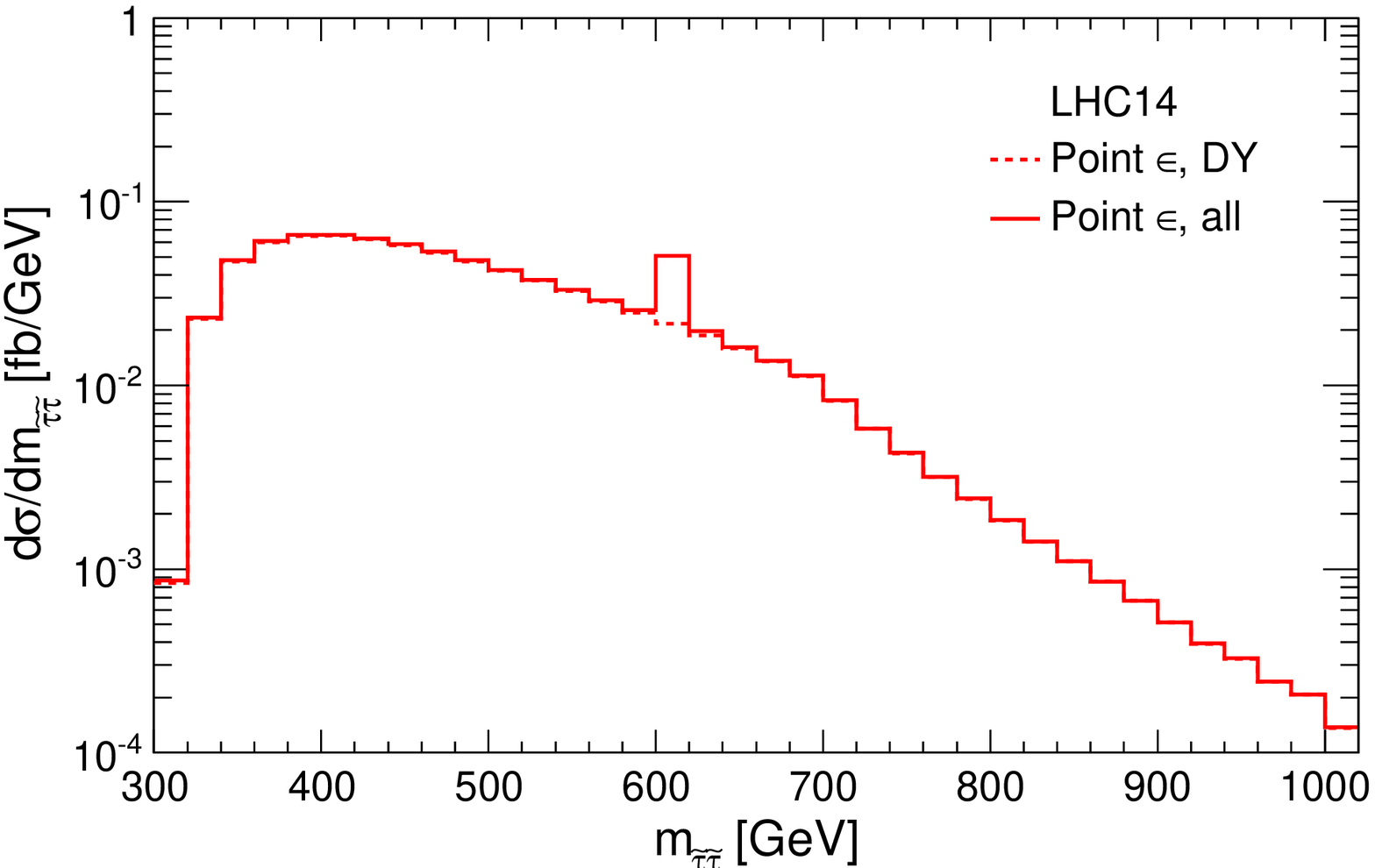}\\[.5ex]
{\small \hspace*{2cm} (c) \hspace*{.45\linewidth} (d)} \\[-1ex]
\caption{Invariant mass distributions of directly produced $\stau\stau^*$ pairs 
  at the LHC with $\SqrtS=14~\TeV$. \DY\ predictions are shown by the
  dashed lines and the full cross sections, including the $\bbar$ and
  $gg$ channels, by the solid lines. The distributions are shown for
  the benchmark points (a)~$\alpha$, (b)~$\beta$, (c)~$\gamma$, and
  (d)~$\epsilon$, which are defined in \tabref{tab:compare_cascade}.
}
\label{fig:distributions_minv}
}
%
The kinematical cuts~(\ref{eq:cuts}) are applied, with the requirement
$\vert\eta\vert < 2.4$ for both staus. The invariant mass
distributions show a resonance peak at the mass of the \HH~boson on
top of the \DY\ continuum in all four considered scenarios. For point
$\alpha$ considered in \figref{fig:distributions_minv}\,(a), the peak
is close to the threshold, $\mstaustau\approx2\mstau$, and the
distribution falls off steadily at higher invariant masses. Such a
Higgs resonance at the beginning of the invariant mass distribution is
a strong hint towards efficient stau annihilation in the early
Universe, as further discussed in section~\ref{sec:cmssm_exceptional}
below. In the invariant mass distribution for point $\gamma$ shown in
\figref{fig:distributions_minv}\,(c), the \HH resonance lies on top of
the maximum of the \DY\ contribution and the peak is very pronounced.
For points $\beta$ and $\epsilon$, the \DY\ continuum is dominant and
the resonance appears as a small bump at the tail of the distribution
as can be seen in figures~\ref{fig:distributions_minv}\,(b) and~(d).
Note that although the $\bbar$ and $gg$ channels do not distort much
the shape of the \DY\ curve in scenario $\beta$, they increase the
differential cross section sizeably. This typically happens for large
left-right mixing and large \TB even in the case of an heavy decoupled
\HH~boson, see figure~\ref{fig:crosssections2}.

Both the $\alpha$ and $\gamma$ scenarios would allow for a
determination of the mass \mHH and also the width $\GammaHH$
(especially in case of point $\gamma$) with a few $\fbarn^{-1}$ of
data at the LHC with $\SqrtS = 14~\TeV$. Nevertheless, also in
scenarios with a rather heavy \HH~boson (such as point~$\beta$) and in
scenarios with small \TB (such as point~$\epsilon$), the LHC might
eventually be able to determine the mass $\mHH$ from the invariant
mass distribution of the directly produced long-lived $\stau\stau^*$
pairs. Such a procedure only requires $2\mstau<\mHH$ and is thus
generic in large parts of the MSSM parameter space and within the
CMSSM. Here a future study of theoretical and experimental
uncertainties including detector effects and the possible
contamination with SUSY backgrounds from cascade decays is necessary
to allow for more precise predictions.

\subsection{Prospects for the stopping of staus}
\label{sec:stopping}

Very slow moving CHAMPs might loose all their momentum and get stopped
within the main detector or in some additional stopping detector. The
analysis of their subsequent decays may then help to identify the LSP
in an R-parity conserving setting (\cf
section~\ref{sec:longlivedstau}) or to probe the size of the R-parity
violating coupling. In such a way, \eg, the gravitino or axino mass
and also its couplings might be tested in the future. In general, as a
rule of thumb, CHAMPs with $\beta\gamma < 0.45$, as stated
in~(\ref{eq:betagammacut}), are expected to get stopped in the
detectors at the LHC. In the following, we show that staus produced
directly via the $\bbbar$ and $gg$ channels provide a large additional
source of potentially stopped objects, especially in the
cosmologically motivated scenarios discussed in
section~\ref{sec:exceptionalYstau}.

In \figref{fig:distributions_bg} we give the number of directly
produced staus for an integrated luminosity of $L=1~\fbarn^{-1}$ at
the LHC with $\SqrtS=14~\TeV$ as a function of
$\beta\gamma=\vert\mathbf{p}\vert/m$ for the benchmark points $\alpha$
and $\beta$ introduced above and defined in
\tabref{tab:benchmarks}.
%
\FIGURE[t]{
\includegraphics[width=.49\textwidth,]{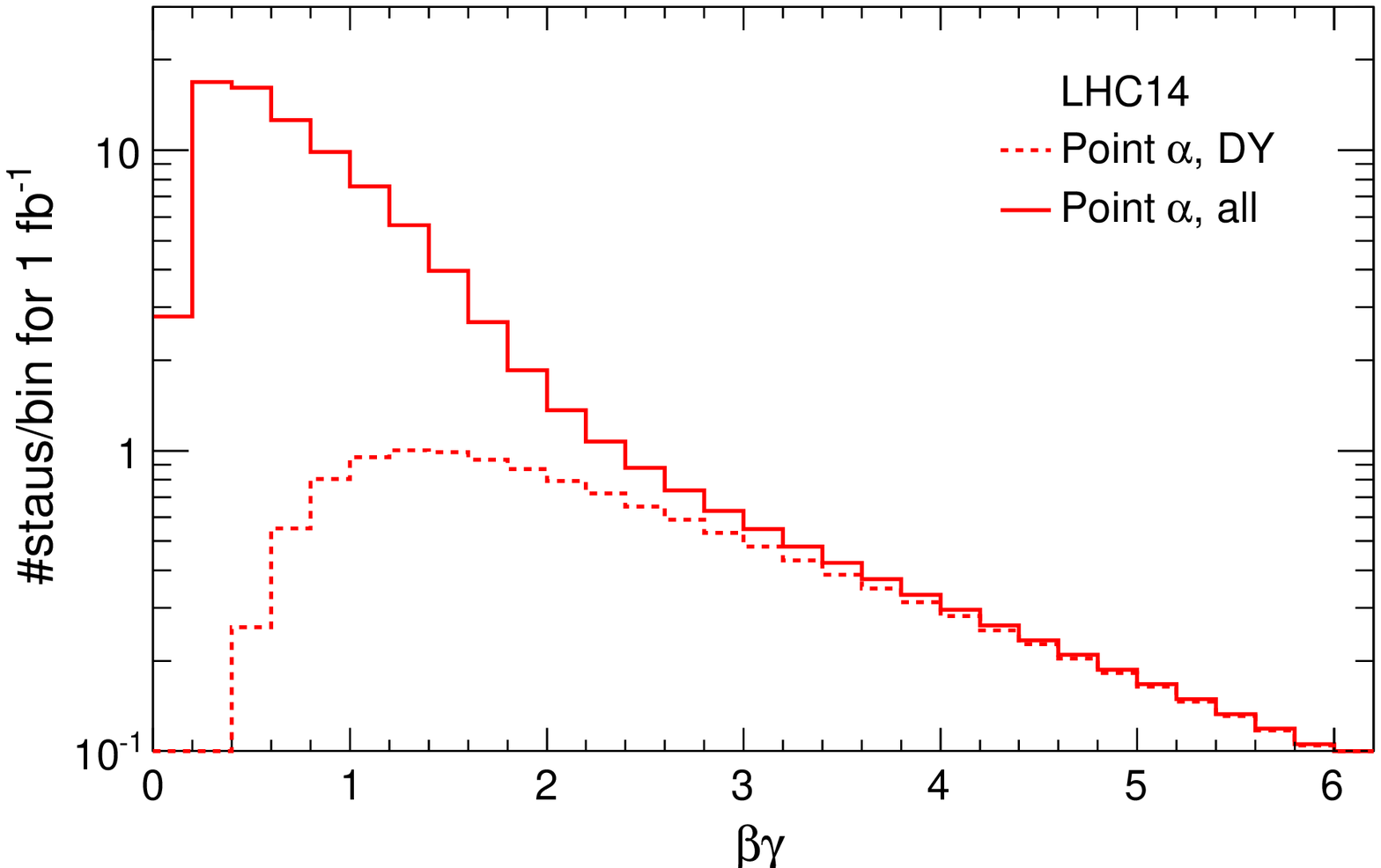}
\includegraphics[width=.49\textwidth,]{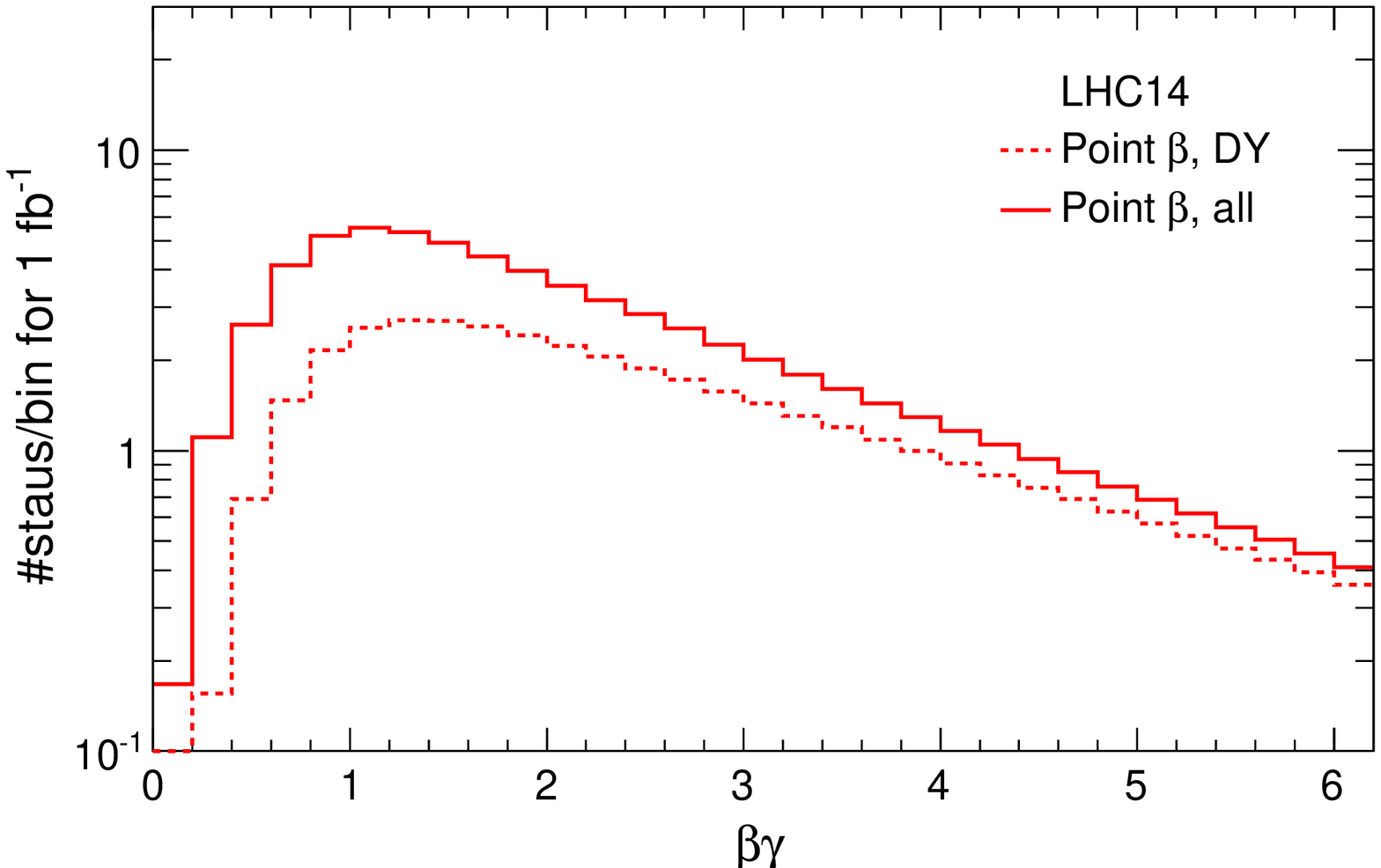}\\[.5ex]
{\small \hspace*{2.17cm} (a) \hspace*{.455\linewidth} (b)} \\[-1ex] 
\caption{Directly produced staus as a function of 
  $\beta\gamma=\vert\mathbf{p}\vert/m$ at the LHC with
  $\SqrtS=14~\TeV$ for an integrated luminosity of $L=1~\fbarn^{-1}$.
  The full results, including $\bbar$ and $gg$ channels, are shown by
  the solid lines and the \DY\ predictions by the dashed lines.  Staus
  with $\beta\gamma \lesssim 0.45$ are expected to get stopped in the
  LHC detectors. Distributions are shown for the benchmark points
  (a)~$\alpha$ and (b)~$\beta$, which are defined in
  \tabref{tab:benchmarks}.}
\label{fig:distributions_bg}
}
%
We only require the pseudo-rapidity of a stau to fulfill
$\vert\eta\vert < 2.4$ to be included in the shown histograms. No
other kinematical cuts are imposed as we are especially interested in
very slow moving objects. We count each produced stau individually.

In both scenarios $\alpha$ and $\beta$, the number of potentially
stopped staus, \ie, those with $\beta\gamma<0.45$, is enlarged when
the $\bbar$ and $gg$ channels are included in the cross section
prediction. This enhancement is particularly substantial in scenario
$\alpha$ where $2\mstau\approx\mHH$ and where the staus at the
\HH~boson resonance are thus produced almost at rest in the center of
mass frame. Here, a large sample of stopped staus could be collected
already with a rather small integrated luminosity. For an integrated
luminosity of $L=1~\fbarn^{-1}$, about $N\approx25$ staus would be
stopped, which is encouraging. Indeed, from the \DY\ process alone,
one would expect only $N\lesssim1$.  For an account of the
experimental prospects, it is thus crucial to consider also the
$\bbar$ and $gg$ channels.

\section{Direct stau production within the CMSSM}
\label{sec:cmssm}

In this section we investigate the direct pair production cross
section of light staus within the CMSSM. Our aim is to provide a
precise prediction for the $\staustaubar$ cross section, taking the
\DY\ process as well as the $\bbar$~annihilation and $gg$~fusion
channels into account. We are interested in parameter regions with a
quasi-stable \stau LOSP, where the $\staustaubar$ cross section is of
particular importance. We consider the \mzero-\mhalf plane of the
CMSSM with $A_0=2\mzero$, $\TB=55$, and $\mu > 0$, which is
cosmologically motivated by the possible occurance of exceptional
small stau yields~\cite{Pradler:2008qc}. In this plane, \TB is large
and the \stau prefers to be right-handed, \ie, $\thetastau >\pi/4$;
this is generic in the CMSSM due to the different running of the
left-handed and right-handed soft masses. Thus, the $\bbar$ and $gg$
channels can give large contributions to the stau production cross
section. For the CMSSM benchmark scenarios introduced in
section~\ref{sec:phenolonglived}, we compare our results for direct
stau production with indirect stau production mechanisms via the
production and decay of other heavier SUSY particles. We find that
direct stau production is often the dominant source of staus at
colliders, in particular, at lower center-of-mass energies when the
production of other SUSY particles is suppressed by their heavier
masses.

\subsection{CMSSM scans of the direct stau pair production cross section}

\FIGURE[t]{%
\includegraphics[width=.95\textwidth]{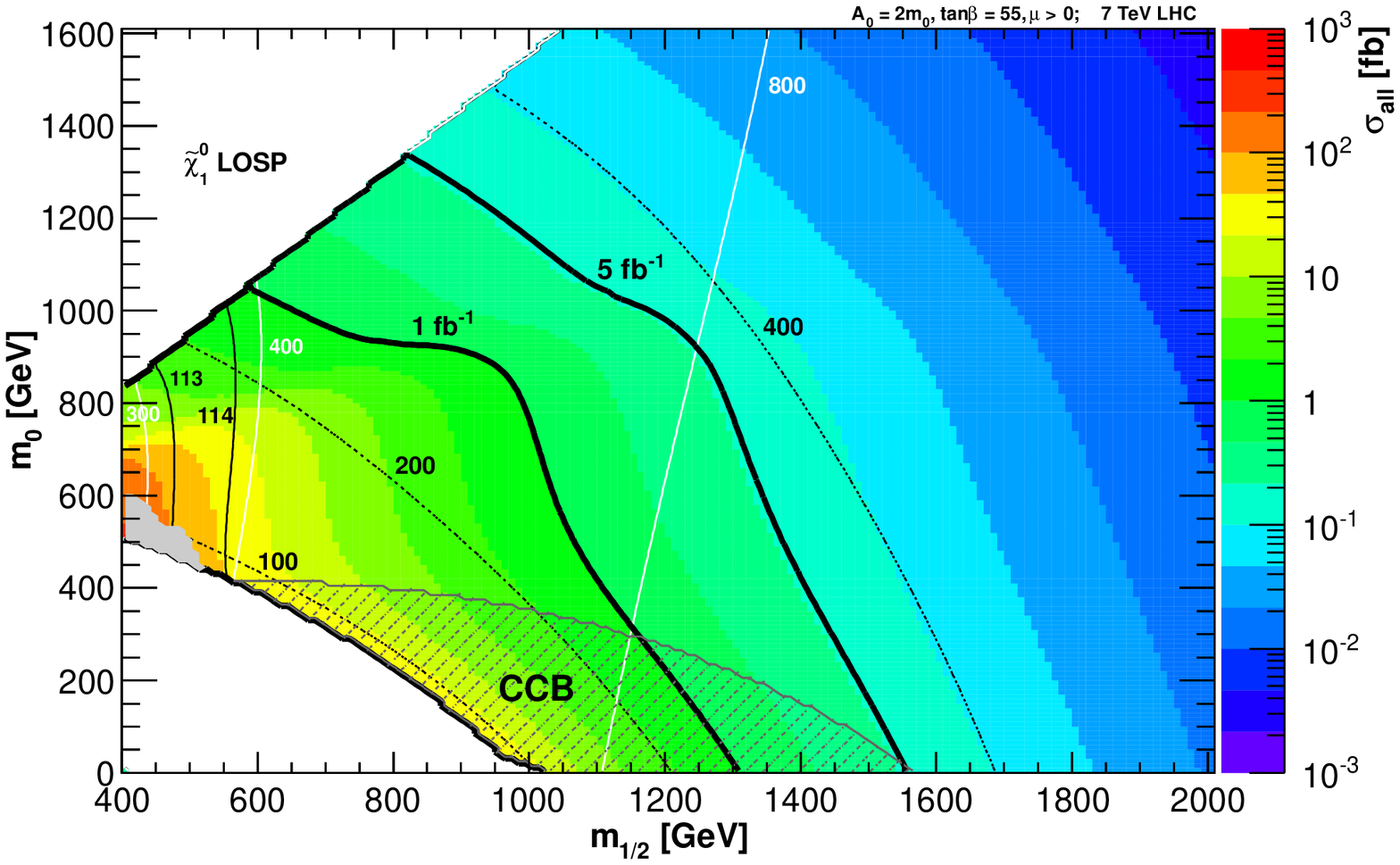}\\%
\includegraphics[width=.95\textwidth]{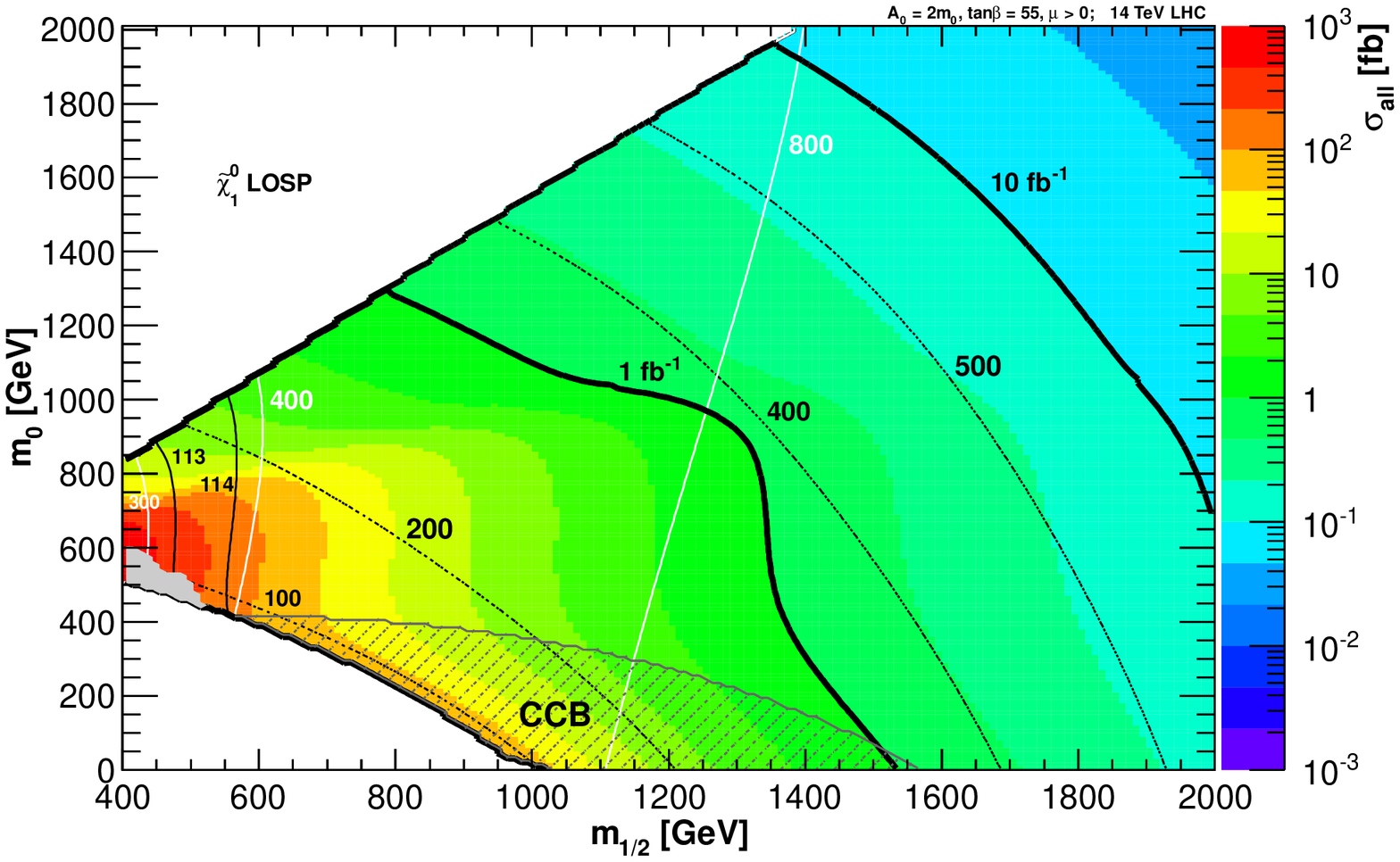}%
\caption{Contours of the total direct $\stau\staubar$ 
  production cross section (shaded, colored) at the LHC with
  $\SqrtS=7~\TeV$ (top) and $14~\TeV$ (bottom) after the
  cuts~(\ref{eq:cuts}) in the CMSSM $m_0$-$m_{1/2}$ plane with
  $\TB=55$, $A_0=2\mzero$, and $\mu>0$.  The white region in the lower
  left is excluded because of a tachyonic spectrum, impossible EWSB,
  or $\mstau\le82~\GeV$. In the upper white area, the lightest
  neutralino $\neu_{1}$ is the LOSP.  Parameter points in the gray
  area around $\mzero\sim\mhalf\sim500~\GeV$ do not respect the
  constraint $\BR(B_s^0\to\mu^+\mu^-)<4.3\times 10^{-8}$. The lower
  hatched area is in tension with the CCB constraint~(\ref{eq:CCB}).
  The thin solid black lines indicate $\mh$, the dashed black lines
  $\mstau$, and the white lines $\mHH$, where the mass values are
  given in units of $\GeV$ at the respective contour.  For a naive
  estimate of the discovery potential, thick black lines show regions
  in which at least one $\stau\staubar$-pair-production event is
  expected for integrated luminosities of $\mathcal{L}=1~\fbarn^{-1}$
  and $5~\fbarn^{-1}$ at $\SqrtS=7~\TeV$ and
  $\mathcal{L}=1~\fbarn^{-1}$ and $10~\fbarn^{-1}$ at
  $\SqrtS=14~\TeV$.}
\label{fig:cmssm_LHC}
}
%
\afterpage{\clearpage}
%
\FIGURE[h]{
\includegraphics[width=.95\textwidth]{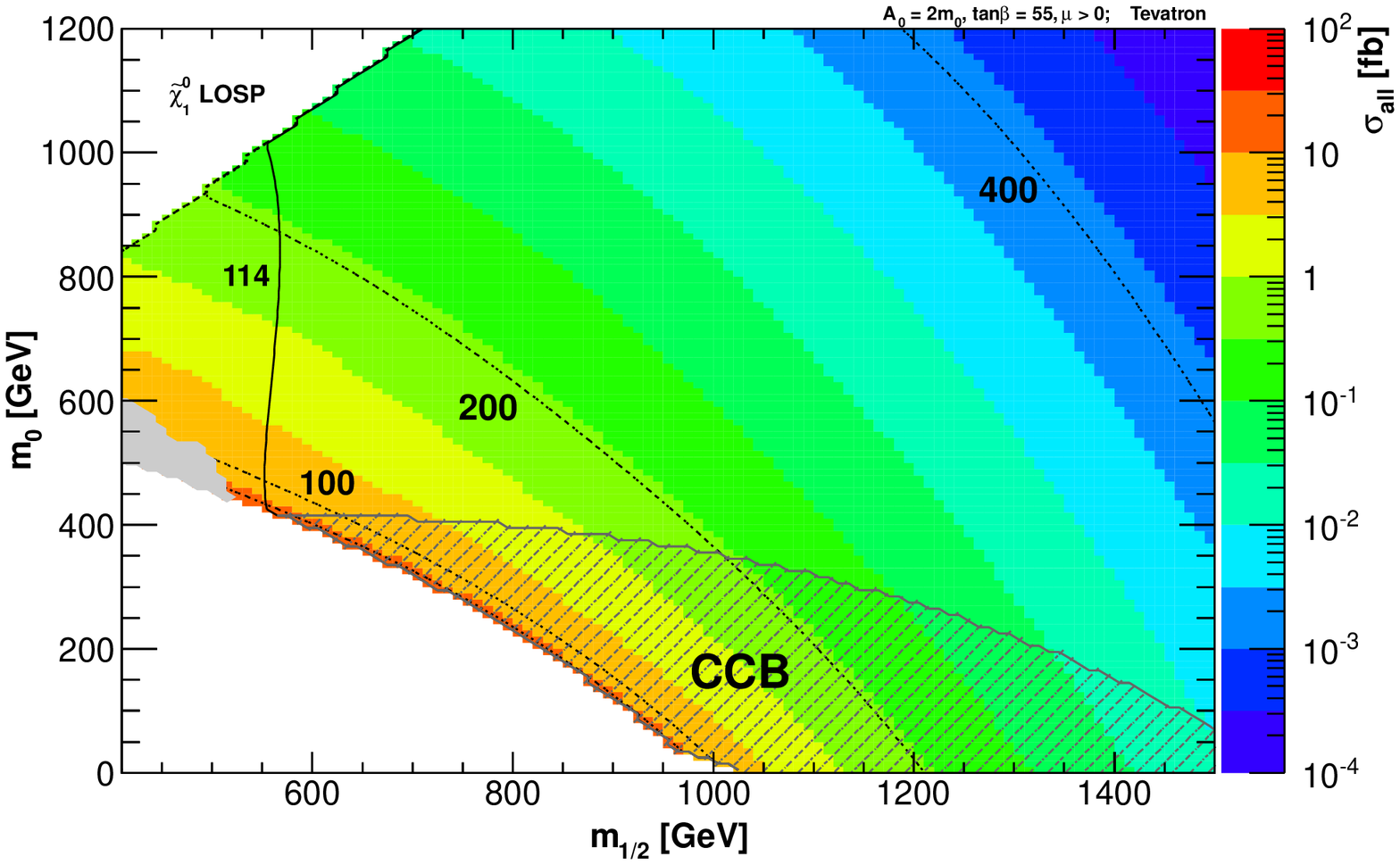}
\caption{Contours of the total direct $\stau\staubar$ 
  production cross section (shaded, colored) at the Tevatron with
  $\SqrtS=1.96~\TeV$ after the cuts~(\ref{eq:cuts}) in the CMSSM
  $m_0$-$m_{1/2}$ plane with $\TB=55$, $A_0=2\mzero$, and $\mu>0$.
  All shown contours and regions are as in \figref{fig:cmssm_LHC}. A
  tiny (dark orange) strip with $\sigma>10~\fbarn$ is in tension with
  searches for CHAMPs at the Tevatron~\cite{Aaltonen:2009kea}.}
\label{fig:cmssm_tevatron}
}
%
In \figsref{fig:cmssm_LHC}{fig:cmssm_tevatron} we show in a
\mzero-\mhalf plane of the CMSSM the direct stau production cross
section at the LHC and at the Tevatron, as well as mass contours and
excluded/disfavored parameter regions. For $A_0=2\mzero$, $\TB=55$ and
$\mu > 0$, we consider the region in which the \stau is the LOSP.
Again, we compute the low-energy SUSY spectrum with \SPheno, while the
Higgs sector is reevaluated with \FeynHiggs. Flavor constraints are
evaluated using \SuperISO. The cross section prediction includes the
\DY\ channels with NLO $K$-factors, the $\bbar$~annihilation and
$gg$~fusion contributions, and the cuts~(\ref{eq:cuts}) adequate for a
long-lived stau are applied. The white area in the lower left is
excluded by a tachyonic spectrum, impossible electroweak symmetry
breaking (EWSB), or a stau mass $\mstau \le 82~\GeV$ below the
conservative LEP limit~(\ref{eq:mstauLEPlimit}). In the upper left
white area, the LOSP is the lightest neutralino $\neu_{1}$ (as
indicated). In the gray area around $\mzero\sim\mhalf\sim500~\GeV$,
$\BR(B_s^0\rightarrow\mu^+\mu^-)$ exceeds the upper limit given
in~(\ref{eq:Bmumu}). The hatched area is disfavored by the CCB
constraint~(\ref{eq:CCB}). Contours for a constant Higgs mass of
$\mh=113~\GeV$ and $114~\GeV$ are shown as thin solid black lines.
%
%
The dashed black lines are $\mstau$ contours and the solid white lines
$\mHH$ contours, where the associated mass values are indicated on the
respective contours in units of $\GeV$.

The cross section depends mainly on \mstau and \mHH and varies over
several orders of magnitude in the given parameter ranges. At the LHC
with $\SqrtS =14~\TeV$, it reaches $10^3~\fbarn$ in the region with
$\mzero\lesssim 700~\GeV$ and $\mhalf\lesssim 500~\GeV$ and drops to
$\lesssim 2\times10^{-2}~\fbarn$ for, \eg,
$\mzero\sim\mhalf\sim2~\TeV$, as can be seen in the lower panel of
\figref{fig:cmssm_LHC}. When going down from $\SqrtS=14~\TeV$ to
$7~\TeV$, considered in the upper panel of \figref{fig:cmssm_LHC}, we
observe a decrease of the cross section by up to about a factor of 5
(see also figures~\ref{fig:inclusive_compare}
and~\ref{fig:cuts_compare}). To give a naive estimate of the discovery
potential, we also show contours (thick lines) on which one
$\stau\staubar$-pair-production event is expected for integrated
luminosities of $\mathcal{L}=1~\fbarn^{-1}$ and $5~\fbarn^{-1}$ at
$\SqrtS=7~\TeV$ and $\mathcal{L}=1~\fbarn^{-1}$ and $10~\fbarn^{-1}$
at $\SqrtS=14~\TeV$. The parameter regions to the left of these lines
could thus be accessible by the experiments at the LHC already in the
very near future, in particular, since the SM backgrounds are expected
to be well under control with the kinematical cuts discussed above.
However, a more realistic determination of the discovery reach and/or
the exclusion limits should be performed in the context of a detailed
study including detector effects. Such a study has been performed
recently in the case of direct \DY\ production \cite{Heisig:2011dr}.

At the Tevatron, the overall cross section for the direct production
of staus is much smaller than at the LHC. As can be seen from
\figref{fig:cmssm_tevatron}, it ranges typically from some tenths to a
few femtobarns. (Here we should note that the shading/color scales in
\figref{fig:cmssm_tevatron} differ from the ones in
\figref{fig:cmssm_LHC}). Larger values in the considered CMSSM plane
are only found very close to the region excluded by the LEP mass
limit~(\ref{eq:mstauLEPlimit}) in the lower left corner of
\figref{fig:cmssm_tevatron}. Here, in a tiny strip of the parameter
plane, the cross section exceeds $10~\fbarn$ (as indicated by the dark
orange color coding) and thus challenges the current cross section
limit~(\ref{eq_tevatronlimit}) from the CDF
experiment~\cite{Aaltonen:2009kea}.
 
The actual shape of the cross section and mass contours in the
\mzero-\mhalf planes can easily be understood. We focus here on
parameter regions with a \stau LOSP, where typically $\mhalf>\mzero$
in the CMSSM. For $\TB>20$, the $A^0$ mass can approximately be
written as $m_{A}^2\approx m_0^2+2.5 m_{1/2}^2$ (neglecting Yukawa
interactions) \cite{Drees:1995hj} and thus $m_{A}$ is mainly
determined by $\mhalf$. For $\mhalf\gg 0$, the Higgs sector is then
typically in the decoupling limit, where $\mHH \approx m_{A}$, and
thus \mHH also mainly determined by $\mhalf$. The dependence of
$\mstau$ on $\mzero$ and $\mhalf$ is less intuitive in the discussed 
parameter range (\stau LOSP, large \TB, sizeable Yukawa couplings),
but the usual relation $\mstau^2\propto\mzero^2+0.15 \mhalf^2$
\cite{Drees:1995hj} can still be considered. Thus, one finds for $\mhalf\gg\mzero$
that it is always $\mHH>2\mstau$. For the stau production cross
section, this means that there can be important contributions from the
$\bbar$ and $gg$ channels in the region $\mhalf\gg\mzero$, where the
\HH~boson can go on-shell. Towards large \mzero and \mhalf, however,
the \stau and (even faster) the \HH become heavy and contributions
from the $\bbar$ and $gg$ channels become less important compared to
the \DY\ channel.  If the \DY\ contributions dominate, the overall
production cross section is basically a function of the \stau mass and
decreases strongly for higher \mstau.

Close to the boundary of the $\neu_{1}$ LOSP region, where
$\mzero\approx\mhalf$, the \stau gets heavier relative to the
\HH~boson so that $2\mstau>\mHH$, which means that the direct stau
production via an on-shell \HH~boson is no longer possible. However,
the position of the transition at $2\mstau=\mHH$ depends
strongly on \TB. For large \TB, bottom and tau Yukawa couplings can be
sizable and drive down the masses of the heavy Higgses.  For small
\TB, this transition lies mostly within the $\neu_1$ LOSP region and
only for very large values of \mhalf within the \stau LOSP region.
Thus, for smaller values of \TB, contributions from on-shell \HH boson
exchange are a generic feature of the CMSSM, as one usually finds
$2\mstau<\mHH$.

Let us emphasize that even if the $\bbar$ and $gg$ contributions are
not necessarily large in parameter regions with $\mHH>2\mstau$, \ie,
when the \HH~boson is heavy, this configuration still opens the
channels for stau production via on-shell \HH~exchange. As discussed
in section~\ref{sec:parameter_determination}, this could allow us to
determine the \HH~boson width and mass by investigating the
$\staustaubar$ invariant mass distribution.

\subsection{Direct stau production vs.\ staus from cascade decays}
\label{sec:cascade_decays}

So far we have focused on the analysis of direct stau production. But
in a \stau LOSP scenario, staus will also be generated in any SUSY
particle production process where heavier sparticles are produced that
cascade down to lighter ones and eventually decay into the LOSP, with
SM particles emitted along the SUSY decay chain. At hadron colliders,
the largest contribution to the overall SUSY cross section is usually
expected to originate from the production and subsequent decay of
color-charged SUSY particles, \ie, squarks and gluinos.  However, also
direct production of neutralinos and charginos (\eg, $\neu_i\cha_j$)
and associated production of a neutralino or chargino with a gluino or
squark can give sizeable contributions in large parts of the allowed
SUSY parameter space.

In \tabref{tab:compare_cascade}
%
\newcolumntype{L}[1]{>{\raggedright\arraybackslash}p{#1}} 
\newcolumntype{C}[1]{>{\centering\arraybackslash}p{#1}}
\TABULAR[t]{L{2cm}cC{1.8cm}C{1.8cm}C{1.9cm}C{2cm}}{
\hline\hline
\multicolumn{2}{c}{Benchmark point} & $\alpha$ & $\beta$ & $\gamma$ & $\epsilon$  \\
\hline\hline    
\multicolumn{2}{l}{LHC 7 TeV}           &&&& \\
$\sigma(\stau\stau^*)_{\text{DY}}$ &[$\fbarn$]  & $3.2 (2.3)$  & $12.5 \, (7.3) $
& $9.0
\, (5.6)$  & $7.95 \, (5.00)$   \\
$\sigma(\stau\stau^*)_{\text{\bbar}}$   &[$\fbarn$]     & $9.8 \, (5.1)$ & $0.03 \,
(0.02)$&
$19.2 \, (16.5)$ & $0.07 \, (0.06)$     \\
$\sigma(\stau\stau^*)_{\text{gg}}$      &[$\fbarn$]     & $0.1 \, (0.1)$  & $3.3 \, (2.4)$
&
$0.32 \, (0.25)$  & $0.01 \, (0.01)$  \\
$\sigma(\stau\stau^*)_{\text{all}}$     &[$\fbarn$]     & $13.1 \, (7.5)$  & $15.8 \,
(9.7)$ &
$28.5 \, (22.4)$ & $8.03 \, (5.07)$ \\
$\sigma(\tilde{g}\tilde{g})$    &[$\fbarn$]     & $0.05$ & $10^{-6}$    & $0.06$        &
$2.57$ \\
$\sigma(\tilde{g}\tilde{q})$    &[$\fbarn$]     & $0.63$ & $4\times 10^{-4}$ & $0.99$   &
$37.36$  \\
$\sigma (\tilde{q}\tilde{q})$   &[$\fbarn$]     & $1.18$  & $0.006$              & $2.41$
& $77.25$               \\
$\sigma(\tilde{\chi}\tilde{q})$+$\sigma (\tilde{\chi}\tilde{g})$&[$\fbarn$]     &  $0.481$
& $0.007$       & $0.72$        & $12.77$  \\ 
$\sigma(\tilde{\chi}\tilde{\chi})$ &[$\fbarn$] &        $20.4$   & $0.29$       & $19.8$ &
$91.78$ \\
\hline
\multicolumn{2}{l}{LHC 14 TeV}           &&&& \\        
$\sigma(\stau\stau^*)_{\text{DY}}$ &[$\fbarn$]  & $11.2 \, (5.64)$ & $37.5 \, (15.9)$ &
$28.0 \, (12.4)$        & $24.7 \, (11.2)$   \\ 
$\sigma(\stau\stau^*)_{\text{\bbar}}$   &[$\fbarn$]     & $58.4 \, (27.0)$ & $0.7 \,
(0.2)$ &
$113.3 \, (87.1)$& $0.5 \, (0.4)$ \\ 
$\sigma(\stau\stau^*)_{\text{gg}}$      &[$\fbarn$]     & $0.7 \, (0.4)$ & $17.4 \,
(11.1)$ &
$1.8 \, (1.3)$ & $0.07 \, (0.05)$  \\ 
$\sigma(\stau\stau^*)_{\text{all}}$     &[$\fbarn$]     & $70.3 \, (33.1)$ & $55.6 \,
(27.2)$ &
$143.1 \, (100.8)$ & $25.3 \, (11.6)$    \\ 
$\sigma (\tilde{g}\tilde{g})$   &[$\fbarn$]     & $20.2$        & $0.12$ & $20.8$ &
$232.19$         \\ 
$\sigma (\tilde{g}\tilde{q})$   &[$\fbarn$]     & $104.4$       & $2.46$ & $133.2$ &
$1328.4$ \\ 
$\sigma (\tilde{q}\tilde{q})$   &[$\fbarn$]     & $92.5$        & $6.46$ & $139.0$ &
$1301.1$ \\     
$\sigma (\tilde{\chi}\tilde{q})$+$\sigma (\tilde{\chi}\tilde{g})$&[$\fbarn$]    & $16.9$ 
& $1.08$        &  $22.4$ & $175.12$     \\ 
$\sigma (\tilde{\chi}\tilde{\chi})$ &[$\fbarn$] & $134.5$       & $6.40$ & $131.1$ &
$422.2$ 
\\ 
\hline\hline
}
{
  Hadronic cross sections for various SUSY pair production processes
  at the LHC with $\SqrtS=7~\TeV$ and $14~\TeV$.  For direct
  $\stau\staubar$-pair production, we list our cross section results
  before and after applying the kinematical cuts~(\ref{eq:cuts}),
  where the latter are given in parantheses. The other cross sections
  are inclusive NLO results obtained with \Prospino, where no
  kinematical cuts have been considered.
\label{tab:compare_cascade}
}
%
we compare our predictions for the direct $\stau\stau^*$-pair
production cross sections contributions from the \DY,
$\bbar$~annihilation, and $gg$~fusion channels with the inclusive
cross section for other SUSY particle production cross sections,
calculated at NLO with \Prospino, for the benchmark points defined in
\tabref{tab:benchmarks}.  (Note that an average squark mass
$m_{\tilde{q}}$ is listed in \tabref{tab:benchmarks}.) We sum over all
possible combinations of squark, neutralino, and chargino eigenstates.
For simplicity, we consider inclusive cross sections without any
kinematical cuts. Only for the direct stau production channels, cross
sections after applying the cuts~(\ref{eq:cuts}) are additionally
given in parentheses.  Considering the LHC, each cross section is
listed for $\SqrtS=7~\TeV$ and $14~\TeV$.

From a comparison of the inclusive production cross sections, we can
see that direct stau production is an important source of staus for
the benchmark points $\alpha$, $\beta$, and $\gamma$ at the LHC with
$\SqrtS=7~\TeV$. Only electroweak neutralino/chargino pair production
($\tilde{\chi}\tilde{\chi}$) can give comparable contributions. We
even find that direct stau production can constitute the dominant part
of the overall SUSY cross section together with
$\tilde{\chi}\tilde{\chi}$ production. The other cross sections are
suppressed at $\SqrtS=7~\TeV$ by the heavier masses of squarks and
gluinos. The situation changes for $\SqrtS = 14~\TeV$, where the
center-of-mass energy is high enough so that strongly interacting SUSY
particles can be produced copiously.  However, for point $\beta$,
where $\mhalf$ is particularly large, the gluino is so heavy that
direct stau production is always the dominant source for staus at
colliders. Still, we annotate that the LHC at $\SqrtS=7~\TeV$ might in
some scenarios provide a more suitable environment for the study of
direct stau production than the LHC at $\SqrtS=14~\TeV$, where staus
originating from cascade decays would need to be suppressed by
additional cuts.

It is also interesting to look more closely at the composition of the
total direct stau production cross section in these scenarios. Points
$\alpha$ and $\gamma$ have a very similar composition of
$\sigma(\stau\stau^*)_{\text{all}}$ (but a very different stau yield,
see also section~\ref{sec:cmssm_exceptional}) and the \bbar
annihilation channel is the dominant stau production mechanism for
both points.

Benchmark point $\beta$ is considered to illustrate the impact of a
large $\stau\stau^*h^0$ coupling. Here, large values of $\mu$ and \TB
together with a relatively large mixing result in this large coupling
(\cf\ section~\ref{sec:exceptionalYstau} and
appendix~\ref{app:stauhiggs_couplings}). However, such very large
couplings are in strong conflict with CCB constraint~(\ref{eq:CCB}).
For this point, the \HH~boson is very massive and \bbar annihilation
and gluon fusion into an intermediate \HH~boson is suppressed by the
heavy particle's propagator. Stau production is thereby dominated by
the \DY\ channel and gets sizeable contributions from the gluon fusion
channel, where especially processes mediated by the \hh~boson are
important.

Finally, benchmark scenario $\epsilon$ differs from the above
scenarios by much smaller values of $\mhalf$ and $\TB$. It could be
considered a `typical' \stau LOSP scenario, without an exceptional
stau yield. In this case, the direct stau production cross section is
well described by the \DY\ process, whereas the $\bbar$ and $gg$
channels are basically negligible. Moreover, the indirect stau
production mechanisms are much more efficient than the direct ones.

\section{Collider tests of an exceptionally small relic stau abundance}
\label{sec:cmssm_exceptional}

In the preceding sections we have considered various aspects of stau
production at hadron colliders with emphasis on parameter regions that
allow for an exceptionally small yield of a long-lived
stau~(\ref{eq:exceptionalYstau}). Because of the appealing features
described in section~\ref{sec:exceptionalYstau}, we now discuss the
testability of such an exceptional yield in collider experiments.
While related prospects for collider phenomenology were already
addressed in
refs.~\cite{Ratz:2008qh,Pradler:2008qc,Endo:2010ya,Endo:2011uw}, we
show here for the first time that contributions from $\bbbar$
annihilation and gluon fusion to direct stau production can play a
particularly important role for this testability at hadron colliders.

Assuming a standard cosmological history with a reheating temperature
$\TR>\mstau/20$, the key requirements for an exceptionally small
thermal relic stau yield~(\ref{eq:exceptionalYstau}) are (i)~a
relatively small stau mass of $\mstau\lesssim 200~\GeV$, (ii)~the mass
pattern $2\mstau\simeq\mHH$, which allows for primordial $\stau$
annihilation via the $\Hhiggs$ resonance~\cite{Pradler:2008qc}, and/or
(iii)~enhanced stau-Higgs couplings, which are often associated with a
sizeable stau-left-right mixing~\cite{Ratz:2008qh,Pradler:2008qc}, as
described in section~\ref{sec:exceptionalYstau} and
appendix~\ref{app:stauhiggs_couplings}. Now, our studies of direct
stau production in the previous sections demonstrate clearly that the
contributions from $\bbbar$ annihilation and gluon fusion are
sensitive to all three of these requirements. In contrast, the \DY\ 
process is sensitive to $\mstau$ and the stau-mixing angle
$\thetastau$ only.

\subsection*{\boldmath Excess of direct stau production cross sections
over \DY\ predictions}

Based on our results in sections~\ref{subsec:numerical}
and~\ref{sec:cmssm}, we already know that $\bbbar$-annihilation and
gluon-fusion processes can lead to direct stau production cross
sections that exceed the \DY\ predictions significantly, in
particular, for $2\mstau\lesssim\mHH$ and/or enhanced stau-Higgs
couplings. This motivates us to explore the ratio
\begin{align}
 R=\sigma(\stau\stau^*)_{\text{all}}/\sigma(\stau\stau^*)_{\text{DY}} 
\label{eq:R}
\end{align}
as a potential indicator for a cosmological scenario with an
exceptionally small stau yield. Here the total direct stau production
cross section, $\sigma(\stau\stau^*)_{\text{all}}$, and the \DY\ 
prediction, $\sigma(\stau\stau^*)_{\text{DY}}$, are considered after
applying the cuts~(\ref{eq:cuts}) for scenarios with a long-lived
stau.

Before presenting and discussing our theoretical results for $R$, let
us comment on its experimental determination which will have to rely
on measurements of $\mstau$ and $\sigma(\stau\stau^*)_{\text{all}}$.
While an accuracy of $<1\%$ is expected for a $\mstau$ determination
at the LHC~\cite{Ambrosanio:2000ik,Ellis:2006vu},
$\sigma(\stau\stau^*)_{\text{all}}$ measurements may be more
difficult, as mentioned in section~\ref{sec:phenolonglived}.  If
indirect stau production is significant, they will require jet/lepton
vetos and/or additional kinematical cuts. With a precisely known
$\mstau$, $\sigma(\stau\stau^*)_{\text{DY}}$ can be calculated
theoretically with an uncertainty of about a factor of 2 that is
related to its dependence on $\thetastau$; see
figures~\ref{fig:crosssections1}\,(b)
and~\ref{fig:crosssections2}\,(a). To obtain a conservative estimate
of $R$, one will then evaluate~(\ref{eq:R}) with the maximum
$\sigma(\stau\stau^*)_{\text{DY}}$ at $\thetastau=0$. In fact, as
already addressed in section~\ref{sec:parameter_determination}, a
measurement of $\thetastau$ based on direct stau production is
conceivable only if the \DY\ contribution dominates, \ie, for
$R\simeq1$. On the other hand, studies of staus produced in cascade
decays may help to indirectly probe $\thetastau$~\cite{Kitano:2010tt}
and thereby to determine $R$ more precisely. Thus, we present in the
following results for $R$ that are not conservative estimates but
theoretical predictions taking into account
$\sigma(\stau\stau^*)_{\text{DY}}$ with its full $\thetastau$
dependence. For an unknown $\thetastau$, $R>2$ will then be a required
indication for sizeable contributions of the $\bbbar$ and $gg$
channels.

In figure~\ref{fig:ratio7tevyield} 
%
\FIGURE[h!]{
\includegraphics[width=0.97\textwidth]{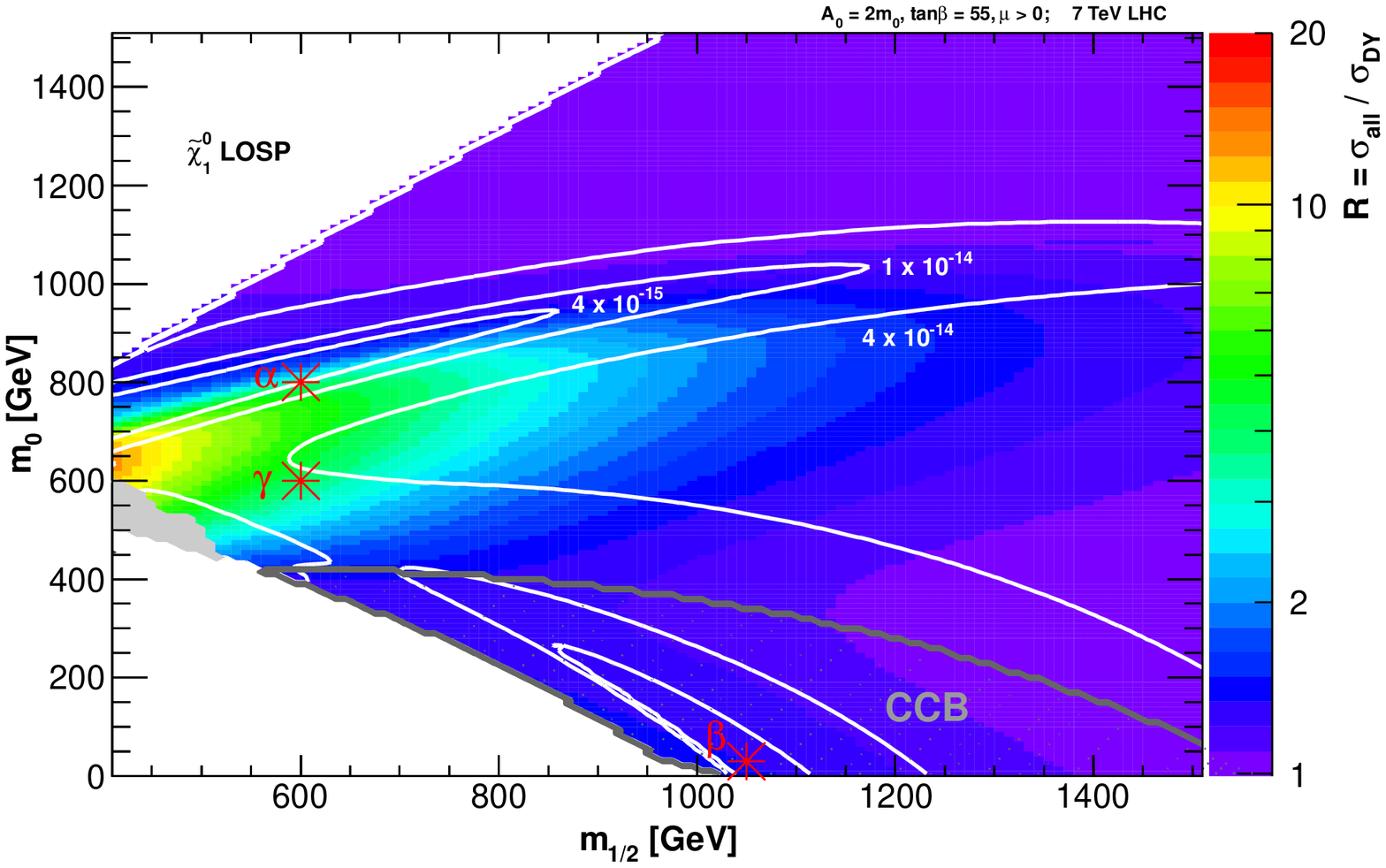}\\%
\includegraphics[width=0.97\textwidth]{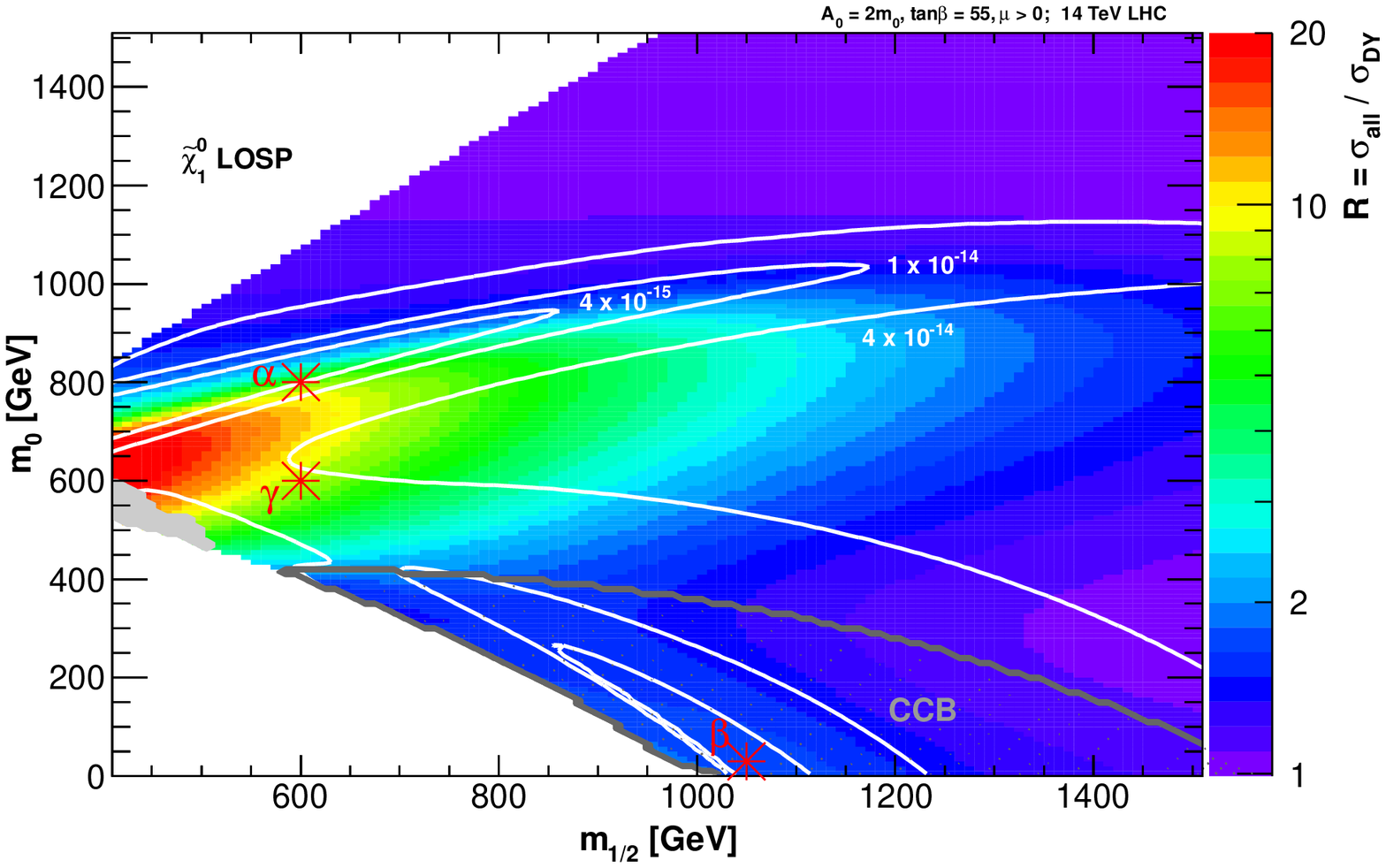}
\caption{The ratio $R=\sigma(\stau\stau^*)_{\text{all}}/\sigma(\stau\stau^*)_{\text{DY}}$ 
  (shaded contours, colored) for the LHC with $\SqrtS = 7~\TeV$ (top
  panel) and $14~\TeV$ (bottom panel) and $\Ystau=4\times10^{-15}$,
  $10^{-14}$, $4\times10^{-14}$ (white lines) in the \mzero-\mhalf
  CMSSM plane with $A_0=2\mzero$, $\tanbeta=55$, and $\mu>0$. The CCB
  constraint (gray hatched region) and excluded/unconsidered regions
  are as in figure~\ref{fig:cmssm_LHC}. The labeled (red) stars
  indicate the location of the benchmark points $\alpha$, $\beta$, and
  $\gamma$, defined in table~\ref{tab:benchmarks}.}
\label{fig:ratio7tevyield}
}
%
the shaded (colored) contours indicate our theoretical predictions for
$R$ at the LHC with $\SqrtS=7~\TeV$ (top panel) and $14~\TeV$ (bottom
panel) in the \mzero-\mhalf CMSSM plane with $A_0=2\mzero$,
$\tanbeta=55$, and $\mu>0$. The white lines show contours of
$\Ystau=4\times10^{-15}$, $10^{-14}$, $4\times10^{-14}$ as obtained
with \micromegas~\cite{Belanger:2010gh}. The labeled (red) stars
indicate the location of the benchmark points $\alpha$, $\beta$, and
$\gamma$, defined in table~\ref{tab:benchmarks}. Excluded and
unconsidered regions are as in figure~\ref{fig:cmssm_LHC}, and also
the region disfavored by the CCB constraint~(\ref{eq:CCB}) is
indicated but now by the gray hatched region.

The $R$ contours show very explicitly that the \DY\ prediction can
underestimate the direct $\stau\stau^*$ production cross section by up
to a factor of 10 (20) for $\SqrtS=7~\TeV$ ($14~\TeV$). This
demonstrates again the potential importance of the $\bbbar$ and $gg$
channels included in our calculations. In fact, their effect with
respect to the \DY\ predictions is more relevant at $\SqrtS = 14~\TeV$
than at $7~\TeV$. This results from the $\bbbar$ and gluon
luminosities in the proton that benefit more strongly from the higher
$\SqrtS$ than those of the lighter quarks. On the other hand, also
indirect stau production can be much more efficient for
$\SqrtS=14~\TeV$; cf.\ table~\ref{tab:compare_cascade}. Thereby, it
may even be more difficult to identify direct stau production events
at $\SqrtS = 14~\TeV$ despite potentially larger values of $R$.

In table~\ref{tab:YstauR}
%
\TABULAR[t]{lcC{1.6cm}C{1.6cm}C{1.6cm}C{1.6cm}}{
\hline\hline
\multicolumn{2}{c}{Benchmark point} & $\alpha$& $\beta$ &$\gamma$ & $\epsilon$  
\\
\hline\hline    
$R_\mathrm{LHC7}$       &       & $3.3$ & $1.3$ & $4.0$ & $1.01$
\\
$R_\mathrm{LHC14}$      &       & $5.8$ & $1.7$ & $8.1$ & $1.04$
\\
\hline
\stauY\,\,[$10^{-15}$]  & &$3.5$         & $2.5$        & $37.7$        & $164$  
\\
\hline\hline
}{
  The stau yield $\Ystau$ and
  $R=\sigma(\stau\stau^*)_{\text{all}}/\sigma(\stau\stau^*)_{\text{DY}}$
  at the LHC with $\SqrtS = 7~\TeV$ ($R_\mathrm{LHC7}$) and $14~\TeV$
  ($R_\mathrm{LHC14}$) for the benchmark scenarios $\alpha$, $\beta$,
  $\gamma$ and $\epsilon$, defined in table~\ref{tab:benchmarks} and
  partially indicated in figure~\ref{fig:ratio7tevyield}. The stau
  yield is obtained from \micromegas and the $R$ values from the
  respective cross sections after kinematical cuts~(\ref{eq:cuts})
  given in parantheses in table~\ref{tab:compare_cascade}.
\label{tab:YstauR}
}
%
we list the $R$ values at the LHC with $\SqrtS = 7~\TeV$
($R_\mathrm{LHC7}$) and $14~\TeV$ ($R_\mathrm{LHC14}$) and the stau
yield $\Ystau$ for the benchmark points $\alpha$, $\beta$, $\gamma$,
and $\epsilon$. We see again that $R$ increases when going from
$\SqrtS=7~\TeV$ to $14~\TeV$. This effect is most pronounced for the
points~$\alpha$ and~$\gamma$ for which $\bbbar$ annihilation dominates
the direct stau production cross section. A considerable 30\% increase
of $R$ is predicted also for point~$\beta$ for which gluon fusion
contributes up to about $40\%$ of $\sigma(\stau\stau^*)_{\text{all}}$;
\cf~table~\ref{tab:compare_cascade}.

To understand the shape of the $R$ contours in
figure~\ref{fig:ratio7tevyield}, it is instructive to look at the
lines of constant $\mstau$ and $\mHH$, which are shown for the same
CMSSM parameter choice in figure~\ref{fig:cmssm_LHC}.  By
interpolating between the intersections of the $\mstau=200~\GeV$ and
$\mHH=400~\GeV$ and of the $\mstau=400~\GeV$ and $\mHH=800~\GeV$
contours, on can infer the location of the line with $2\mstau=\mHH$.
Only below this line, on-shell \HH exchange is possible and can lead
to $R\gg 2$ in the vicinity of this line, where a smaller $\mHH$
allows for a larger $R$. By going along a contour on which $\mstau$
does not change (such as the $\mstau=200~\GeV$ contour) from the
region with smaller $\mHH$ into the direction with larger $\mHH$, one
encounters the qualitative behavior that is illustrated in
figure~\ref{fig:crosssections1}\,(c).

Let us now turn to the main aspects of the $\Ystau$ contours in
figure~\ref{fig:ratio7tevyield}; for additional details we refer
to~\cite{Pradler:2008qc} in which those contours were studied in the
same CMSSM plane. An exceptionally small
yield~(\ref{eq:exceptionalYstau}) can be found in the two separate
regions enclosed by the $\Ystau=4\times 10^{-15}$ contour. The region,
to which point $\alpha$ belongs, allows for primordial $\stau$
annihilation via the $\Hhiggs$ resonance, and the region, to which
point $\beta$ belongs, for efficient $\stau$ annihilation via enhanced
stau-Higgs couplings. While the latter region is in conflict with the
CCB contstraint~(\ref{eq:CCB}), we still include this point in our
discussion of the testability of an exceptionally small $\Ystau$ at
colliders.

Comparing regions with $R\gtrsim2$ to those with
$\Ystau<4\times10^{-15}$, we find that there is no one-to-one link
between a sizeable $R$ value and an exceptionally small yield. For
example, while the point $\beta$ is associated with such an
exceptional yield, we obtain $R<2$, even at the LHC with $\SqrtS =
14~\TeV$, as can be seen in table~\ref{tab:YstauR}. Moreover, in
figure~\ref{fig:ratio7tevyield}, also moderate values of $R\simeq
2$--$3$ occur in the $2\mstau\simeq\mHH$ region enclosed by the
$\Ystau< 4\times 10^{-15}$ contour. On the other hand, point $\gamma$
is associated with $R\simeq 4$--$8$ while the yield at this point
exceeds the limit~(\ref{eq:exceptionalYstau}) by more than one order
of magnitude. Only for points such as $\alpha$, one finds both an
exceptionally large $R\simeq3$--$6$ and an exceptionally small stau
yield. Thus, a sizeable $R$ could very well be a first hint for the
possibility of efficient stau-annihilation in the early Universe but
additional investigations will be crucial to clarify the situation.

\subsection*{\boldmath Exceptional $\Ystau$ and $R$ 
in models with non-universal Higgs masses}

Before proceeding, we would like to illustrate more clearly that a
region with resonant primordial stau annihilation and an exceptionally
small $\Ystau$ can be associated with different values of $R$. The
resonance condition $2\mstau \approx \mHH$ is found in
$\mzero$--$\mhalf$ CMSSM planes only in a small horizontal `funnel'
region and for specific combinations of parameters. In less
constrained models with non-universal Higgs masses (NUHM) this mass
pattern occurs in a more generical way
and already for smaller values of \TB. 

In figure~\ref{fig:nuhm}
%
\FIGURE[t]{
\includegraphics[width=.7\textwidth]{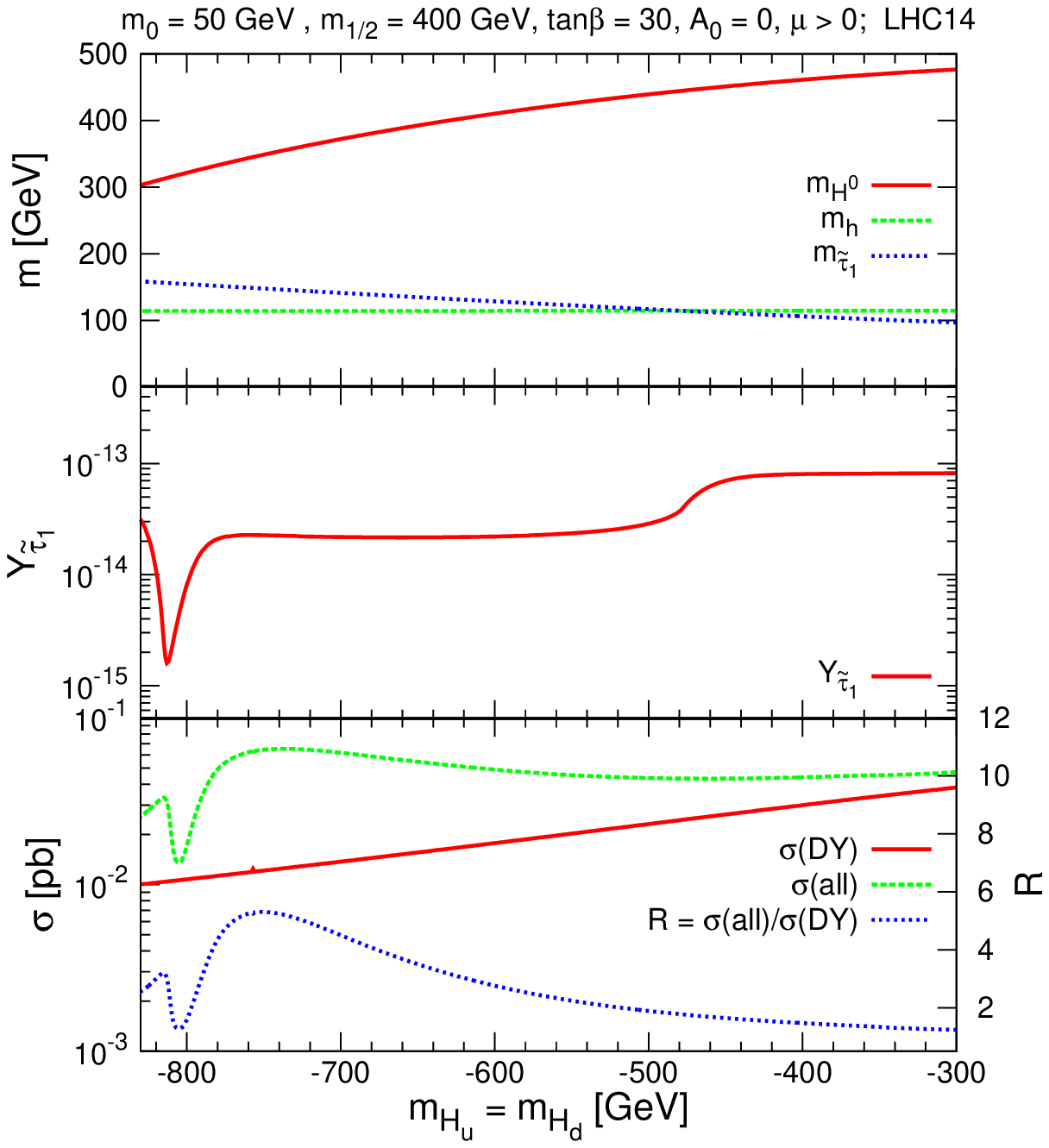}
  \caption{The masses $\mstau$, $\mh$, and $\mHH$ (top panel), 
    the stau yield $\Ystau$ (middle panel), the cross sections
    $\sigma(\stau\stau^*)_{\text{all}}$ and
    $\sigma(\stau\stau^*)_{\text{DY}}$ and $R$ at the LHC with
    $\SqrtS=14~\TeV$ (bottom) as a function of the Higgs-mass
    parameter $m_{H_u}=m_{H_d}=m_0^H$, defined at the high scale, in
    the NUHM1 model with $\mhalf=400~\GeV$, $\mzero=50$, $\TB=30$,
    $A_0=0$.}
\label{fig:nuhm}
}
%
we consider a scenario in which the two Higgs mass parameters are
equal (and negative) but different from the other high scale scalar
mass parameter,
$m_{H_u}=m_{H_d}=m_0^H\ne m_0$,
which is a framework denoted as NUHM1 model. We vary $m_0^H$ and set
the other parameters to
$\mzero=50~\GeV$, $\mhalf=400~\GeV$, $\tanbeta=30$, $A_0=0$, and $\mu>0$,
which are defined as in the CMSSM. In the top panel, \mHH is indicated
by the solid (red) line, \mh by the dashed (green) line, and \mstau by
the dotted (blue) line. The middle panel shows $\Ystau$. In the bottom
panel, $\sigma(\stau\stau^*)_{\text{DY}}$ is shown by the solid (red)
line, $\sigma(\stau\stau^*)_{\text{all}}$ by the dashed (green) line,
and $R$ by the dotted (blue) line. Note that the negative mass
parameter $m_0^H$ has to be understood as the root of the modulus of a
negative squared soft-mass of the Higgses. For a more negative
$m_0^H$, the \stau is no longer the LOSP and eventually (\ie, for an
even more negative $m_0^H$) the stability bound
$m_{H_{u/d}}^2 + \vert\mu\vert^2 \geq 0$
can be violated at $\MGUT$. In such cases, there might be a vacuum
instability leading to electroweak symmetry breaking already at
$\MGUT$.

Considering the top panel, one sees that $\mHH$ gets smaller and
$\mstau$ larger towards smaller values of $m_0^H$. For
$m_0^H\simeq-815~\GeV$, one finds the resonance condition
$2\mstau=\mHH$. In a narrow region around this point, a significant
depletion of the stau yield $\Ystau$ by about one order of magnitude
down to an exceptional value of $\lesssim 2\times 10^{-15}$ can be
seen in the middle panel. To the right of this resonance point, \ie,
for $m_0^H>-815~\GeV$, $2\mstau<\mHH$ so that on-shell $\HH$-boson
exchange can contribute to direct stau production. Thereby, this leads
to a significant contribution of the $\bbbar$ and $gg$ channels to
$\sigma(\stau\stau^*)_{\text{all}}$ and $R$ increases
significantly up to about 5. However, the maximum of
$\sigma(\stau\stau^*)_{\text{all}}$ and of $R$ is shifted away from
the resonance point at which $\Ystau$ approaches its minimum, and
towards this resonance point, $\sigma(\stau\stau^*)_{\text{all}}$ and
$R$ decrease significantly and show even a local minimum. This
behavior is qualitatively different from the one shown in
figure~\ref{fig:crosssections1}\,(c) where the maximum of
$\sigma(\stau\stau^*)_{\text{all}}$ is very close to the threshold for
on-shell Higgs exchange, $2\mstau=\mHH$. The difference results from
the kinematical cuts~(\ref{eq:cuts}) applied in figure~\ref{fig:nuhm}.
As shown in figure~\ref{fig:distributions_pt_beta}, the $\bbbar$ and
$gg$ channels lead to a significant excess of staus with low $\pT$ and
low $\beta$ compared to the \DY\ channel in the vicinity of the
resonance.  Thus, by imposing the $\pT$ and $\beta$
cuts~(\ref{eq:cuts}), one loses a significant amount of direct stau
production events near the resonance condition $2\mstau=\mHH$.

For the CMSSM scans in figure~\ref{fig:ratio7tevyield}, the above
explains also that maximum values of $R$ are shifted to some extend
away from the $2\mstau\approx\mHH$ region and deeper into the
$2\mstau<\mHH$ region when moving along contours of constant $\mstau$.
In fact, the dip at the resonance is present also in those scans but
not visible due to a limited resolution.

Based on this finding and as already addressed in
section~\ref{sec:kinematicalcuts}, we would thus recommend a shifting
of the $\pT$ and $\beta$ cuts to include events with smaller $\pT$ and
$\beta$.  If this is feasible such that direct stau production events
can still be identified confidently and if close to the resonance, one
will find $R$ values that increase substantially when lowering the
cuts on $\pT$ and $\beta$.  Such a behavior can then provide a hint
for resonant stau annihilation in the early Universe and thereby for
the possibility of an exceptionally small $\Ystau$.  For further
clarification, we propose in the following studies of differential
direct stau production cross sections.

On the side, we remark that the mass scale of the heavy Higgs bosons
is, by renormalization group running, fixed by \mhalf in NUHM1 models.
Although we can always realize $2\mstau\approx\mHH$ in NUHM1 models,
these parameter points have a light \stau only for small \mhalf.
Thus, one does not find scenarios with efficient direct stau
production and subdominant indirect stau production from cascade
decays. However, in the alternative less constrained NUHM2 models,
$m_{H_u}$ and $m_{H_d}$ are chosen independently. These two parameters
can be traded for the parameters $m_A$ and \TB at the low scale. In
the context of direct stau production, this setup can effectively be
described by the low scale parameters $\mstau$, $m_A$, and $\TB$ with
implications shown in section~\ref{sec:production}. Now, the different
stau production mechanism might `decouple' and there is the
possibility that direct production remains as the only relevant source
for stau pairs at hadron colliders.

\subsection*{\boldmath With a little help from differential 
direct $\stau\stau^*$ production cross sections}

So far our discussion in this section has focussed on the quantity $R$
(and thereby on the integrated direct stau production cross section)
and its possible dependence on the $\pT$ and $\beta$ cuts. Now, with
long-lived staus, it will be a realistic possibility to measure also
differential direct $\stau\stau^*$ production cross sections such as
the ones discussed in section~\ref{sec:phenolonglived}.

For a situation with a sizeable value of $R>3$, as encountered for
points $\alpha$ or $\gamma$, the situation is most promising since the
differential distributions differ substantially from the \DY\ 
predictions and provide valuable additional information. Our results
show that such a large $R$ results from contributions of the $\bbbar$
channel that become substantial near $2\mstau=\mHH$, \ie, the
threshold for on-shell $\HH$ exchange. Manifestations of an on-shell
$\HH$ exchange that leads to the large $R$ value will then show up in
the invariant mass distribution of the directly produced stau pair in
the form of a resonance peak at $\mstaustau=\mHH$. Considering those
$\mstaustau$ distributions for the points $\alpha$ and $\gamma$ in
figures~\ref{fig:distributions_minv}\,(a) and~(c), respectively, this
feature is clearly visible. As already explained in
section~\ref{sec:parameter_determination}, the associated $\mHH$ can
then be extracted and compared to $2\mstau$ which marks also the
minimum value of $\mstaustau$ in such a distribution. In fact, the
$\HH$ resonance at point $\gamma$ shows that $\mHH$ is still close to
$2\mstau$ but too large to allow for highly efficient resonant
primordial stau annihilation. This is different for point $\alpha$
where the $\mHH$ resonance peak sits much closer to the minimum
$\mstaustau$ value. This tells us immediately that this is a scenario
with $2\mstau\approx\mHH$ and thus with the possibility of efficient
resonant primordial stau annihilation. Moreover, the $\mstaustau$
distribution provides us also with the width $\Gamma_{\HH}$ which is
an important input in calculations of $\Ystau$ in that region.

For more moderate values of $R\simeq2$--$3$, it may still be possible
to clarify the situation. If such an $R$ value results from on-shell
$\HH$ exchange, again a $\HH$ resonance peak will show up in the
corresponding $\mstaustau$ distribution so that the situation can be
resolved as explained above. In the possible case that no resonance
peak shows up in the $\mstaustau$ distribution (even under excellent
experimental conditions), the excess over the \DY\ prediction may be
due to a $2\mstau$ value that is slightly above $\mHH$ so that
on-shell $\HH$ is just not possible. This is the situation
encountered, \eg, for $m_0^H\simeq-820~\GeV$ in the NUHM1 model
considered in figure~\ref{fig:nuhm}. Here we still expect an excess of
events over the \DY\ predicition towards the minimum $\mstaustau$
value in the invariant mass distribution but---without a resonance
peak---it will not be possible to determine $\mHH$ or $\Gamma_{\HH}$
in experimental studies of direct stau production. Nevertheless, those
quantities may still be accessible at the LHC, \eg, via studies of
associated $\bbbar\hh/\HH$ production with $\hh/\HH\to\mu^+\mu^-$,
which will remain conceivable also for very heavy colored sparticles.
In fact, these reactions are considered to be very promising for
$\mHH$ measurements at the LHC, despite the relatively small
$\BR(\hh/\HH\to\mu^+\mu^-)$~\cite{Ball:2007zza}. If this is indeed
feasible, the described shape of the $\mstaustau$ distribution
together with a finding of $2\mstau\approx\mHH$ will then point to the
possibility of an exceptionally small $\Ystau$. In fact, also in case
of a more sizeable $R$, a second independent determination of $\mHH$
in studies of other processes will provide an important consistency
check and test whether the observed resonance is indeed associated
with the $\HH$ boson.

For a scenario with resonant primordial stau annihilation and the
associated mass pattern $2\mstau\approx\mHH$, the differential
distributions with respect to $\pT$ and $\beta$ will also provide
valuable information, which is already evident from our discussion on
the $\pT$ and $\beta$ cut dependence of $R$ and from
section~\ref{sec:kinematicalcuts}. However, we would like to stress
once more that these distributions are very different from the \DY\ 
prediction towards low $\pT$ and low $\beta$ values for both the
$2\mstau<\mHH$ case and the $2\mstau>\mHH$ case. This can be seen
explicitly in figure~\ref{fig:distributions_pt_beta} for two different
scenarios with $|2\mstau-\mHH|=10~\GeV$.

The most challenging situation with respect to the testing of the
viability of an exceptionally small $\Ystau$ is encountered for
scenarios such as point $\beta$. Here the exceptional yield results
only from enhanced stau-Higgs couplings. Again the $\mstaustau$
distribution shows differences with respect to the \DY\ prediction as
can be seen in figure~\ref{fig:distributions_minv}\,(b). These are
manifestations of the $gg$ channel contribution: While the
$\HH$-mediated processes lead to the $\HH$ resonance peak at
$\mstaustau\simeq 760~\GeV$, the excess that shows up over a wide
$\mstaustau$ range (and in particular towards lower $\mstaustau$)
results mainly from the $\hh$-mediated processes that benefit from the
significantly enhanced $\stau\stau^*\hh$ coupling. However, from this
information alone, it will be very difficult to infer confidently that
an exceptional $\Ystau$ is possible. Moreover, in our numerical
studies, we find that the $gg$ channel contribution usually does not
lead to $R>2$ (\cf\ point $\beta$ in table~\ref{tab:YstauR}). Thus,
one will have to rely on other investigations that are sensitive to
large values of $\tanbeta$, $|\mu|$, and/or $|A_\tau|$. As outlined in
section~9 of ref.~\cite{Pradler:2008qc}, some of those investigations
provide additional motivation for a future linear collider at which
processes such as $e^+e^-\to\stau\stau^*\hh/\HH$ and
$\gamma\gamma\to\stau\stau^*\hh/\HH$ can indeed allow for direct
experimental determinations of the stau-Higgs couplings.

Experimental insights into $\tanbeta$, $|\mu|$, and/or $|A_\tau|$ will
be relevant also for the scenarios with larger $R$ discussed above
since larger stau-Higgs couplings are associated with more efficient
resonant primordial stau annihilation and thereby with smaller
$\Ystau$. Here analyses of the decays $\HH\to\tau\bar{\tau}/\bbbar$
and associated limits/findings in the $m_A$--$\tanbeta$ plane will be
highly interesting~\cite{Schumacher:2011jq,Chatrchyan:2011nx}. Equally
exciting will be the outcome of the ongoing searches for the decay
$B_s\to\mu^+\mu^-$, \eg, at the LHCb experiment~\cite{Aaij:2011rj}. As
can be seen in table~\ref{tab:benchmarks}, with a sensitivity to a
$\BR(B_s \to \mu^+\mu^-)$ as small as $10^{-8}$, one will be able to
test points such as $\alpha$ and $\beta$ which allow for an
exceptionally small stau yield.

\section{Conclusions}
\label{sec:conclusion}

We have studied the direct hadronic production of a pair of staus
$\stau\stau^*$ within the MSSM. In addition to the well-known \DY\ 
process, we have considered $\stau\stau^*$ production processes
initiated by $\bbar$~annihilation and gluon fusion, with all
third-generation mixing effects taken into account. This allows us to
provide reliable predictions of hadronic slepton production at
$\ord(\alpha_s^2 \alpha^2)$. These predictions are independent of the
stau lifetime and applicable in $\neu_1$ LSP scenarios with the
$\stau$ NLSP as well as in settings in which the $\stau$ is
long-lived.

In considerable parts of the MSSM parameter space, we find that the
additional $\bbar$ and $gg$ channels lead to a substantial enhancement
of the direct stau production cross section over the NLO-QCD \DY\ 
prediction. This enhancement can be even larger than one order of
magnitude. Particularly significant corrections are found when direct
stau production can proceed via the exchange of an on-shell heavy
CP-even Higgs boson \HH and when the left-right-stau mixing is
sizeable. Moreover, the contributions of the $\bbar$ and $gg$ channels
are enhanced also in the case of large stau-Higgs couplings which are
associated with large values of $\TB$, $|\mu|$, and/or $|A_{\tau}|$
and thereby again with a sizeable left-right-stau mixing.

In cosmologically motivated scenarios with gravitino or axino dark
matter, the stau can be the lightest SUSY particle within the MSSM. In
an R-parity conserving setting, the stau will then typically be
long-lived since it can only decay into the extremely weakly
interacting gravitino or axino. Such long-lived staus can lead to the
striking collider signature of a charged massive particle, \ie, a
slowly moving charged object with large transverse momentum. SM
backgrounds to this signature originate only from slow moving muons
and kinematical cuts on the velocity $\beta$ and $p^T$ are required to
separate these backgrounds. For such scenarios, we have investigated
differential distributions of the directly produced staus and the
associated integrated cross sections after application of the
kinematical cuts. Our findings show that staus from the $\bbar$ and
$gg$ channels are often softer and slower than those produced in the
\DY\ channel. We thus recommend that experiments should try to soften
their cuts to improve sensitivity to these additional channels. Here,
a detailed study including detector effects should be performed to
investigate the possible discovery reach and exclusion limits.

Once long-lived staus are observed at the LHC, one will be able to
measure the stau mass $\mstau$ accurately, \eg, in TOF measurements. A
measurement of the direct stau production cross section---if governed
by \DY\ processes---will then probe the stau mixing
angle~$\thetastau$. However, we have shown that there is also the
possibility of an early observation of a significant excess of the
direct stau production cross section over the \DY\ prediction.
Indeed, such a finding can be a first hint for Higgs physics at the
LHC. Moreover, our results demonstrate that measurements of the
distribution of direct stau production events as a function of the
invariant stau-antistau mass $\mstaustau$ can give $\mstau$,
independently, and, more importantly, the mass of the heavy CP-even
Higgs boson $\mHH$. Although challenging, with precise measurements of
the invariant mass $\mstaustau$, we find that these distributions may
provide us also with the Higgs width $\GammaHH$. In fact, for large
event samples, both the $\mHH$ and the $\GammaHH$ determination might
even be possible in parameter regions in which direct stau production
is governed by the \DY\ channels.

We have also presented results that are encouraging for the stopping
of long-lived staus in the collider
detectors~\cite{Martyn:2006as,Asai:2009ka,Pinfold:2010aq,Freitas:2011fx}
or in additional surrounding
material~\cite{Goity:1993ih,Hamaguchi:2004df,Feng:2004yi,Hamaguchi:2006vu}. 
With a large number of stopped staus, such experiments could allow for
analyses of their late decays. Those analyses may give unique insights
into the nature of the LSP into which the stau decays and into the
vertex that governs this decay. In this way, it could be possible to
use collider experiments to probe physics at scales as high as the
Peccei--Quinn scale~\cite{Brandenburg:2005he,Freitas:2011fx} or the
Planck scale~\cite{Buchmuller:2004rq}. A crucial criterium for the
stopping of large numbers of staus is that a large fraction of them is
produced with relatively slow initial velocities. This is exactly what
we find if the staus are directly produced via the $\bbbar$ and $gg$
channels in the appealing scenarios in which the thermal relic stau
abundance can be exceptionally small. Here the number of stopped
directly produced staus may exceed expections based on the \DY\ 
channels by more than one order of magnitude.

Within the CMSSM, we have provided cross section predictions for
direct stau production at the LHC with $\SqrtS=7~\TeV$ and $14~\TeV$
and at the Tevatron. Here our focus has been on a particular
\mzero-\mhalf plane in which the long-lived staus can have an
exceptionally small thermal relic stau abundance. For the considered
scenarios with $\TB=55$, we predict substantial contributions from the
$\bbar$ and $gg$ channels in large areas of the \mzero-\mhalf plane.
By comparing our results for the Tevatron with the associated existing
upper limit of $10~\fbarn$, a small strip along the conservative lower
stau mass limit from LEP of $82~\GeV$ is found to be disfavored in
that particular plane. On the other hand, our cross section
predicitions show that it will be difficult to discover directly
produced staus with $\mstau\gtrsim 100~\GeV$ at the Tevatron. This is different 
for the LHC with $\SqrtS=7~\TeV$ where tests of direct stau production will 
be possible in the very near future. In particular, the CMSSM parameter 
region in which an exceptionally small stau yield is possible because of 
resonant primordial stau annihilation will be tested very soon.

To address the relative importance of direct stau production with
respect to indirect stau production in cascade decays, we have
considered four CMSSM benchmark points. Our calculations show that
direct stau production can be one of the dominant contributions
especially in the cosmologically motivated scenarios with an
exceptionally small stau yield. Moreover, we find that the early LHC
with $\SqrtS=7~\TeV$ may provide a better environment for the study of
direct stau production than the LHC with $\SqrtS=14~\TeV$ at which
indirect stau production is often expected to become dominant.

Finally, we have explored the testability of the conditions that allow
for an exceptionally small stau yield at the LHC. Within the CMSSM and
for a NUHM1 scenario, we have studied whether an excess of the direct
stau production cross section over the \DY\ prediction can be used as
an indicator for the possibility of an exceptional yield. Although no
one-to-one link is found, a large excess over the \DY\ prediction can
very well be a first hint of efficient stau annihilation in the early
Universe. Additional investigations---especially in the Higgs
sector---will still be crucial to clarify the situation. Important
additional insights can be provided by studying the differential
distributions of the directly produced $\stau\stau^*$ pairs. In
particular, the differential distribution as a function of the
invariant mass $\mstaustau$ may clarify the situation in a striking
way: If the mentioned large excess is observed and if one is in the
region that allows for efficient resonant primordial stau
annihilation, in which $2\mstau\approx\mHH$, this $\mstaustau$
distribution will show a pronounced \HH resonance peak right at the
beginning.

In summary, direct stau production including \bbar and $gg$ channels
can be probed in the very near future or even with data already
available. Once discovered, this process might shed light on SUSY
parameters and important cosmological questions soon.

\section*{Acknowledgments} 
We are grateful to T.~Hahn, W.~Hollik, J.~Germer, P.~Graf, and A.~Landwehr
for valuable discussions.
This work was supported in part 
by the Cluster of Excellence 
``Origin and Structure of the Universe''
and by the US DOE under contract No.~DE-FG02-95ER40896.

\appendix

\section{Stau sector in the MSSM}
\label{app:staus}

In this appendix we introduce the notation that we use to describe the
stau sector in the MSSM, including the left-right mixing of the two
stau masss eigenstates and stau-Higgs couplings.

\subsection{Stau mixing and mass eigenstates}
\label{app:stau_mixing}

After electroweak symmetry breaking the soft-breaking terms in the
MSSM Lagrangian induce mixing between the left- and right-handed gauge
eigenstates in the sfermion sector.  Under the assumption of minimal
flavor violation, the sfermion mass matrices and trilinear couplings
are diagonal in family space and no mixing occurs amongst different
flavors. Furthermore, we assume all parameters to be real.  Including
all F-term , D-term and soft-term contributions, the stau-mass-squared
matrix then reads in the basis of gauge eigenstates $(\stauL,
\stauR)$:
\begin{equation}
    \label{eq:stau-mass-matrix}
    \stauMAT = 
    \begin{pmatrix}
    \mtau^2 + \mLL^2 & \mtau \Xtau \\
             \mtau \Xtau      & \mtau^2 + \mRR^2      
    \end{pmatrix} = 
    (\stauROT)^\dagger
    \begin{pmatrix}
    \mstau^2 & 0\\
    0 & \mstautwo^2   
    \end{pmatrix}
    \stauROT,
\end{equation}
with
\begin{align}
\begin{split}
    \label{eq:M-mstau-entries}
    \mLL^2 & =  m^2_{\tilde{L}_3} +  \left( - \frac{1}{2} + \sin^2{\theta_W}  \right)
\mZ^2 \cos{2\beta},\\
    \mRR^2 & =  m^2_{\tilde{E}_3} - \sin^2{\theta_W}  \mZ^2 \cos{2\beta}, \\
    \Xtau & = \Atau - \mu \TB \ .
\end{split}
\end{align}
Here, $m_{\tilde{L}_3}$ and $m_{\tilde{E}_3}$ are the left-handed and
right-handed stau soft-breaking masses and \Atau\ is the trilinear
coupling in the stau sector, $\mu$ the Higgs-higgsino mass parameter,
and $\TB=v_2/v_1$ the ratio of the two Higgs vacuum expectation
values. As indicated in (\ref{eq:stau-mass-matrix}), the stau mixing
matrix can be diagonalized by an orthogonal $2\times2$ matrix
\stauROT, parametrized by the stau mixing angle \thetastau,
\begin{equation}
  \label{eq:staurot}
      \stauROT =
    \begin{pmatrix}
      \cos{\thetastau} & \sin{\thetastau} \\
      -\sin{\thetastau} & \cos{\thetastau}
    \end{pmatrix},
\end{equation} 
and the stau mass eigenvalues squared are given by
\begin{equation}
   \label{eq:stau-masses}  
m_{\supertau_{1},\supertau_{2}}^2 = \mtau^2 + \frac{1}{2} \left[\mRR^2+\mLL^2 \mp
      \sqrt{(\mLL^2-\mRR^2)^2 + 4\mtau^2\Xtau^2 } \right].
\end{equation}
By convention \stauROT is chosen such that \stau is the lighter of the
two eigenstates. Imposing this requirement and choosing $0\leq
\thetastau < \pi $, the mixing angle is determined by
\begin{equation}
    \label{eq:thetastau}
    \tan{2 \thetastau}  =  \frac{ 2 \mtau \Xtau }
    {  \mLL^2 - \mRR^2 } \ , \qquad 
\text{and}\qquad
    \sin{2\thetastau} = \frac{ 2 \mtau \Xtau}{\mstau^2 - \mstautwo^2} .
\end{equation}
For a mixing angle of $\thetastau=\pi/2 \, (0)$, \stau is purely
right(left)-handed, while maximal mixing occurs for $\thetastau=\pi/4$
and $3\pi/4$.

Equation~(\ref{eq:thetastau}) gives a direct relation between the
mixing angle, the off-diagonal parameter $X_{\tau}$ and the gauge
eigenstates ($\supertau_L,\supertau_R$) or mass eigenstates
($\stau,\supertau_2$).  Thus, in section \ref{sec:production} and
partly in section \ref{sec:phenolonglived}, we use $\mstau$ and
$\thetastau$ as input parameters, together with $\Atau$, $\mu$, and
$\TB$, to determine the heavier stau mass $\mstautwo$ and then compute
$m_{\tilde{L}_3}$ and $m_{\tilde{E}_3}$
from~(\ref{eq:stau-mass-matrix}).  Furthermore, by $SU(2)_L$
invariance, $m_{\tilde{L}_3}$ then sets the mass of the tau-sneutrino,
$m_{\tausneutrino}$.

\subsection{Stau-Higgs couplings}
\label{app:stauhiggs_couplings}

In the minimal flavor violating MSSM, the sfermions couple directly to
the Higgs fields via dimensionful parameters. The stau-Higgs couplings
are given by \cite{Haber:1997dt}
\begin{equation}
    \label{eq:L-stau-stau-Higgses}
    \Lagrangian_{\tilde{\tau}\tilde{\tau}\mathcal{H}} = \frac{g}{\mW} \sum_{I,J= \L , \R}
\supertau^*_{I} \,
    \couptriLR{\supertau^*_{I}}{\supertau_{J}}{ \mathcal{H}} \,
    \supertau_{J} \,  \mathcal{H} \, , 
\end{equation}
where $\mathcal{H}$ stands for any of the neutral Higgs and Goldstone
bosons, $\mathcal{H}=h^0,H^0,A^0,G^0$. In the basis of the gauge
eigenstates $(\supertau_{L},\supertau_{R})$, the reduced coupling
$\couptriLR{\supertau^*_{I}}{\supertau_{J}}{ h^0}$ reads
\begin{align}
    \label{eq:lighthiggs-stau-stau-couplings}
     \couptriLR{\tilde{\tau}^*_I}{\tilde{\tau}_J}{\hhiggs} & = 
    \begin{pmatrix}
      \displaystyle
      -\frac{\cos2\theta_W}{2} \mZ^2 \sapb + \mtau^2
      {\frac{\sa}{\cb}} 
      &
      \displaystyle
      \frac{\mtau}{2} \left( \Atau \frac{\sa}{\cb}
        + \mu \frac{\ca}{\cb} \right) 
      \\ 
      \displaystyle
      \frac{\mtau}{2} \left( \Atau
        \frac{\sa}{\cb} + \mu \frac{\ca}{\cb} \right) 
      & 
      \displaystyle
      - \sin^2\theta_W \mZ^2
      \sapb + \mtau^2 {\frac{\sa}{\cb}}
    \end{pmatrix}\ ,
\end{align}
and $\couptriLR{\tilde{\tau}^*_I}{\tilde{\tau}_J}{\Hhiggs}$ can be
obtained upon the replacement $\alpha \rightarrow \alpha-\pi/2$, where
$\alpha$ is the Higgs scalar mixing angle. Here and below, the
shorthand notation $c_{\gamma}\equiv\cos\gamma$ and
$s_{\gamma}\equiv\sin\gamma$ (with $\gamma=\alpha,\beta,\thetastau$)
is used. The reduced coupling for the CP-odd Higgs boson $A^0$ reads
\begin{align}
    \label{eq:cpoddhiggs-stau-stau-couplings}
     \couptriLR{\tilde{\tau}^*_I}{\tilde{\tau}_J}{A^0} & = 
    \begin{pmatrix}
      \displaystyle
                0
      &
      \displaystyle
      +i~\frac{\mtau}{2} \left( \Atau \TB
        + \mu \right) 
      \\ 
      \displaystyle
      -i~\frac{\mtau}{2} \left( \Atau
        \TB + \mu \right) 
      & 
      \displaystyle
                0
    \end{pmatrix}\ ,
\end{align}
and $\couptriLR{\tilde{\tau}^*_I}{\tilde{\tau}_J}{G^0}$ can be
obtained upon the replacement: $\Atau\TB+\mu\rightarrow\mu\TB-\Atau$.

We are particularly interested in the couplings between the lighter
mass eigenstate \stau and the CP-even Higgs bosons \hhiggs and
\Hhiggs.  These can be found by diagonalizing the coupling matrix
in~(\ref{eq:lighthiggs-stau-stau-couplings}) with the orthogonal
matrix \stauROT defined in~(\ref{eq:staurot}):
  \begin{align}
    \label{eq:higgses-stau-stau-couplings-PHYS}
     \couptri{\stau^*}{\stau}{\hhiggs}& =
      \displaystyle
      \left( -{\frac{1}{2}} \csqth + \sin^2\theta_W \ctwoth \right) \mZ^2 \sapb + \mtau^2
      {\frac{\sa}{\cb}} + 
       \frac{\mtau}{2} \left( \Atau \frac{\sa}{\cb}
        + \mu \frac{\ca}{\cb} \right) \stwoth \ .
\end{align}
The coupling $\couptri{\stau^*}{\stau}{\Hhiggs}$ can be read
from~(\ref{eq:higgses-stau-stau-couplings-PHYS}) after the replacement
$\alpha \rightarrow \alpha-\pi/2$. Thus, the SUSY parameters $\mu$ and
\Atau enter proportional to $\sin 2\thetastau$ and the coupling peaks
at $\thetastau=\pi/4$. This holds analogously for the
$\stau^*\stau\Hhiggs$ coupling, whereas the diagonal coupling to the
CP-odd Higgs boson \Ahiggs vanishes: \couptri{\stau^*}{\stau}{A^0}=0.

\section{Resummation in the bottom sector}
\label{app:resum_b}

The Higgs sector in the MSSM corresponds to a type-II two-Higgs
doublet model, where the down-type quarks couple to $H_1$ and the
up-type quarks to $H_2$. After spontaneous symmetry breaking, the up-
(down-)type quarks gain masses by coupling to the non-zero $H_2$
($H_1$) vacuum expectation values $v_2$ ($v_1$). At tree-level, the
bottom-quark mass $m_b$ is given by
\begin{equation}
m_b = \lambda_b v_1,
\label{eq:treeMBLB}
\end{equation}
where $\lambda_b$ is the $\bbar H_1$ Yukawa coupling.  However
radiative corrections induce an effective $\bbar H_2$ coupling that
can significantly alter the tree-level
relation~\cite{Hall:1993gn,Hempfling:1993kv,Carena:1994bv,Pierce:1996zz,Carena:1999py}.
These higher-order contributions do not decouple at low energies and
are enhanced by a factor $\tan \beta =v_2 /v_1$,
\begin{align}
\begin{split}
m_b &= \lambda_b v_1 + \Delta \lambda_b v_2 
= \lambda_b v_1 \, \left( 1+ \frac{\Delta \lambda_b}{\lambda_b} \TB \right)
\\
& \equiv \lambda_b v_1 \left( 1 + \Delta m_b \right).
\end{split}
\end{align}
As shown in ref.~\cite{Carena:1999py}, the leading $\tan \beta$
enhanced terms can be resummed to all orders and easiest be
implemented by using an effective bottom-quark Yukawa coupling
$\lambda_b^{\text{eff}} \equiv m_b^{\text{eff}}/v_1$.  The dominant
contributions to $\Delta m_b$ arise from gluino-sbottom and
chargino-stop loops,
\begin{align}
  \label{eq:eff2}
  \Delta m_b &=  \frac{2\alphas}{3\pi} M_{3} \mu \tan\beta\;
  I(m_{\Sb_1}, m_{\Sb_2}, m_{\gluino}) + \frac{\lambda_t^2}{16\pi^2} \mu A_t \tan \beta\;
  I(m_{\St_1}, m_{\St_2},\mu),
\end{align}
with the gluino mass $M_{3}$, the lighter (heavier) sbottom mass
$m_{\Sb_1}$ ($m_{\Sb_2}$) the trilinear coupling $A_t$ in the stop
sector, the lighter (heavier) stop mass $m_{\St_1}$ ($m_{\St_2}$), and
the loop function
\begin{align}
  I(a,b,c) &= \frac{1}{(a^2-b^2)(b^2-c^2)(a^2-c^2)}
  \left[ a^2b^2 \log\frac{a^2}{b^2} + b^2c^2 \log\frac{b^2}{c^2} +
    c^2a^2 \log\frac{c^2}{a^2} \right]. 
\end{align}
Further neutralino-sbottom contributions are proportional to the weak
coupling $g^2$ and subdominant only.

Here we follow refs.~\cite{Heinemeyer:2004xw,Germer:2011an} and use a
$\drbar$ definition of the effective mass to take large logarithms
from the running Yukawa coupling into account.  The effective Yukawa
coupling $\lambda_b^{\text{eff}}$ is then defined as follows,
\begin{equation}
  \label{eq:eff1}
\lambda_b^{\text{eff}}  = \frac{1}{v_1}\frac{m_b^{\drbar}(\mu_R) + m_b \, \Delta
m_b}{1+\Delta m_b} \equiv \frac{m_b^{\drbar,\text{eff}}}{v_1}  \, ,
\end{equation}
where the second term in the numerator is necessary to avoid double
counting of the resummed terms.  The $\drbar$ bottom mass at the
renormalization scale $\mu_R$ can be obtained from the on-shell
bottom-quark mass, $m_b^{\text{OS}}$, and the UV-finite parts of the
bottom-quark self-energy (here in Lorentz decomposition),
\begin{equation}
 m_b^{\drbar}(\mu_R) =m_b^{\text{OS}} +\frac{m_b}{2} \left ( 
\Sigma^{\text{fin.}}_{L}(m_b)+ 
 \Sigma^{\text{fin.}}_{R}(m_b) + 2\,  \Sigma^{\text{fin.}}_{S}(m_b) \right ),  
\end{equation}
with 
\begin{equation}
m_b^{\text{OS}} = m_b^{\msbar}(M_Z) \, b^{\text{shift}},
\qquad
 b^{\text{shift}} = 1+ \frac{\alpha_s}{\pi} \left(\frac{4}{3} -\log
\frac{[m_b^{\msbar}(M_Z)]^2}{M_Z^2} \right).
\end{equation}

Further $\tan\beta$ enhancement effects arise in the trilinear
couplings involving Higgs--bottom interactions. They can be resummed
and taken into account by modifying the $\bbar \mathcal{H}$ coupling
$g_{\bbar \mathcal{H}}$. The combined effect of the resummation in the
relation between $\lambda_b$ and $m_b$ and of the resummation in the
Higgs--bottom vertices is accounted for by performing the following
substitutions in the couplings,
\begin{align}
\begin{split}
  g_{bbh^0} &\rightarrow g_{bbh^0}\big|_{\lambda_b \rightarrow \lambda_b^{\text{eff}}} 
        \left(1-\frac{\Delta m_b}{\tan\beta\tan\alpha}\right) , 
  \qquad
   g_{bbA^0} \rightarrow g_{bbA^0}\big|_{\lambda_b \rightarrow \lambda_b^{\text{eff}}} 
        \left(1-\frac{\Delta m_b}{\tan^2\beta}\right ) ,
\\
  g_{bbH^0} &\rightarrow g_{bbH^0}\big|_{\lambda_b \rightarrow \lambda_b^{\text{eff}}}
        \left(1+\Delta m_b\frac{\tan\alpha}{\tan\beta}\right) ,
  \qquad
  g_{bbG^0} \rightarrow g_{bbG^0}\,.
\end{split}
\end{align}
The coupling involving the Goldstone boson $G^0$ is not modified since
the contribution from the vertex corrections exactly compensates the
contribution of the bottom-Yukawa coupling resummation.

\bibliographystyle{JHEP}

\providecommand{\href}[2]{#2}\begingroup\raggedright\endgroup


\end{document}